\documentclass[prd,superscriptaddress,nofootinbib,twocolumn,floatfix]{revtex4-2} 
\usepackage[utf8]{inputenc}

\usepackage{graphicx}
\usepackage{amsmath}
\usepackage{amssymb}
\usepackage{amsfonts}
\usepackage{bm}
\usepackage{booktabs}
\usepackage{multirow}
\usepackage{url}
\usepackage{tabularx}
\usepackage{array}
\usepackage{siunitx}
\usepackage{xcolor}
\usepackage{placeins}
\usepackage[colorlinks=true, allcolors=blue]{hyperref}
\newcommand{\orcid}[1]{\texorpdfstring{\begingroup
  \hypersetup{hidelinks}\href{https://orcid.org/#1}{\includegraphics[width=10pt]{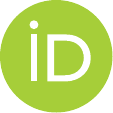}\,}
\endgroup}{}}
\definecolor{osu}{HTML}{ba0c2f}

\begin{document}

\title{Exact spherical-wave forward model for radio reflection from stratified media and implications for the anomalous-polarity events observed by ANITA}

\author{Paramita Dasgupta\,\orcid{0000-0002-9559-4803}}
\email{dasgupta.80@osu.edu}

\affiliation{Center for Cosmology and AstroParticle Physics (CCAPP), \href{https://ror.org/00rs6vg23}{\color{osu}Ohio State University}, Columbus, OH 43210}
\affiliation{Department of Physics, \href{https://ror.org/00rs6vg23}{\color{osu}Ohio State University}, Columbus, OH 43210}
\affiliation{Department of Astronomy, \href{https://ror.org/00rs6vg23}{\color{osu}Ohio State University}, Columbus, OH 43210}

\date{\today}

\begin{abstract}
Radio detection of ultra-high energy particles (energies $\gtrsim10^{18}$~eV) relies on the propagation and reflection of broadband radio pulses at boundaries between natural media. In the Sommerfeld--Weyl treatment, a spherical wave is decomposed into plane-wave components and their reflection from a single homogeneous interface is calculated exactly. We extend that treatment to an arbitrary number of laterally uniform spherical layers. Both the spherical-wave decomposition of the source and the spherical geometry of the boundary are retained, while the reflection and transmission coefficients of the plane-wave components are replaced by the exact characteristic-matrix coefficients of the layered medium, evaluated at the local incidence angle on the spherical boundary. When the layer contrast is removed, the formalism recovers the single-boundary result to machine precision, and it reproduces the published spherical-surface reflectivity calculation to better than $1.1\%$ at ten HiCal-2 elevation angles, with a mean deviation of $0.6\%$. As a check of the implementation, we reproduce the measured HiCal-1 reflected pulses from their measured direct partners, with a best signed correlation of $0.83$ and a median of $0.70$ across $106$ pairs, of which $101$ show the expected polarity inversion. Applied to the six reported ANITA anomalous-polarity event geometries, the buried-layer refractive index required for a sign change of the reflection coefficient ranges from $1.68$ at the steepest event to $3.8$--$5.4$ at the four near-horizon events. The full waveform calculation gives no non-inverted reflected pulse at any of these angles, showing that shallow, laterally uniform firn layering does not account for the polarity of the anomalous ANITA events. Because the formalism depends only on the complex refractive index of the medium, it applies more generally to isotropic, nonmagnetic stratified media, including ice, lunar regolith, and conducting layers.
\end{abstract}

\maketitle

\section{Introduction}
\label{sec:intro}

Radio detection of ultra-high energy (UHE) neutrinos and cosmic rays, here referring to primaries above roughly $10^{18}$~eV, relies on coherent radio emission from particle cascades~\cite{Askaryan1962,Askaryan1965,Huege2016}. For UHE cosmic rays (UHECRs), radio emission from air showers is dominated by the geomagnetic mechanism~\cite{HuegeFalcke2005,Scholten2008,deVries2010,Huege2016} and can be detected after reflection from the Antarctic ice surface. In-ice neutrino interactions produce Askaryan emission~\cite{Askaryan1962,Askaryan1965,Saltzberg2001,Gorham2007}, which can be observed after refraction through the ice--air boundary. We focus here on the reflection of radio signals from spherical boundaries in both single-boundary and layered media. The same formalism also gives the corresponding transmission coefficient, but transmission from in-medium sources is beyond the scope of this work.

The Antarctic Impulsive Transient Antenna (ANITA) provides the main high-altitude data set for studying reflected UHECR signals. At an altitude of about $37$~km, ANITA viewed $\mathcal{O}(10^{6})~\mathrm{km}^{2}$ of Antarctic ice with dual-polarization antennas over roughly $200$--$1200$~MHz~\cite{Gorham2009}. ANITA observed UHECR air showers through their geomagnetic radio emission reflected from the ice surface~\cite{Hoover2010,Schoorlemmer2016}. For the horizontally polarized component, reflection from the higher-index ice reverses the pulse polarity. The non-inverted polarity observed in a small number of below-horizon events therefore remains unexplained~\cite{ANITA2016,ANITA2018,ANITA2021}. These anomalous-polarity events motivate a more complete treatment of the reflected field, including the effects of subsurface structure. We apply the model developed here to the two steep ANITA-I/III events and the four near-horizon ANITA-IV events reported in Refs.~\cite{ANITA2016,ANITA2018,ANITA2021}.

Surface reflectivity is also an important component of balloon-borne UHECR measurements, as it enters directly into the energy reconstruction of the primary cosmic ray. ANITA flew with the High-Altitude Calibration (HiCal) radio-frequency (RF) transmitters, which emitted calibration pulses that were recorded by the ANITA payload both directly and after reflection from the Antarctic surface. The reflected-to-direct power ratio, denoted by $r/d$, provides a measure of the surface power reflection coefficient, with each direct-reflected (D-R) pulse pair providing one measurement at its corresponding reflection geometry. HiCal-1 flew with ANITA-3~\cite{Gorham2017}, while HiCal-2 flew with ANITA-4 during the 2016--2017 austral flight season and provided data for detailed studies of Antarctic surface reflectivity over a range of elevation angles, measured above the local surface at the reflection point~\cite{ProhiraHiCal2,Prohira2018}. The spherical single-boundary calculation was developed and validated using HiCal data~\cite{Prohira2018} and subsequently refined in Refs.~\cite{DasguptaThesis2020,DasguptaJain2021}. We use the HiCal-2 reflectivity results from Ref.~\cite{Prohira2018} to validate the single-boundary limit and a subset of 106 HiCal-1 direct/reflected pulse pairs~\cite{Gorham2017}, previously used in Refs.~\cite{DasguptaThesis2020,DasguptaJain2021}, to validate the reflected-pulse calculation in Sec.~\ref{sec:hical}.

The HiCal geometry cannot be adequately described by a flat, smooth Antarctic surface. At shallow incidence angles, the region coherently contributing to the reflected field can extend over hundreds of meters, making both Earth curvature and the root-mean-square (rms) surface-height variations relevant to the signal measured at the high-altitude ANITA payload~\cite{Gorham2017,Prohira2018}. The spherical single-boundary treatment of Refs.~\cite{Prohira2018,DasguptaJain2021} accounts for these effects and has been used to reproduce both the HiCal-2 reflectivity and reflected-pulse waveforms~\cite{DasguptaThesis2020,DasguptaJain2021}.

The same boundary response is relevant to in-ice radio-neutrino detectors. The Askaryan Radio Array (ARA) is an in-ice radio array at the South Pole consisting of five independent stations, with antennas deployed at depths of up to about $200$~m~\cite{Allison2012ARA,ARA2021Calibration}. Its event reconstruction and neutrino searches account for both refracted and reflected propagation through the ice~\cite{ARA2020TwoStation,KH_PA_analysis,ARAfulllivetime2023}, and its fifth station includes a low-threshold phased array designed to improve sensitivity to low signal-to-noise-ratio (SNR) signals~\cite{DasguptaARENA2024,ARA2025NeutrinoSearch}. Askaryan emission from cosmic-ray showers in ice has also been observed with ARA~\cite{Alden2026}. The Payload for Ultrahigh Energy Observations (PUEO), the balloon-borne successor to ANITA, completed its Antarctic flight in 2025--2026 and achieves improved sensitivity with a phased-array trigger~\cite{PUEO2021,PUEOtrigger2026}. An accurate treatment of the boundary response is therefore relevant to both balloon-borne and in-ice radio detectors.

Antarctic firn provides a natural example where the single-boundary treatment must be extended. Its radio-frequency refractive index rises from about $1.3$ near the surface toward the deep-ice value $n\simeq1.78$~\cite{Kovacs1995}. In the South Pole parameterization used in ARA analyses, the near-surface index deficit has an $e$-folding depth of about $50$~m~\cite{ARA2020TwoStation,KH_PA_analysis,ARAfulllivetime2023}. Smooth firn profiles are used in ray-tracing reconstructions, while discrete density contrasts from wind crusts, ice lenses, refrozen melt layers, and buried surfaces can produce additional internally reflected components. Such structures have been proposed as possible contributors to the anomalous-polarity ANITA events~\cite{Shoemaker2020}. A subsequent experimental study using ANITA and HiCal data constrained these scenarios~\cite{Smith2021}. That study tested the proposal against measured waveform properties, while the model developed here predicts the reflected field of a given layered column from first principles, so that the polarity condition can be evaluated as a function of elevation, buried-layer index, and frequency. An exact spherical-wave treatment of a rough spherical surface overlying laterally uniform stratified media therefore provides a way to study these effects while retaining the geometry of the existing single-boundary calculation.

Several established approaches are useful for describing radio-wave propagation in stratified media, with each emphasizing different aspects of the problem. Ray tracing through a smooth $n(z)$ profile provides the propagation path and arrival time, while parabolic-equation methods describe wave propagation through depth-dependent media, including near-field and shadow-zone effects~\cite{Prohira2021}. For reflection from layered media, the characteristic-matrix method provides the coherent reflection and transmission coefficients of the layers~\cite{Wait1956,Wait1970,Chew1995,BornWolf,Abeles1950}. Here, we combine the exact spherical-wave decomposition with the characteristic-matrix treatment of layered media to obtain an exact forward model for reflection from a spherical, stratified surface.

Related spherical-wave reflection problems have also been studied in acoustics. The Weyl decomposition has been used to describe reflection of spherical waves from layered media~\cite{BrekhovskikhGodin}, including rough interfaces through tangent-plane constructions~\cite{Pinson2015}. Those treatments are scalar, so a single coefficient fixes the reflection of each plane-wave component. The electromagnetic problem is a vector one. Each plane-wave component must be resolved onto the local $s$ (perpendicular to the plane of incidence) and $p$ (parallel to the plane of incidence) polarization directions at its point of incidence, reflected with the corresponding polarization-dependent coefficient, and projected back onto the observed field direction. We construct this from Maxwell's boundary conditions at every interface, retaining the full multiple-reflection response of the layered medium and evaluating its coefficient at the local incidence angle of each component on a rough spherical boundary.

We extend the validated spherical single-boundary formalism~\cite{Prohira2018,DasguptaJain2021} to laterally uniform stratified media. For each plane-wave component of the spherical wave, the Fresnel reflection coefficient is replaced by the characteristic-matrix reflection coefficient of the layered medium. The resulting coefficient includes the coherent contributions from the uppermost boundary and from internal reflection and transmission through the spherical layers. The formalism reduces to the single-boundary result when the layer contrast is removed and retains the spherical geometry and surface-roughness treatment of Ref.~\cite{Prohira2018}. Because the formulation depends only on the complex refractive index of the layers, it also applies to other isotropic, nonmagnetic stratified media, including lunar regolith, planetary ice, and lossy or conducting subsurface layers~\cite{CoRaLS,RomeroWolf2015,REASON2024,Bruzzone2013}.

This paper has four main results. First, we develop a spherical-wave forward model for reflection of electromagnetic radiation from spherical stratified media and show that it recovers the validated single-boundary calculation at machine precision when the layer contrast is removed. Second, we determine the elevation-dependent conditions under which shallow firn layering can produce a polarity reversal for the ANITA geometries considered, providing a quantitative test of the proposed shallow-layering explanation for the anomalous-polarity events observed by ANITA. Third, we test the widely used specular factorization against the full angular calculation of the total reflected field. We find that, for balloon-altitude geometries, the two agree at the $0.2\%$ level, while the full angular integral becomes important for sources close to the boundary. Fourth, we validate the reflected-pulse calculation directly against measured HiCal-1 direct/reflected pairs, finding the expected polarity inversion in $101$ of $106$ cases and reproducing the measured reflected pulses with a best signed correlation of $0.83$.

Section~\ref{sec:formalism} reviews and validates the single-boundary formalism. Section~\ref{sec:layered} derives its extension to stratified media. Section~\ref{sec:validity} examines the specular factorization, and Sec.~\ref{sec:results} presents the numerical results. Section~\ref{sec:applications} discusses applications of this general framework beyond Antarctic ice, including planetary radar sounding and the lunar-reflection problem in 21-cm cosmology, and Sec.~\ref{sec:summary} summarizes the results.

\section{Spherical-wave reflection from a spherical boundary}
\label{sec:formalism}

\begin{figure*}[!t]
\centering
\includegraphics[width=0.750\linewidth]{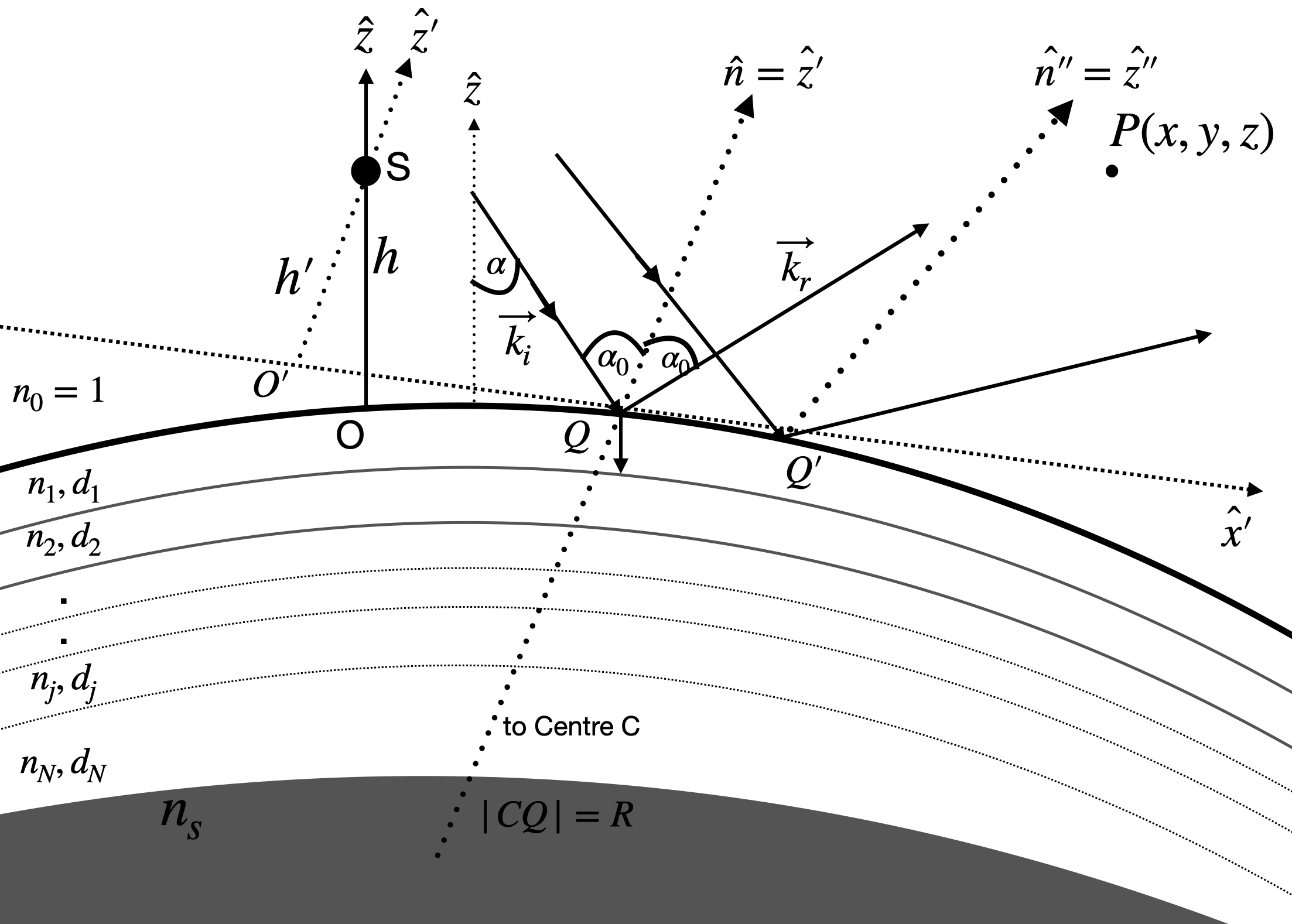}
\caption{Geometry of the reflection calculation for a spherical boundary
(following Fig.~12 of Ref.~\cite{Prohira2018}). The spherical surface has center $C$ (not shown) and radius $R$. A Hertz dipole source is located at $S$, at height $z_0=h$ above the surface point $O$, and the reflected field is evaluated at the observer position $P(x,y,z)$. The spherical wave is expanded by the Sommerfeld--Weyl identity, Eq.~(\ref{eq:weyl}), into plane-wave components. In the plane shown, one component has incident wave vector
$\vec{k}_i$ at angle $\alpha$ from the downward vertical ($-\hat{z}$) and
intersects the surface at $Q$. The local normal at point $Q$ is $\hat{n}=\hat{z}'$, and the local tangent plane defines the incidence angle $\alpha_0$ of Eq.~(\ref{eq:alpha0}). The quantity $h'$ is the perpendicular distance from
$S$ to this tangent plane. A different plane wave intersects the surface at
$Q'$, where the local normal $\hat{n}''=\hat{z}''$ and tangent plane are
different, so its local incidence angle is evaluated separately. The reflected
field is therefore described locally by plane waves at each incident point,
but is not a single global plane wave. Beneath the surface are concentric
layers, the $j$th with refractive index $n_j$ and thickness $d_j$, above a
semi-infinite substrate $n_s$. Maxwell's boundary conditions are satisfied at
each incident point, and the reflection and transmission coefficients are
evaluated there. The azimuth $\beta$ of each plane wave (not shown) rotates
the plane of incidence about the local normal.}
\label{fig:geometry}
\end{figure*}

We first review the framework for calculating the reflection of a spherical electromagnetic wave radiated by a dipole source above a spherical boundary between two homogeneous media, following Refs.~\cite{Prohira2018,DasguptaJain2021}. We also validate the numerical implementation against the spherical-surface reflectivity calculation of Ref.~\cite{Prohira2018}. Section~\ref{sec:layered} extends this framework to a stratified medium by replacing the single-boundary reflection coefficient with the corresponding layered-medium coefficient. The framework is motivated by radio waves produced by cosmic ray interactions in the atmosphere and observed by balloon-borne experiments such as ANITA and PUEO. The spherical wave is decomposed into plane-wave components, and each component is reflected from the spherical surface using the Fresnel coefficients evaluated at its local incidence angle. Each component is treated as locally planar at the surface, while the spherical geometry of the source wave is retained in the angular decomposition and that of the boundary in the coherent sum that gives the total reflected and transmitted fields.

We use the following terminology throughout this paper. The \emph{boundary} is the upper surface where air meets the underlying medium. A \emph{layer} is a homogeneous slab beneath the boundary, specified by its refractive index and thickness. The ordered set of layers is the \emph{stack}, and the semi-infinite medium below the deepest layer is the \emph{substrate}. In this section, the stack is empty, so the boundary separates air directly from the substrate. We take the substrate to be Antarctic ice and use this single-boundary case to validate the calculation against the HiCal data~\cite{Prohira2018}.

The relevant geometry is shown in Fig.~\ref{fig:geometry}. The calculation involves two independent spherical geometries. The first defines the source field. A Hertz dipole radiates a spherical wave, which Eq.~(\ref{eq:weyl}) decomposes into plane-wave components. The second is the spherical reflecting surface, a convex sphere of radius $R$ centered at $C$, whose local geometry is described by the tangent plane at the point where each plane-wave component intersects the surface. 

We consider the unprimed $(x,y,z)$ axes of Fig.~\ref{fig:geometry} as the global coordinate system. The origin $O=(0,0,0)$ lies on the surface directly below the source $S$. In this coordinate system, the source is at $S=(0,0,z_0=h)$ and the observer is at $P=(x,y,z)$. At each incidence point $Q$, we introduce a local coordinate system whose $\hat{z}'$ axis aligns with the local surface normal. Boundary conditions are applied in this local coordinate system, while the angular integration and the final reflected and transmitted fields are computed in the global coordinate system, following Ref.~\cite{Prohira2018}.

The sources considered here, including HiCal calibration pulses~\cite{Gorham2017,ProhiraHiCal2} and radio emission from UHECR air showers~\cite{Hoover2010,Schoorlemmer2016,Huege2016}, are broadband. Because Maxwell's equations and the associated boundary conditions are linear, each frequency component can be treated independently. The total reflected and transmitted fields are then computed by calculating the field at each frequency and summing over the entire frequency band. All quantities below therefore refer to a single frequency $f$, with $\omega=2\pi f$ and an implicit time dependence $e^{-i\omega t}$.

Above the spherical surface, we express the total field as the sum of the direct and reflected fields, which we write in terms of Hertz potentials,
\begin{equation}
\vec{\Pi}=\vec{\Pi}_{\rm dir}+\vec{\Pi}_{\rm ref}.
\label{eq:hertztotal}
\end{equation}
For a horizontal dipole at $S$, polarized along $\hat{y}$, the direct potential is
\begin{equation}
\vec{\Pi}_{\rm dir}=\frac{e^{ikr}}{4\pi\epsilon r}\,\hat{y}.
\label{eq:hertz}
\end{equation}
Here $k=2\pi f/c$ is the free-space wavenumber, $c$ is the speed of light in vacuum, $\epsilon$ is the permittivity of the air above the surface, and 
\begin{equation}
r=\sqrt{x^2+y^2+(z-z_0)^2}
\label{eq:rdist}
\end{equation}
is the distance from the source to the field point. The electric and magnetic fields follow from
\begin{equation}
\vec{E}=\vec{\nabla}(\vec{\nabla}\cdot\vec{\Pi})+k^2\vec{\Pi},
\qquad
\vec{H}=\frac{k^2}{i\omega\mu}\,\vec{\nabla}\times\vec{\Pi},
\end{equation}
with $\mu$ the permeability~\cite{Stratton,Prohira2018}. The reflected term in Eq.~(\ref{eq:hertztotal}) is determined by the boundary conditions. Following Refs.~\cite{Prohira2018,DasguptaJain2021}, we construct it here.

A spherical wave cannot be reflected from the boundary by applying Fresnel coefficients directly, because those coefficients are defined for plane waves. The spherical wave can, however, be written exactly as a superposition of plane waves through the Sommerfeld--Weyl representation~\cite{Weyl1919,Sommerfeld1949,Stratton},
\begin{equation}
\frac{e^{ikr}}{r}=\frac{ik}{2\pi}\int_{0}^{2\pi}\!\!\!\int_{0}^{\pi/2-i\infty}
\!\!\!\!e^{i\Phi_0}\,\sin\alpha\,d\alpha\,d\beta ,
\label{eq:weyl}
\end{equation}
with phase
\begin{equation}
\Phi_0=k\big[x\sin\alpha\cos\beta+y\sin\alpha\sin\beta+(z_0-z)\cos\alpha\big].
\label{eq:weylphase}
\end{equation}
The boundary problem is then solved component by component. Each plane wave component is matched to the local tangent plane at its point of incidence on the spherical surface. The appropriate reflection and transmission coefficients are obtained from Maxwell's boundary conditions in that local frame, and the reflected contributions are then added coherently in the field integral. Following Refs.~\cite{Prohira2018,DasguptaJain2021}, $\alpha$ is measured from the downward vertical $-\hat{z}$, so $\alpha=0$ corresponds to a vertically down-going plane wave, and $\beta$ is the azimuth about $\hat{z}$. For a flat surface, $\alpha$ is also the incidence angle. For the spherical surface, the local incidence angle is instead $\alpha_0(\alpha)$, defined below. Equation~(\ref{eq:weyl}) is an identity and therefore introduces no approximation.
The representation is valid for $0\le z\le z_0$.

The upper limit $\pi/2-i\infty$ of the polar angle integral in Eq.~(\ref{eq:weyl}) carries the contour into imaginary values of $\alpha$. These are the evanescent components, which decay rather than propagate away from the source. They are negligible in the far-field balloon geometry and were omitted in Ref.~\cite{Prohira2018}. We quantify this contribution in Sec.~\ref{sec:layered}, and treat separately the evanescent behavior inside the medium, which the reflection and transmission coefficients for layered media retain automatically.

This treatment is valid for the far zone, $r\gg\lambda$, where $\lambda=c/f$ is the free-space wavelength. The near-field terms of the dipole field are then negligible, and the plane wave components computed below propagate over the distances of interest. For the balloon geometry, $r$ is tens of kilometers while $\lambda$ is of order half a meter, so this condition is well satisfied.

Each component of the Sommerfeld--Weyl integral is specified by its propagation direction. In the global coordinate system this incident direction is
\begin{equation}
\hat{k}_i=(\sin\alpha\cos\beta,\ \sin\alpha\sin\beta,\ -\cos\alpha).
\label{eq:khat}
\end{equation}
On a flat surface, $\alpha$ would also be the incidence angle. But on a spherical surface, the incident angle is defined by $\alpha_0(\alpha)$ for each plane wave of Eq.~(\ref{eq:weyl}). The wave reaches the surface at $Q$, whose normal $\hat{n}=\hat{z}'$ is tilted relative to the global $\hat{z}$ axis (Fig.~\ref{fig:geometry}), so the relevant incidence angle is the local angle $\alpha_0$ measured from that normal. For a spherical surface the relation is fixed by geometry~\cite{Prohira2018},
\begin{equation}
\sin\alpha_0=\frac{R+h}{R}\,\sin\alpha ,
\label{eq:alpha0}
\end{equation}
and is rederived in Appendix~\ref{app:geometry}. Thus $\alpha_0>\alpha$ for a convex surface, while $\alpha_0\to\alpha$ as $R\to\infty$.

Equation~(\ref{eq:alpha0}) also determines which components of the angular spectrum reach the spherical surface. An incident component must satisfy $\sin\alpha\le R/(R+h)$, or equivalently $\alpha_0\le90^\circ$. Components outside this range do not intersect the surface and contribute only to the direct field in Eq.~(\ref{eq:hertztotal}). Throughout this work, elevation is defined as $90^\circ-\alpha_0$, measured above the local tangent plane at the point where the plane wave component strikes the spherical surface. The specular point is where the path $S\to Q\to P$ is stationary, and the region around it dominates the coherent reflected field.

The reflected and transmitted wave vectors corresponding to an incident plane wave with $\hat{k}_i$ given by Eq.~(\ref{eq:khat}) in the global coordinate system are
\begin{align}
\hat{k}_r&=(\sin\theta_r\cos\beta,\ \sin\theta_r\sin\beta,\ \cos\theta_r),
\label{eq:khatr}\\[3pt]
\hat{k}_t&=(\sin\theta_t\cos\beta,\ \sin\theta_t\sin\beta,\ -\cos\theta_t),
\label{eq:khatt}
\end{align}
with
\begin{equation}
\theta_r=2\alpha_0-\alpha ,
\qquad
\theta_t=\alpha+\alpha_t-\alpha_0 .
\label{eq:thetart}
\end{equation}
Here $\alpha_t$ is the angle of refraction in the medium below the boundary. These relations follow from reflection and refraction at the local tangent plane at $Q$. The incident, reflected, and transmitted directions lie in the same plane of incidence, whose azimuth is $\beta$. In the flat-surface limit $\alpha_0\to\alpha$, these expressions reduce to the usual relations $\theta_r=\alpha$ and $\theta_t=\alpha_t$.

At each incident point $Q$, the reflected plane wave component is determined by solving the boundary conditions and evaluating the Fresnel coefficient at the local incidence angle. For a boundary between air, $n_0=1$, and a homogeneous medium of index $n_1$, taken here as $n_1=1.4$ for Antarctic ice following Ref.~\cite{Prohira2018}, the reflection coefficients are
\begin{align}
f_r^{\,s}&=\frac{n_0\cos\alpha_0-n_1\cos\alpha_t}
                {n_0\cos\alpha_0+n_1\cos\alpha_t},
\label{eq:fs}\\[3pt]
f_r^{\,p}&=\frac{n_1\cos\alpha_0-n_0\cos\alpha_t}
                {n_1\cos\alpha_0+n_0\cos\alpha_t},
\label{eq:fp}
\end{align}
where $\alpha_t$ is the transmitted angle in the local tangent plane (Fig.~\ref{fig:geometry}), fixed by Snell's law, $n_0\sin\alpha_0=n_1\sin\alpha_t$. The labels $s$ and $p$ denote electric field components perpendicular and parallel to the plane of incidence, respectively. The single-index approximation is replaced by the stratified treatment in Sec.~\ref{sec:layered}.

Curvature does not change the form of the Fresnel coefficients. They are still the plane wave coefficients evaluated at the local tangent plane. Curvature enters through the local incidence angle $\alpha_0$, the reflected direction $\hat{k}_r$, and the propagation phase along the path $S\to Q\to P$. Plane wave components that reach the surface at different points, shown as $Q$ and $Q^{\prime}$ in Fig.~\ref{fig:geometry}, therefore leave in different directions. The reflected field is locally a plane wave near each incident point, but it is not a plane wave globally.

The coefficients in Eqs.~(\ref{eq:fs})--(\ref{eq:fp}) do not require the lower medium to be denser than the upper medium, nor do they require either medium to be lossless. This formalism is general. It follows from Maxwell's boundary conditions and applies, with the same branch convention, to real or complex refractive indices in non-magnetic media. The refractive indices determine the coefficient values but do not change the underlying boundary-matching calculation. Three consequences of this are useful here.

First, for air-to-ice reflection, $n_1>n_0$, the $s$-polarized coefficient is
negative at every incidence angle:
\[
n_1\cos\alpha_t
=\sqrt{n_1^2-n_0^2\sin^2\alpha_0}
>n_0\cos\alpha_0 .
\]
Thus, the $s$-polarized reflected field is polarity-inverted at all incidence angles. The $p$-polarized coefficient, by contrast, changes sign at the Brewster angle, $\arctan(n_1/n_0)=54.5^\circ$ for $n_1=1.4$, so its sign must be evaluated separately from that of the $s$ coefficient. The two conventions also differ at normal incidence, where $f_r^{\,p}=-f_r^{\,s}$. The projection onto the horizontal field direction provides a compensating geometric sign, so the physical reflected field remains single-valued there (Appendix~\ref{app:fields}).

Second, if the source is in the denser medium and the wave travels toward the air boundary, the same equations describe total internal reflection. Above the critical angle
\[
\alpha_c=\arcsin(n_0/n_1),
\]
which is $45.6^\circ$ for ice, $\cos\alpha_t$ becomes imaginary and $|f_r|=1$. The reflected field acquires a phase shift, while the field in air is evanescent. The sign of the imaginary root is fixed by the $e^{-i\omega t}$ convention, with the decaying solution chosen so that $\mathrm{Im}\,k_z\ge0$, where $k_z$ is the component of the wavevector normal to the boundary. This regime is relevant to transmission from in-ice sources but is outside the far-field reflection calculation considered here.

Third, for a lossy or conducting medium, the refractive index is complex and the reflection coefficients are generally complex at all incidence angles. There is then no sharp critical angle, and the reflected field can be both
attenuated and phase shifted. Section~\ref{sec:applications} applies the same formalism to lunar regolith and seawater. The calculation itself is unchanged, with the properties of the medium entering only through its refractive index and conductivity. These properties therefore remain free parameters throughout the formalism.

A real ice surface is not smooth on the scale of the wavelengths considered here. We include surface roughness through the validated weight
\begin{equation}
\begin{gathered}
F_{\rm rough}=\exp\!\big[-2k^2\sigma_h^2(\rho_\perp)\cos^2\alpha_0^{\rm spec}\big],\\[2pt]
\sigma_h(\rho_\perp)=\sigma_h(L_0)\left(\rho_\perp/L_0\right)^{\mathcal{H}},
\end{gathered}
\label{eq:rough}
\end{equation}
where $\alpha_0^{\rm spec}$ is the local incidence angle of Eq.~(\ref{eq:alpha0}) evaluated at the specular point, denoted $\theta_z$ in Ref.~\cite{Prohira2018}. The superscript indicates that this is a fixed value, in contrast to $\alpha_0(\alpha)$, which varies across the angular integral in Eq.~(\ref{eq:weyl}). Here $\sigma_h$ is the root-mean-square (rms) surface height variation over a horizontal baseline, $L_0$ is the reference baseline at which it is quoted, $\mathcal{H}$ is the Hurst parameter, written as a script symbol to distinguish it from the H-Pol label, and $\rho_\perp$ is the transverse distance from the specular point in the tangent plane~\cite{Prohira2018,DasguptaJain2021}. We use $L_0=150$~m, $\sigma_h(150~\mathrm{m})=0.041$~m, and $\mathcal{H}=0.65$ from the validated HiCal-2 treatment~\cite{Prohira2018,DasguptaJain2021,DasguptaThesis2020}. A second parameter set with $\sigma_h(150~\mathrm{m})=0.051$~m and the same $\mathcal{H}$ gives comparable agreement with the HiCal-2 data~\cite{DasguptaThesis2020}.

Two features of Eq.~(\ref{eq:rough}) are important here. First, because $\alpha_0^{\rm spec}$ is measured from the local normal, the effect of surface roughness decreases toward grazing incidence. For a given height variation,
the resulting path difference becomes smaller as the incidence angle approaches $90^\circ$. Second, $F_{\rm rough}$ varies across the angular integral. It is unity at the specular point and decreases away from it as the transverse distance increases and larger surface-height variations are sampled. Roughness therefore further confines the reflected field to the specular region.

The framework can also accommodate other surface models without changing the rest of the calculation. Reference~\cite{DasguptaThesis2020} considered anisotropic roughness and local surface slopes motivated by Antarctic sastrugi, using the Reference Elevation Model of Antarctica (REMA)~\cite{REMA2019}. We leave these extensions to future work. Here we use the isotropic form of Eq.~(\ref{eq:rough}), with parameters fixed by the HiCal-2 reflectivity data~\cite{Prohira2018}.

Combining the spherical-wave decomposition, the local reflection coefficient obtained from the wave-optics boundary conditions for each plane wave, and the roughness factor, the horizontally polarized (H-Pol) reflected field at the observer is
\begin{equation}
E^{H}_{\rm ref}(f)=\frac{ik}{2\pi}
\int_{0}^{2\pi}\!\!\!\int_{0}^{\pi/2-i\infty}\!\!\!\!
W(\alpha,\beta)F_{\rm rough}\,e^{i\Phi_r}\sin\alpha\,d\alpha\,d\beta ,
\label{eq:eref}
\end{equation}
with the polarization and geometric weight that incorporates the spherical geometry of the boundary,
\begin{equation}
W(\alpha,\beta)=f_r^{\,s}\cos^2\!\beta
-f_r^{\,p}\cos\alpha\,\cos(2\alpha_0-\alpha)\sin^2\!\beta .
\label{eq:polweight}
\end{equation}
Here $\Phi_r=k(|SQ|+|QP|)$ is the geometric phase along the reflected path, where $|SQ|$ and $|QP|$ are the lengths of the two legs $S\to Q$ and $Q\to P$. In the flat limit it reduces to Eq.~(\ref{eq:weylphase}) with $(z_0-z)$ replaced by $(z_0+z)$, the usual image-source result. Equation~(\ref{eq:polweight}) follows from projecting the $\hat{y}$-oriented dipole field onto the local $s$ and $p$ directions at $Q$ and reflecting each component with its corresponding coefficient. Two angles must be distinguished in Eq.~(\ref{eq:eref}). The boundary coefficients are evaluated at each plane wave's own $\alpha_0(\alpha)$ inside the integral, while $\sigma_h$ varies with $\rho_\perp$ across the footprint. The roughness factor is the only factor evaluated at the fixed angle $\alpha_0^{\rm spec}$. Appendix~\ref{app:geometry} gives $|SQ|$ and the position of $Q$, and Appendix~\ref{app:fields} derives the projection and the corresponding transmitted field. The transmitted field is not needed for the balloon-borne reflection geometry considered here, where the formalism is validated against the HiCal-2 data~\cite{Prohira2018}, but is needed for sources inside the ice or regolith. It is therefore included in Appendix~\ref{app:fields}.

In the numerical spherical-boundary calculation, the real propagating part of the angular spectrum is restricted to the components that intersect the sphere of radius $R$, as given above. Here $R$ is a free parameter. For Antarctic ice, we take $R=R_\oplus=6371$~km, the mean radius of the Earth. For the far-field balloon geometry considered here, the evanescent part of the contour is negligible, as discussed above and in Sec.~\ref{sec:layered}.

The quantity compared with the Antarctic measurements is the H-Pol component, which matches the HiCal transmitter and the predominantly horizontal geomagnetic emission from Antarctic UHECR air showers. The relative strength of the two polarizations depends on the observation geometry. The geomagnetic emission is polarized along $\vec{v}\times\vec{B}$, where $\vec{v}$ is the shower propagation direction and $\vec{B}$ is the geomagnetic field. Because $\vec{B}$ is steeply inclined over Antarctica, the vertical component of the emission is suppressed~\cite{ANITA2018}. We nevertheless compute both $f_r^{\,s}$ and $f_r^{\,p}$, so the same calculation applies when the other polarization is significant.

Initial numerical estimates of the reflectivity as a function of incidence angle at an interface between two media were presented in Ref.~\cite{Gorham2017}, and the spherical-surface treatment including Earth curvature and surface roughness was developed in Ref.~\cite{Prohira2018}. We reproduce that calculation before extending it. Specifically, we recompute the reflected-to-direct power ratio $r/d$ for Antarctic ice, including the roughness factor of Eq.~(\ref{eq:rough}), averaged over $200$--$650$~MHz as in the HiCal-2 comparison of Ref.~\cite{Prohira2018}. The ratio $r/d$ denotes the surface-reflected power divided by the directly received power for the same calibration pulse geometry. The corresponding calculated quantity is defined in Appendix~\ref{app:fields}.

In Sec.~\ref{sec:layered}, the single-boundary coefficients $f_r^{\,s,p}$ of Eqs.~(\ref{eq:fs})--(\ref{eq:fp}) are replaced by the corresponding coefficients for a layered medium. The spherical geometry, polarization decomposition, and roughness factor remain unchanged. We first validate the spherical single-boundary calculation against the HiCal-2 reflectivity
results~\cite{Prohira2018}, before introducing the layered-medium formalism.

The horizontal Hertz dipole is used for the power reflection ratio $r/d$ and for the HiCal-1 reflected pulse validation in Sec.~\ref{sec:hical}. The boundary treatment is independent of this source choice. It acts on each plane-wave component separately, so the same construction applies to any source that can be represented as an angular spectrum of plane waves. A higher multipole changes the weights of these components, while an extended source such as a cosmic-ray air shower requires an additional integration over the emitting current distribution. In both cases, the boundary operator remains unchanged.

Let $\mathcal{T}_{\rm ref}^{H}(f)$ denote Eq.~(\ref{eq:eref}), the fixed-frequency response of the
source--boundary--observer geometry to unit spectral amplitude at the source, including the angular integral and the polarization projection. For a pulse with source spectrum $\widetilde A(f)$, the time-domain reflected field is then
\begin{equation}
E_{\rm ref}^{H}(t)=
\mathrm{Re}\int_{f_{\min}}^{f_{\max}}
\widetilde A(f)\,\mathcal{T}_{\rm ref}^{H}(f)\,
e^{-2\pi i f t}\,df ,
\label{eq:freqrecon}
\end{equation}
where $f_{\min}$ and $f_{\max}$ are the limits of the analysis band, or the corresponding discrete Fourier sum for sampled data. Equation~(\ref{eq:freqrecon}) therefore involves an integral over frequency, polar angle, and azimuth. We evaluate the angular integral at each frequency and then perform the frequency sum. The layer coefficient remains inside the frequency integral. It is not replaced by a band-averaged value. A band-averaged
reflectivity is a different quantity, evaluated only after the full frequency
dependence has been calculated.

The final frequency combination depends on the observable. For the band-averaged reflectivity used in Table~\ref{tab:validation}, the power ratio is evaluated at each frequency across the HiCal-2 band and then averaged, following the comparison in Ref.~\cite{Prohira2018}. For the time-domain pulses in Sec.~\ref{sec:hical}, the complex reflected field is evaluated at each frequency and then Fourier transformed, preserving the relative phases of the Fourier components. In both cases, the reflected field is first calculated at each frequency, including the $s$- and $p$-polarized plane-wave contributions. The results are then either averaged over frequency or combined to reconstruct the pulse. This order is important because a frequency-dependent boundary response can contain structure that is not
visible in the band-averaged power.

Band averaging is also useful because the spherical surface calculation varies rapidly with elevation at a fixed frequency. These oscillations arise from interference between contributions with different path lengths in the angular integral. For a smooth boundary, their angular scale is about $0.05^\circ$ (see Fig.~13 of Ref.~\cite{Prohira2018}). A comparison at a single frequency and elevation can therefore depend sensitively on the angular grid used in the calculation. Averaging over the HiCal-2 frequency band smooths these oscillations and provides a stable benchmark for the stratified extension.

Table~\ref{tab:validation} compares the two calculations at ten HiCal-2 elevation angles. Reference values are extracted from the published spherical, rough-surface reflectivity curve of Ref.~\cite{Prohira2018}. Because this comparison relies on extracted central values, the third decimal place should not be strictly interpreted. The calculations agree to better than $1.1\%$ at every angle, with a mean deviation of $0.6\%$. The deviations all share the same sign, pointing to a small systematic offset, consistent with the numerical truncation of the finite integration window, rather than random statistical scatter. Importantly, the overall normalization contains no free parameters. As a baseline consistency check, in the flat, smooth, single-boundary limit, the ratio $r/d$ must reduce to the frequency-independent Fresnel power reflection coefficient for an air--ice interface ($n_0=1.0$ and $n_1=1.4$). Our numerical implementation recovers this analytical limit to $1.5\%$, limited by the finite integration window.

\begin{table}[!tb]
\caption{Validation of the numerical pipeline against the spherical-surface power reflection ratio $r/d$ of Ref.~\cite{Prohira2018}. Both calculations include Earth curvature, the roughness correction of Eq.~(\ref{eq:rough}), and are averaged over $200$--$650$~MHz.}
\label{tab:validation}
\begin{ruledtabular}
\begin{tabular}{cccc}
Elevation (deg) & Ref.~\cite{Prohira2018} & This work & Ratio \\ \hline
5  & 0.191 & 0.191 & 1.000 \\
6  & 0.228 & 0.229 & 1.004 \\
7  & 0.256 & 0.258 & 1.008 \\
10 & 0.290 & 0.293 & 1.010 \\
12 & 0.285 & 0.288 & 1.011 \\
15 & 0.259 & 0.260 & 1.004 \\
17 & 0.238 & 0.240 & 1.008 \\
20 & 0.204 & 0.205 & 1.005 \\
22 & 0.184 & 0.184 & 1.000 \\
25 & 0.157 & 0.158 & 1.006 \\
\end{tabular}
\end{ruledtabular}
\end{table}


\section{Reflection from a spherical stratified boundary}
\label{sec:layered}
Section~\ref{sec:formalism} gives the reflected field of a dipole above a single spherical boundary. We now develop a first-principles framework for a stratified medium beneath that boundary. The only change is the coefficient that describes the reflection of each plane wave in Eq.~(\ref{eq:weyl}). The Sommerfeld--Weyl decomposition of the source, the tangent-plane geometry of Eq.~(\ref{eq:alpha0}), the polarization weights, and the roughness factor remain unchanged. The Fresnel coefficients of Eqs.~(\ref{eq:fs})--(\ref{eq:fp}) are replaced by the characteristic-matrix coefficients of the layered medium~\cite{BornWolf}, obtained from the same Maxwell boundary conditions.

The layered medium, or stack, consists of $N$ homogeneous layers beneath the boundary. The $j$th layer, with $j=1,\dots,N$, has refractive index $n_j$ and thickness $d_j$, and the stack lies above a semi-infinite substrate with index $n_s$. The case $N=0$ gives the single boundary considered in Sec.~\ref{sec:formalism}. The indices may be complex, so the same formalism applies to dielectric, lossy, and conducting media, as demonstrated in
Sec.~\ref{sec:applications} for lunar regolith and seawater. The substrate represents the material below the structure of interest, such as deep glacial ice beneath a firn column, seawater beneath an overlying layer, or bedrock beneath lunar regolith. Since the substrate has no lower boundary in the model, a wave transmitted into it does not return. A continuous depth
profile $n(z)$ can be approximated by a stack of sufficiently thin layers, as discussed at the end of this section.

We treat the reflection of the dipole field one plane wave at a time. Each plane wave in Eq.~(\ref{eq:weyl}) is labeled by $(\alpha,\beta)$ and reaches the spherical boundary at an incident point, denoted by $Q$, $Q'$, and so on for different plane waves (Fig.~\ref{fig:geometry}). At each
incident point, we introduce the plane tangent to the spherical surface (Fig.~\ref{fig:geometry}). Within the region around the incident point that contributes coherently, the spherical shells can be treated as parallel layers. The incident field is then locally planar, and the response of the stack is that of a plane wave incident on a plane layered medium. This response is evaluated at the local incidence angle $\alpha_0$ of Eq.~(\ref{eq:alpha0}).

\begin{figure*}[!t]\centering
\begin{minipage}[t]{0.50\textwidth}
\centering
\includegraphics[width=0.99\linewidth]{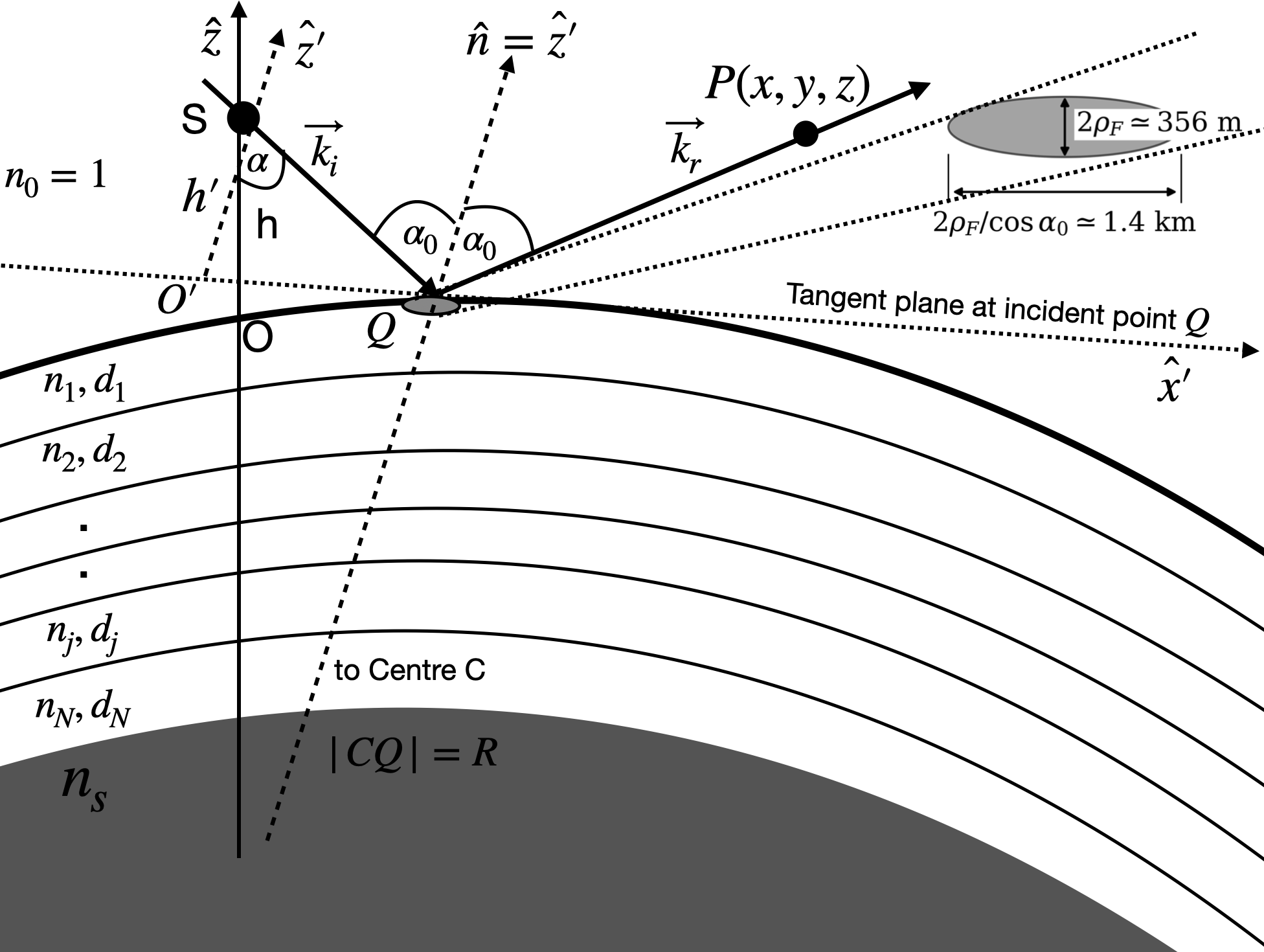}
\par\smallskip(a)
\end{minipage}\hfill
\begin{minipage}[t]{0.50\textwidth}
\centering
\includegraphics[width=0.99\linewidth]{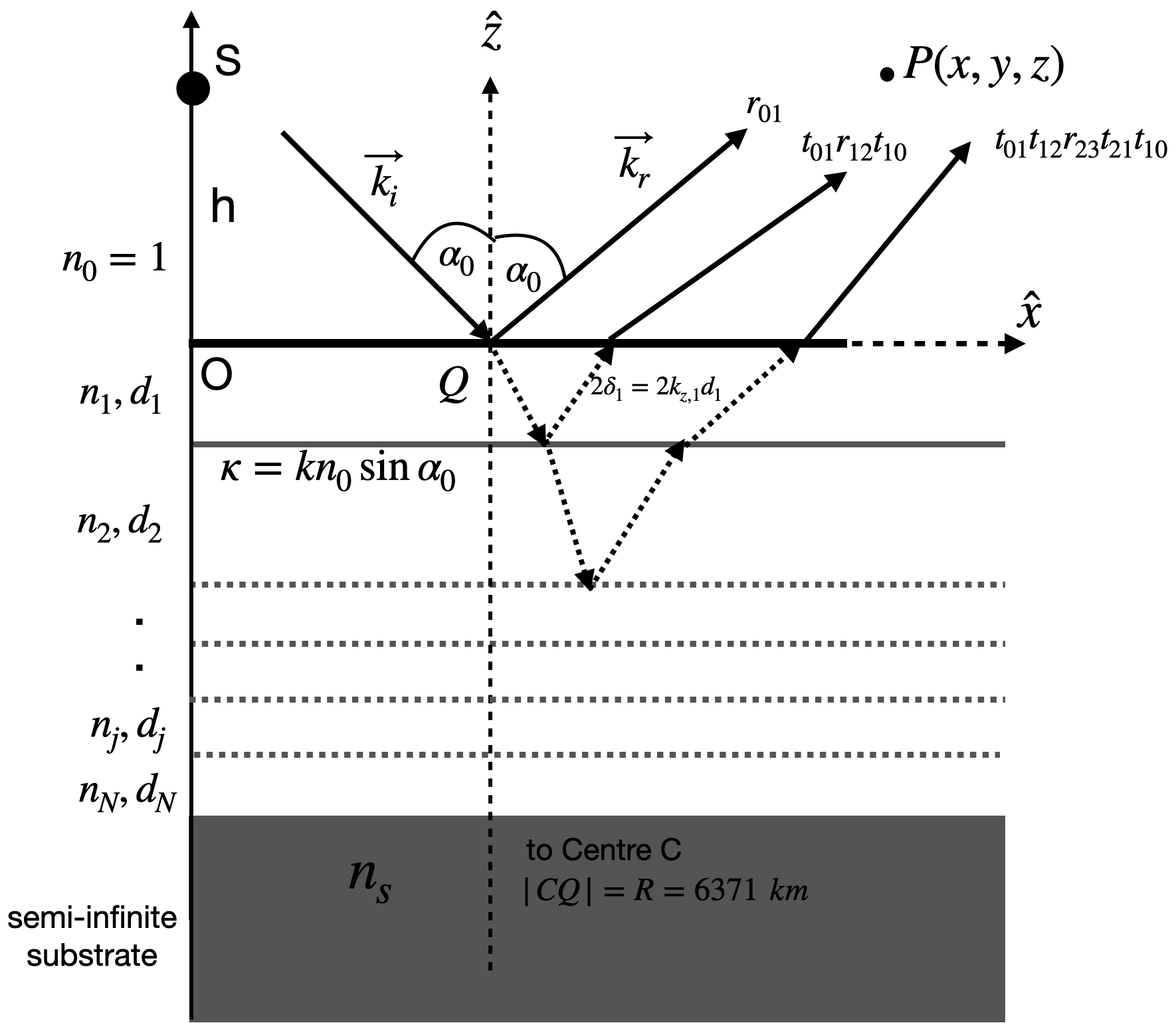}
\par\smallskip(b)
\end{minipage}
\caption{Geometry of the spherical reflection problem and the local
layered-medium treatment. (a): A plane-wave component from the source
$S$ reaches the spherical surface at $Q$ and contributes to the reflected
field at the observer $P(x,y,z)$. The local outward normal is
$\hat{n}=\hat{z}'$, and $h'$ is the perpendicular distance from $S$ to the
tangent plane at $Q$. The coordinate system is rotated so that $O'$ is the
origin of the local frame. Only a finite region around $Q$ contributes
coherently. That region is the shaded patch at $Q$, enlarged in the inset at
upper right, where $\rho_F$ is the first Fresnel radius of
Eq.~(\ref{eq:fresnelrad}); the values shown are for the ANITA/HiCal balloon
geometry drawn here, with $h=37$~km, $f=650$~MHz, and a local elevation of
$15^\circ$. The layers are treated as locally parallel across the cross-track
footprint, while curvature in the plane of incidence is retained through the
local angle $\alpha_0(\alpha)$.
(b): In the local tangent-plane description, the surface reflection
has amplitude $r_{01}$. A wave entering the first layer and returning from the
$1$--$2$ interface has amplitude
$t_{01}r_{12}t_{10}e^{2i\delta_1}$, where
$\delta_1=k_{z,1}d_1$. A return from the next interface has amplitude
$t_{01}t_{12}r_{23}t_{21}t_{10}
e^{2i(\delta_1+\delta_2)}$. If layer 2 borders the substrate directly,
$r_{23}$ is replaced by $r_{2s}$. All higher-order internal reflections are
included coherently through the characteristic matrix of
Eq.~(\ref{eq:rstack}). The substrate $n_s$ is semi-infinite, so waves
transmitted into it do not return.}
\label{fig:stack}
\end{figure*}

The tangent plane and the angle $\alpha_0$ are different for each plane wave component because each component meets the spherical boundary at a different point (see Fig.~\ref{fig:geometry}). The reflected field therefore remains globally non-planar, as in Sec.~\ref{sec:formalism}. Panel~(a) of Fig.~\ref{fig:stack} shows the size of the coherent region relative to the surface radius, while panel~(b) shows the local tangent-plane description and the internal reflections within the stack. The quantitative condition for treating the shells as parallel is given later in this section and derived in Appendix~\ref{app:geometry}.

At the incident point $Q$, it is convenient to resolve the wavevector into components normal and tangential to the local interface. In the local frame at $Q$, with $\hat{n}(Q)$ denoting the outward normal, the arriving plane wave
has
\begin{equation}
\vec{k}_0=\big(\kappa\cos\beta,\ \kappa\sin\beta,\ -k_{z,0}\big),
\qquad
\kappa^2+k_{z,0}^2=k^2n_0^2 ,
\label{eq:kvector}
\end{equation}
with tangential magnitude $\kappa=k\,n_0\sin\alpha_0$ and normal component
$k_{z,0}=k\,n_0\cos\alpha_0$. Because every layer is uniform in the two directions parallel to the boundary, only the normal component can change on
crossing an interface. The tangential part takes the same value throughout the stack,
\begin{equation}
\kappa=k\,n_0\sin\alpha_0=k\,n_j\sin\alpha_j=k\,n_s\sin\alpha_s ,
\qquad \beta_j=\beta ,
\label{eq:transverse_conservation}
\end{equation}
which is Snell's law written in wavevector form and the same continuity
condition used in Refs.~\cite{Prohira2018,DasguptaJain2021}.
Equation~(\ref{eq:transverse_conservation}) also implies that the refracted wavevector
stays in the plane of incidence of that plane wave, and that for isotropic
layers the stack coefficient depends on the direction of $\vec{\kappa}$ only
through $\alpha_0$, not on $\beta$. The azimuth still enters the full
calculation, through the $s$ and $p$ basis vectors at $Q$ and the geometric phase in Eq.~(\ref{eq:eref}).

Two points about Eq.~(\ref{eq:transverse_conservation}) are useful to mention. First, $\kappa$ is conserved across interfaces for a given frequency and plane wave direction. It is not fixed throughout the calculation. It depends on both frequency and direction of the wavevector, $\kappa=\kappa(f,\alpha)=k\,n_0\sin\alpha_0$, with $k=2\pi f/c$ and $\alpha_0=\alpha_0(\alpha)$ from Eq.~(\ref{eq:alpha0}). The angular integrations in Eq.~(\ref{eq:eref}) therefore sweep $\kappa$ over $0\le\kappa\le k n_0$ for the propagating plane waves, with larger values corresponding to the evanescent part of the contour. The Fourier sum over the
frequency band similarly sweeps over $k$.

Second, the normal component of the wavevector is not conserved. Instead, it takes a different value in each layer,
\begin{equation}
k_{z,j}=\sqrt{k^2n_j^2-\kappa^2}=k\,n_j\cos\alpha_j ,
\label{eq:kz}
\end{equation}
and this variation determines the phase accumulated by each plane wave as it propagates through the layers.  A single traversal of layer $j$ advances the phase by $\delta_j=k_{z,j}d_j$, so a round trip through the layer contributes $2\delta_j=2k_{z,j}d_j=2k\,n_jd_j\cos\alpha_j$.

The response of a single buried interface is fixed by Maxwell's boundary conditions at that interface. Continuity of the tangential electric and magnetic fields determines the reflected and transmitted amplitudes. The same conditions apply at each of the $N+1$ interfaces in the stack. It is convenient
to write the coefficients for the two polarizations in a common form. We
define
\begin{equation}
q_j=
\begin{cases}
n_j\cos\alpha_j, & s\ \text{polarization},\\[2pt]
\cos\alpha_j/n_j, & p\ \text{polarization},
\end{cases}
\label{eq:qdef}
\end{equation}
where $\alpha_j$ follows from Eq.~(\ref{eq:transverse_conservation}). For $s$ polarization the electric field is perpendicular to the plane of incidence and lies in the plane of the interface, while for $p$ polarization the magnetic field has this property. We use $q_j$ as the standard characteristic variable for the chosen polarization. The coefficients for the interface between media $j-1$ and $j$ are
\begin{equation}
r_{j-1,j}=\frac{q_{j-1}-q_j}{q_{j-1}+q_j},
\qquad
t_{j-1,j}=\frac{2q_{j-1}}{q_{j-1}+q_j},
\label{eq:rhoint}
\end{equation}
for each polarization separately. For $j=1$, the reflection coefficient reduces to the Fresnel coefficients of Eqs.~(\ref{eq:fs})--(\ref{eq:fp}). For $s$ polarization, $q_0=n_0\cos\alpha_0$ and $q_1=n_1\cos\alpha_1$ give $r_{01}=f_r^{\,s}$, while the $p$-polarization form gives $r_{01}=f_r^{\,p}$.

The transmission coefficient has a different meaning for the two
polarizations. For $s$ polarization, $t_{j-1,j}$ is the ratio of electric-field
amplitudes, while for $p$ polarization it is the ratio of magnetic-field
amplitudes. The corresponding electric-field ratio for $p$ polarization is
$(n_{j-1}/n_j)\,t_{j-1,j}$. This factor cancels in the reflection coefficient,
which depends on transmission only through the product
$t_{j-1,j}t_{j,j-1}$. It must, however, be included when calculating the
transmitted field below.

A plane wave incident on a laterally uniform boundary does not couple the
$s$ and $p$ polarizations. We therefore calculate their contributions to the
field independently throughout the formalism, with a superscript $s$ or $p$
denoting the corresponding quantity.

Each layer transfers the wave across its thickness. Continuity of the
tangential electric and magnetic fields at every interface can be written as
the transfer of a pair of boundary variables, and for layer $j$ that transfer
is the characteristic matrix
\begin{equation}
M_j=
\begin{pmatrix}
\cos\delta_j & -i\sin\delta_j/q_j\\[3pt]
-iq_j\sin\delta_j & \cos\delta_j
\end{pmatrix},
\qquad \delta_j=k_{z,j}d_j .
\label{eq:Mj}
\end{equation}
The pair of variables and the derivation of Eq.~(\ref{eq:Mj}) from Maxwell's
boundary conditions are given in Appendix~\ref{app:stack}. The product
\begin{equation}
M=\prod_{j=1}^{N}M_j
=\begin{pmatrix} m_{11} & m_{12}\\ m_{21} & m_{22}\end{pmatrix}
\label{eq:Mprod}
\end{equation}
transfers the same pair across the whole stack, from the top of the substrate
to the upper surface.

A homogeneous layer is not sampled internally. At a given frequency and incidence angle, its full effect is contained in the phase thickness $\delta_j=k_{z,j}d_j$ and the interface coefficients at its two boundaries. Thus, a millimeter-scale ice lens is represented by a single layer matrix with its physical thickness, rather than by a spatial mesh through the lens. Numerical sampling enters only in two places. First, the frequency grid used for broadband pulses, and second, the slab thickness used to approximate a smooth depth-dependent profile $n(z)$ by many thin homogeneous layers.

If the refractive index is complex, then $k_{z,j}$ and $\delta_j$ are complex
as well. The propagation factor is
$e^{i\delta_j}=e^{i\,{\rm Re}(k_{z,j})d_j}
e^{-{\rm Im}(k_{z,j})d_j}$, so a lossy layer attenuates on the skin-depth
scale $\ell_{\rm skin}=1/{\rm Im}(k_{z,j})$. Thus, the thickness of a
lossless layer affects the reflection through its interference phase, while
for a lossy layer it affects both the phase and the attenuation.

Imposing continuity at the two ends of the stack, with an incident and
reflected wave above and a single outgoing wave below, gives the reflection
and transmission coefficients of the complete stack (Appendix~\ref{app:stack}):
\begin{align}
r_{\rm stack}^{\,s,p}
&=\frac{(m_{11}+m_{12}q_s)q_0-(m_{21}+m_{22}q_s)}
       {(m_{11}+m_{12}q_s)q_0+(m_{21}+m_{22}q_s)} ,
\label{eq:rstack}\\[4pt]
t_{\rm stack}^{\,s,p}
&=\frac{2q_0}
       {(m_{11}+m_{12}q_s)q_0+(m_{21}+m_{22}q_s)} .
\label{eq:tstack}
\end{align}
Here $q_0$ and $q_s$ are given by Eq.~(\ref{eq:qdef}) for the upper medium
and the substrate, evaluated at the angles $\alpha_0$ and $\alpha_s$,
respectively. The subscript $s$ on $q_s$ denotes the substrate $n_s$
(Fig.~\ref{fig:stack}), not $s$ polarization. The coefficients in
Eqs.~(\ref{eq:rstack})--(\ref{eq:tstack}) are determined by the layer indices
$n_j$, thicknesses $d_j$, frequency through $k$, and local incidence angle
$\alpha_0$, which incorporates the effect of the spherical boundary
curvature. The incidence angle is set separately for each plane-wave
component by Eq.~(\ref{eq:alpha0}), rather than being a free input as it
would be for a planar interface. The substrate angle $\alpha_s$ then follows
from Eq.~(\ref{eq:transverse_conservation}). Thus, the curvature of the
boundary determines the local incidence angle, while the subsurface structure
determines the corresponding reflection coefficient.

Two properties of Eqs.~(\ref{eq:rstack})--(\ref{eq:tstack}) provide useful numerical checks. Each layer matrix has unit determinant, so $\det M=1$ for any number of layers and any complex indices. For $N=0$, there are no finite layers, so the product in Eq.~(\ref{eq:Mprod}) is the identity:
$m_{11}=m_{22}=1$ and $m_{12}=m_{21}=0$. Equation~(\ref{eq:rstack}) then reduces to
\begin{equation}
r_{\rm stack}^{\,s,p}=\frac{q_0-q_s}{q_0+q_s},
\label{eq:rstackN0}
\end{equation}
which is the Fresnel coefficient for a single boundary between the upper medium $n_0$ and the substrate $n_s$. We use this single-boundary case as a reference when testing the effect of layering, with the substrate index chosen for each case: $n_s=1.4$ in the validation of Sec.~\ref{sec:formalism} and $n_s=1.35$ in the reference stack of Sec.~\ref{sec:results}.

The coefficient in Eq.~(\ref{eq:rstack}) includes all reflections and
transmissions within the stack. A wave arriving at $Q$ is partly reflected
with coefficient $r_{01}$ and partly transmitted into the stack with
$t_{01}$. Once inside the stack, it can reflect from any buried interface, propagate
through the layers, and cross back into the upper medium with $t_{10}$. It
can also undergo further internal reflections, with each reflection and
transmission carrying the corresponding coefficient from
Eq.~(\ref{eq:rhoint}). Maxwell's boundary conditions are satisfied at every
interface. All contributions have the same tangential wavevector and must
therefore be added coherently, with their relative phases determined by the
layer phase thicknesses $\delta_j$ in Eq.~(\ref{eq:Mj}). Appendix~\ref{app:stack}
sums these multiple-reflection and transmission terms explicitly and shows
that the result is identical to Eq.~(\ref{eq:rstack}). The two calculations
agree to machine precision in our implementation.

The construction is general and does not require the refractive index to increase with depth. If a wave enters a lower-index medium, Eq.~(\ref{eq:rhoint}) gives the corresponding sign of $r_{j-1,j}$, which is
retained by the matrix calculation. This is important for determining the
threshold refractive index of a buried layer at which the reflected pulse
changes polarity, as discussed in Sec.~\ref{sec:result_threshold}. The sign
of the reflected amplitude is therefore determined by the full layered
calculation rather than fixed in advance.

The reflected field for a stratified boundary then follows by replacing
$f_r^{\,s,p}$ with $r_{\rm stack}^{\,s,p}$ in Eq.~(\ref{eq:polweight}), while
leaving Eq.~(\ref{eq:eref}) otherwise unchanged:
\begin{equation}
E^{H}_{\rm ref}(f)=\frac{ik}{2\pi}
\int_{0}^{2\pi}\!\!\!\int_{0}^{\pi/2-i\infty}\!\!\!\!
W_{\rm ref}\,F_{\rm rough}\,e^{i\Phi_r}\,\sin\alpha\,d\alpha\,d\beta ,
\label{eq:erefstack}
\end{equation}
\begin{equation}
\begin{split}
W_{\rm ref}={}&r^{\,s}_{\rm stack}(f,\alpha_0)\cos^2\!\beta\\
&-r^{\,p}_{\rm stack}(f,\alpha_0)\cos\alpha\,\cos(2\alpha_0-\alpha)\sin^2\!\beta ,
\end{split}
\label{eq:wrefstack}
\end{equation}
where $\alpha_0=\alpha_0(\alpha)$ is given by Eq.~(\ref{eq:alpha0}).
Equations~(\ref{eq:erefstack})--(\ref{eq:wrefstack}) give the reflected field
and polarization weight used throughout this work. We evaluate them for
different media and geometries, including the interpretation of the ANITA
anomalous-polarity events, and validate the calculation against published
data. Both coefficients are evaluated at the local incidence angle of each
plane-wave component. Thus, $r_{\rm stack}^{\,s,p}$ remains \emph{inside} the
angular integral of Eq.~(\ref{eq:erefstack}) and varies with $\alpha$.

The transmitted field requires no new construction, since the same matrix product already gives $t^{\,s,p}_{\rm stack}$. Repeating the projection of Appendix~\ref{app:fields} for the transmitted direction $\hat{k}_t$ of Eq.~(\ref{eq:khatt}) gives
\begin{equation}
E^{H}_{\rm trans}(f)=\frac{ik}{2\pi}
\int_{0}^{2\pi}\!\!\!\int_{0}^{\pi/2-i\infty}\!\!\!\!
W_{\rm trans}F^{\,t}_{\rm rough}\,e^{i\Phi_t}\sin\alpha\,d\alpha\,d\beta ,
\label{eq:etransstack}
\end{equation}
\begin{equation}
\begin{split}
W_{\rm trans}={}&t^{\,s}_{\rm stack}\cos^2\!\beta\\
&+\frac{n_0}{n_s}t^{\,p}_{\rm stack}\cos\alpha\,\cos(\alpha+\alpha_s-\alpha_0)\sin^2\!\beta ,
\end{split}
\label{eq:wtransstack}
\end{equation}
where the factor $n_0/n_s$ is the conversion noted above. It takes $t^{\,p}_{\rm stack}$ from the magnetic-field ratio returned by Eq.~(\ref{eq:tstack}) to the electric-field ratio that the projection onto $\hat{y}$ requires. Its value is fixed by azimuthal symmetry at normal incidence, where Eqs.~(\ref{eq:qdef}) and~(\ref{eq:tstack}) give $t^{\,s}_{\rm stack}=2n_0/(n_0+n_s)$ and $t^{\,p}_{\rm stack}=2n_s/(n_0+n_s)$, so that only with the factor present does $W_{\rm trans}$ become independent of $\beta$. The remaining quantities are $\Phi_t=k|SQ|+k\,n_s\ell_s$, where $\ell_s$ is the path from the base of the stack to the field point in the substrate. The phase accumulated inside the finite layers is already contained in $t^{\,s,p}_{\rm stack}$ through the $\delta_j$ and must not be added again. The surface weight for the transmitted field is computed as
\begin{equation}
F^{\,t}_{\rm rough}=\exp\!\Big[-\tfrac12\,\sigma_h^2(\rho_\perp)
\big(k_{z,0}-k_{z,s}\big)^2\Big].
\label{eq:roughtrans}
\end{equation}
This is the Kirchhoff coherent-field factor $\exp[-\tfrac12\sigma_h^2(\Delta k_z)^2]$~\cite{BeckmannSpizzichino}, where $\Delta k_z$ is the change in the normal wavenumber across the boundary. For transmission, $\Delta k_z=k_{z,0}-k_{z,s}$. For reflection, taking $k_{z,s}\to-k_{z,0}$ gives $\Delta k_z=2k\cos\alpha_0^{\rm spec}$ and recovers $\exp[-2k^2\sigma_h^2\cos^2\alpha_0^{\rm spec}]$ in Eq.~(\ref{eq:rough}). Equations~(\ref{eq:etransstack})--(\ref{eq:roughtrans}) therefore extend the transmitted-field treatment of Refs.~\cite{Prohira2018,DasguptaJain2021} to an arbitrary layered medium and include the roughness factor. They may be useful for sources within ice or regolith and for sounding pulses entering a layered medium. These cases are outside the scope of the far-field reflection calculation relevant to the
high-altitude balloon geometry considered here, so the numerical results below use only Eq.~(\ref{eq:erefstack}). The transmitted field calculation is included for completeness.

To isolate the effect of the layers from the propagation geometry, we define the \emph{layering ratio} at fixed frequency and polarization,
\begin{equation}
\mathcal{Q}^{\,\chi}(f)=
\frac{E_{\rm layered}^{\,\chi}(f)}
{E_{\rm single}^{\,\chi}(f)} ,
\qquad \chi=s,p .
\label{eq:Qratio}
\end{equation}
This is the reflected field with the layers present divided by the field with the layers removed, using the same geometry, frequency, and polarization in both cases. The same definition applies after recombining the $s$ and $p$ polarization components into any calculated field component, provided the same projection weights are used in the numerator and denominator. Propagation phase and geometric spreading are common to both fields and therefore cancel in the ratio. Thus, $|\mathcal{Q}|$ gives the change in reflected amplitude due to the finite layers, while $\arg\mathcal{Q}$ gives the additional phase relative to the single-boundary case. We omit the superscript when the polarization is clear from context. The ratio is evaluated for the reference firn stack in Sec.~\ref{sec:result_q}.

The roughness factor of Eq.~(\ref{eq:rough}) describes the upper air--medium
interface. It depends on the surface geometry, not on the dielectric structure
below it, so the same validated Antarctic roughness parameters~\cite{Prohira2018,DasguptaThesis2020} apply to the layered and single-boundary
calculations for the same surface. Absolute quantities such as the power ratio
$r/d$ retain this weight, and the results of Sec.~\ref{sec:results} include it.

The default stratified calculation treats the buried interfaces as laterally uniform and smooth. This is a model choice, not a restriction of the characteristic matrix. To test the effect of roughness at a buried interface,
we multiply the reflection coefficient for that interface by
\begin{equation}
F_{{\rm int},j}=\exp[-2k_{z,j}^{\,2}\sigma_{{\rm int},j}^{\,2}],
\label{eq:intrough}
\end{equation}
where $\sigma_{{\rm int},j}$ is the rms height variation of the buried
interface and $k_{z,j}$ is the normal wavenumber on the incident side of
layer $j$. This is the same Gaussian average of the coherent specular field
used for the upper surface, applied here to an internally reflected
amplitude. It does not describe diffuse scattering or lateral correlations
of a rough buried interface. These effects would require a random-interface
calculation, such as the treatment of the analogous acoustic problem in
Ref.~\cite{Pinson2015}. The rough-interface test in Sec.~\ref{sec:thinrough}
shows that this attenuation raises the buried-layer index at which the
reflected pulse loses its polarity inversion. The smooth-interface result
is therefore conservative for the sign-change question.

In the ratio $\mathcal{Q}$, the upper-surface roughness factor appears in the
numerator and denominator and therefore cancels in the specular approximation.
The same cancellation holds to the accuracy of this approximation in the full
angular integral. This separation of the boundary coefficient from the
surface roughness allows the same layered formalism to be used for other
surfaces. For example, lunar or Martian roughness parameters can replace the Antarctic
ice values without changing the layered calculation. Such applications
require roughness parameters appropriate to the surface and wavelength range
of interest and do not rely on the Antarctic validation.

The layers of a planetary surface are concentric shells rather than parallel
planes. With $R$ the radius of the upper surface, the interfaces lie at radii
$R$, $R-d_1$, $R-d_1-d_2$, and so on (Fig.~\ref{fig:geometry}). The
coefficient of Eq.~(\ref{eq:rstack}) is evaluated in the tangent plane at
each incident point, using the local incidence angle $\alpha_0$. Curvature in
the plane of incidence is therefore treated exactly, plane wave by plane
wave. The only additional geometric approximation is transverse to that
plane, across the coherent footprint around $Q$, the shells are treated as
parallel.

The size of this footprint is set by the first Fresnel zone. For a reflected path with legs $L_1$ and $L_2$ on either side of the incident point, the Fresnel radius transverse to the plane of incidence is $\rho_F=\sqrt{\lambda L_1L_2/(L_1+L_2)}$, derived in Appendix~\ref{app:geometry}. At the specular point of the balloon geometry these two paths are equal, each of length $\ell=|SQ|$, so with $L_{\rm sph}=2\ell$ the total reflected path length,
\begin{equation}
\rho_F=\sqrt{\frac{\lambda L_{\rm sph}}{4}} ,
\qquad
L_{\rm sph}=\frac{2h'}{\cos\alpha_0} ,
\label{eq:fresnelrad}
\end{equation}
where $\lambda$ is the wavelength and $L_{\rm sph}$ follows from
Eq.~(\ref{eq:apphprime}). In the plane of incidence the coherent region is
longer by $1/\cos\alpha_0$, so the footprint is an ellipse with semi-axes
$\rho_F$ and $\rho_F/\cos\alpha_0$. For the geometry drawn in
Fig.~\ref{fig:stack}, with $h=37$~km, $f=650$~MHz, and $15^\circ$ local
elevation, $L_{\rm sph}=275$~km and $\rho_F=178$~m, giving a cross-track
footprint diameter of $356$~m and $1.4$~km along the plane of incidence.

The departure of a shell from its tangent plane across this radius is the sagitta
\begin{equation}
\eta=R-\sqrt{R^2-\rho_F^{\,2}}
\simeq\frac{\rho_F^{\,2}}{2R}.
\end{equation}
A displacement $\eta$ along the local normal changes the two-way path by
$2\eta\cos\alpha_0$, giving
\begin{equation}
\Delta\phi
=2k\cos\alpha_0\,\frac{\rho_F^{\,2}}{2R}
=\frac{\pi h'}{R}
\le\frac{\pi h}{R},
\label{eq:sagphase}
\end{equation}
where Eq.~(\ref{eq:fresnelrad}) has been used. The wavelength and the
incidence angle therefore cancel.
For $h=37$~km above the Earth, this gives the frequency-independent bound
$\Delta\phi\le\pi h/R=0.018$~rad (about $1^\circ$) at all elevation angles.
The exact values are $0.0176$~rad at $15^\circ$ and $0.0141$~rad at
$5^\circ$. Since $h'<h$ for a convex surface (Fig.~\ref{fig:stack}), the bound is conservative.

At the same source height, frequency, and $15^\circ$ elevation, the sag is
$4.5$~mm for Mars and $8.4$~mm for the Moon, corresponding to phase errors of
$0.032$ and $0.059$~rad, respectively. The lunar sag is $1.8\%$ of a wavelength.
Since $\eta\propto1/R$ at fixed footprint, the local-parallel-layer
approximation improves for larger bodies. All results in
Secs.~\ref{sec:validity} and~\ref{sec:results} use a spherical Earth with
mean radius $R=R_\oplus=6371$~km. Using the smaller polar radius to account
for Earth's flattening changes these estimates by about $0.2\%$. The flat
limit $R\to\infty$ is retained as a numerical check, for which
$\alpha_0\to\alpha$ and Eq.~(\ref{eq:rstack}) reduces to the planar stack.

A stack need not represent physically distinct slabs. It can also provide a
numerical representation of a laterally uniform, depth-dependent index
$n(z)$, provided the slabs are thin enough to resolve the propagation phase
and the variation of $n(z)$. Convergence is checked by halving the slab
thickness until $r_{\rm stack}^{\,s,p}$ no longer changes. For the smooth firn
profiles considered here, convergence is reached with $N=2000$ slabs over
$60$~m, corresponding to a slab thickness of $3$~cm. A discrete feature
thinner than this grid is not smeared into it. It is instead entered as an
explicit layer with its physical thickness, as computed in Sec.~\ref{sec:thinrough}.

Polar firn provides a relevant example. In-ice radio analyses parameterize
the depth-dependent ice model as
\begin{equation}
n(z)=\mathcal{A}-\mathcal{B}\,e^{-\mathcal{C}z},
\label{eq:icemodel}
\end{equation}
where $z$ is the depth below the surface in meters, $\mathcal{A}$ is the
deep-ice index, $\mathcal{B}$ is the near-surface index deficit, and
$\mathcal{C}^{-1}$ is the $e$-folding depth. The Askaryan Radio Array~\cite{KH_PA_analysis}
at the South Pole fit $\mathcal{A}=1.780$, $\mathcal{B}=0.454$, and
$\mathcal{C}=0.0202~\mathrm{m}^{-1}$ to calibration pulses from a local
transmitter and the SPIceCore borehole~\cite{ARA2021Calibration}, giving a surface index
$n(0)=1.326$ and an $e$-folding depth of about $50$~m
\cite{KH_PA_analysis,ARAfulllivetime2023}. Timing differences between direct
and refracted or reflected ray paths from an englacial transmitter to
receivers at depths up to $200$~m provide further constraints and have been
used to compare Eq.~(\ref{eq:icemodel}) with a glaciologically motivated
three-phase densification model~\cite{Couberly2026}.

Because this profile varies slowly on the scale of a wavelength, it is nearly
impedance matched and reflects weakly. Replacing the discrete reference stack
of Sec.~\ref{sec:results} with a smoothly graded firn column would therefore
reduce, rather than enhance, the layering signature. The discrete stack is
thus conservative. It gives an upper estimate of the modulation that a smooth
firn gradient alone can produce. It is intended to represent discrete density
anomalies superposed on this background profile, such as wind crusts, ice
lenses, and refrozen melt layers.

The effects of birefringence and evanescent waves on the total field are
negligible for the calculations presented here. Polar ice is weakly
birefringent because of its crystal-orientation fabric
\cite{Connolly2022,Jordan2020}. Birefringence gives the two polarization
components slightly different refractive indices and therefore slightly
different propagation phases. To assess its importance here, we compare the
resulting differential phase with the $\pi$ phase change required for a
polarity reversal.

The relevant propagation distance is the path through the buried layers.
For surface reflection, this path is short. A ray travels a distance
$d_j/\cos\alpha_j$ through a layer of thickness $d_j$, where $d_j$ is measured
normal to the interface and $\alpha_j$ is the propagation angle relative to
that normal. Thus, for a ray that reaches the deepest interface and returns,
\begin{equation}
L_{\rm ice}=2\sum_j\frac{d_j}{\cos\alpha_j},
\label{eq:Lice}
\end{equation}
where the sum is over the traversed layers and the angles $\alpha_j$ are
fixed by Eq.~(\ref{eq:transverse_conservation}). For the estimate below, we
use a representative birefringent index splitting $\delta n$. More generally,
the differential delay is obtained by replacing $\delta n L_{\rm ice}$ with
$2\sum_j\delta n_j d_j/\cos\alpha_j$.

For $d_1=d_2=2$~m, $n_1=1.35$, and $n_2=1.75$, the geometries considered below
give $L_{\rm ice}\lesssim11$~m. Taking a representative birefringent index
splitting $\delta n\sim3\times10^{-3}$ for deep South Pole ice
\cite{Jordan2020}, the differential propagation delay is
\begin{equation}
\Delta t\simeq\frac{\delta n\,L_{\rm ice}}{c}\lesssim0.1~{\rm ns}.
\end{equation}
The corresponding phase difference across the $150$--$850$~MHz band is
\begin{equation}
\Delta\phi=2\pi f\Delta t\lesssim0.6~{\rm rad},
\end{equation}
and is about $0.8$~rad at $1200$~MHz, the upper edge of the ANITA band and
the PUEO main-instrument trigger band~\cite{Gorham2009,PUEOtrigger2026}.
Changes in the interface coefficients are smaller, at the relative scale
$\delta n/n\sim2\times10^{-3}$. Thus, for the short paths considered here,
birefringence is too small to produce the $\pi$ phase change required for a
polarity reversal.

Birefringence can become important for kilometer-scale propagation in ice,
as in UHE-neutrino detectors~\cite{Connolly2022,Jordan2020}, or for sources
embedded in the ice. Such cases require an anisotropic treatment and are
outside the scope of the present calculation. The lunar regolith and
seawater examples of Sec.~\ref{sec:applications} are treated as isotropic
effective media.

The second effect is the evanescent part of the contour in
Eq.~(\ref{eq:weyl}), reached at imaginary $\alpha$. In the air above the
surface,
$k_{z,0}=k\cos\alpha=\sqrt{k^2-\kappa^2}$, so a plane-wave component is
evanescent when $\kappa>k$, giving
$k_{z,0}=i\sqrt{\kappa^2-k^2}$ and a factor
$\exp[-\sqrt{\kappa^2-k^2}\,h]$ over a vertical distance $h$. Its decay length
is $(\kappa^2-k^2)^{-1/2}$. For a substantially evanescent component, taking
$\kappa=\sqrt{2}\,k$ gives a decay length of $\lambda/(2\pi)$, or
$5.6$--$32$~cm over the $150$--$850$~MHz band. Compared with
$h=37$~km, no such component of the source field survives propagation to the
surface, which is why these components were omitted in Ref.~\cite{Prohira2018}.

Evanescent behavior inside the medium is different and is handled
automatically by the characteristic-matrix calculation used here. If
$k_{z,j}$ becomes imaginary in a layer, the factor
$e^{ik_{z,j}d_j}$ describes decay or tunneling through that layer. For the
air-to-firn reflections considered here, this does not occur. Propagating
components have $\kappa\le k$, while every layer has $n_j>1$, so $k_{z,j}$
remains real. Evanescent coupling within the medium is therefore relevant to
near-field and transmission problems, not to the far-field reflection
calculation considered here.

A third quantity neglected in this framework is absorption. The refractive
indices used for firn and ice in this work are real, so the layer matrices
contain no attenuation. This is an input choice rather than a restriction of
the formalism. Equation~(\ref{eq:rstack}) allows complex $n_j$, and the
conducting media examples in Sec.~\ref{sec:applications} use this extension.
Appendix~\ref{app:stack} gives the corresponding attenuation through each
layer. The choice of real indices is justified by the short in-medium path.
A ray reaching the deepest interface of the reference stack and returning
travels $L_{\rm ice}\lesssim11$~m. For an amplitude attenuation length
$L_{\rm att}$, the amplitude is reduced by $e^{-L_{\rm ice}/L_{\rm att}}$. For
polar-ice values of order a kilometer at these frequencies, this reduction
is about one percent. Even for the deliberately conservative value
$L_{\rm att}=100$~m, the reduction is about $10\%$. Absorption in shallow firn
therefore does not compete with the interference effect that sets the
polarity threshold. Including absorption would reduce the buried contribution
and hence raise the threshold, in the same direction as the buried-interface
roughness considered in Sec.~\ref{sec:thinrough}.

Finally, the stratified extension builds directly on the validated
single-boundary formalism. It adds multiple-beam interference within the
stack, a standard effect in layered media at optical and radio frequencies
\cite{BornWolf}. In the $N=0$ limit, Eq.~(\ref{eq:rstack}) reduces to the
single-boundary calculation validated against HiCal-2 data in
Sec.~\ref{sec:formalism}. The resulting framework preserves the established
spherical-surface treatment while extending it to stratified boundaries and
provides a basis for studying how subsurface layering and surface roughness
affect signals received after reflection or transmission through natural
media.

\section{Factorization and its validity limits}
\label{sec:validity}

Sections~\ref{sec:formalism} and~\ref{sec:layered} calculate the boundary
coefficient inside the angular integral, at the local incidence angle of each
plane-wave component. Radar-sounding calculations instead evaluate the
coefficient once, at the specular direction, and multiply it by a separately
computed field. This section quantifies when that factorization is adequate.

As defined in Sec.~\ref{sec:formalism}, the specular point is the point on the
surface for which the reflected path $S\to Q\to P$ is stationary. In the
Sommerfeld--Weyl integral of Eq.~(\ref{eq:weyl}), the neighborhood of this
point defines the stationary-phase region. Within this region, the phase
varies slowly enough for the contributions to add coherently, while outside
it, rapid phase variations lead to cancellation. For the balloon-borne
geometries relevant to the ANITA/HiCal measurements and to the PUEO mission
\cite{PUEO2021,PUEOtrigger2026}, the reflected field is therefore dominated by
plane-wave components whose incident points lie close to the specular point on
the spherical surface~\cite{Prohira2018,DasguptaJain2021}.

Appendix~\ref{app:spa} gives the angular half-width of this region as $w\simeq\sqrt{2\pi/(kL_{\rm sph})}$, as shown in Eq.~(\ref{eq:appwindowscale}), where $k=2\pi f/c$ and $L_{\rm sph}$ is the total reflected path length of Eq.~(\ref{eq:apphprime}).
For the $h=37$~km balloon geometry, taking $f=650$~MHz near the upper end of the $200$--$650$~MHz HiCal-2 frequency band gives $w=0.05^\circ$ at
$5^\circ$ elevation and $w=0.07^\circ$ at $15^\circ$ elevation. Because
$w\propto f^{-1/2}$, the corresponding width at the lower end of the band,
$200$~MHz, is larger by $\sqrt{650/200}$, reaching $0.13^\circ$ at
$15^\circ$ elevation. Across this narrow angular interval, the boundary
coefficients vary only weakly. The layered field can therefore be factorized,
to a good approximation, as
\begin{equation}
E_{\rm layered}^{\,\chi}(f)\simeq
E_{\rm single,exact}^{\,\chi}(f)\,
\frac{r_{\rm stack}^{\,\chi}(f,\alpha_0^{\rm spec})}
     {f_r^{\,\chi}(f,\alpha_0^{\rm spec})},
\quad \chi=s,p .
\label{eq:factor}
\end{equation}

This is the standard specular tangent-plane factorization used in radar
sounding~\cite{TangentPlane}, here applied to the layered boundary
coefficient. Both coefficients in the ratio are evaluated at
$\alpha_0^{\rm spec}$, the incidence angle at the specular point, rather than
at the varying angle $\alpha_0(\alpha)$ inside the angular integral. This
allows the ratio to be taken outside the integral. Here
$E_{\rm single,exact}^{\,\chi}$ is the corresponding single-boundary field
component obtained from the full Sommerfeld--Weyl angular integral, while
$f_r^{\,\chi}$ is the Fresnel coefficient for the same polarization. The
H-Pol field used for the HiCal comparison is then obtained by combining the
$s$- and $p$-polarized components with the projection weights given in
Appendix~\ref{app:fields}.

Equation~(\ref{eq:factor}) follows because the Sommerfeld--Weyl representation
of the source field is the same in the layered and single-boundary
calculations. The propagation phase, spherical boundary geometry, polarization
projection, and roughness factor are also identical. The only change at the
boundary is $f_r^{\,\chi}\to r_{\rm stack}^{\,\chi}$. If the ratio
$r_{\rm stack}^{\,\chi}/f_r^{\,\chi}$ varies little across the
stationary-phase window, it can be taken outside the angular integral, giving
Eq.~(\ref{eq:factor}). The factorization is applied frequency by frequency.
Because the layered coefficient can vary strongly with $f$, the
frequency-dependent ratio remains inside the frequency sum or inverse Fourier
transform used to construct a broadband pulse. We test this approximation
against the full Sommerfeld--Weyl calculation, with
$r_{\rm stack}^{\,\chi}$ retained inside the angular integral.

At the balloon altitude used for the HiCal-2 comparison, $h=37$~km, the
factorized and full-angular calculations agree to $0.2\%$ for the reference
two-layer stack. The difference increases as the source approaches the
reflecting surface because $L_{\rm sph}$ decreases and the stationary-phase window
widens. For the same $650$~MHz, $15^\circ$ geometry, $w$ increases from
$0.074^\circ$ at $h=37$~km to $1.98^\circ$ at $h=50$~m and $6.26^\circ$ at
$h=5$~m. The full angular integral must therefore be checked for low-altitude
and in-medium sources, and true near-field sources also require the near-field
source terms omitted in the far-zone treatment of Sec.~\ref{sec:formalism}. For
the balloon reflection geometry, the factorized expression provides a
controlled approximation to the full calculation.

The angular factorization and the tangent-plane treatment involve two
independent approximations. The first is controlled by the stationary-phase
window $w$ and the angular variation of
$r_{\rm stack}^{\,\chi}/f_r^{\,\chi}$. The second is controlled by the
shell sag $\rho_F^{\,2}/2R$ relative to the wavelength. The resulting
sag$/\lambda$ values are given in Table~\ref{tab:bodies}. These two
approximations can behave differently as the source approaches the surface. The coherent footprint $\rho_F=\sqrt{\lambda L_{\rm sph}/4}$ becomes smaller, improving
the tangent-plane approximation, while the angular window $w$ becomes wider,
making the factorization less accurate.

A third condition applies when the layering varies laterally, as in
Appendix~\ref{app:approx}. The lateral scale $L_{\rm lateral}$ over which the
stratification changes must exceed the coherent footprint, so that each
footprint samples essentially one layered column. Thus,
\begin{equation}
\rho_F^{\,2}/2R\ll\lambda
\qquad\text{and}\qquad
L_{\rm lateral}\gg\rho_F ,
\label{eq:hierarchy}
\end{equation}
define the geometric regime in which a curved external surface can be combined with a layered subsurface. All results in this paper are computed for a laterally uniform medium, for which the second condition holds identically.

The calculation contains two distinct sources of oscillation. The first is
geometric. The phase in Eq.~(\ref{eq:weyl}) varies rapidly with the incident
direction $(\alpha,\beta)$, dividing the surface into Fresnel zones whose
contributions interfere constructively and destructively. The reflected field
is the residual after these contributions largely cancel, giving the expected
inverse-distance scaling of a spherical wave. As the elevation changes, the stationary
phase region shifts relative to the Fresnel-zone pattern, producing
oscillations in the smooth-surface reflectivity.

The second source of oscillation is the frequency dependence of the layered
medium. At fixed geometry and polarization, the layering ratio
$\mathcal{Q}^{\,\chi}(f)$ of Eq.~(\ref{eq:Qratio}) varies with frequency, with
a scale set by the optical thickness of the layers. This frequency dependence
is distinct from the geometric oscillation of the angular integral. Surface
roughness suppresses contributions away from the specular region, while
frequency averaging smooths the geometric elevation dependence. A
smooth-surface single-frequency result can therefore look quite different
from a rough-surface band-averaged result. The same applies to
$\mathcal{Q}^{\,\chi}(f)$, whose frequency-dependent structure can be reduced
by band averaging. We therefore examine the layering response at individual
frequencies before forming a broadband pulse or a band-averaged observable.

\section{Numerical results}
\label{sec:results}

All results in this section use a spherical surface with Earth's mean radius,
$R=R_\oplus=6371$~km, and a source height of $h=37$~km, with the local
incidence angle given by Eq.~(\ref{eq:alpha0}). Surface curvature is retained
in the propagation geometry. The remaining geometric approximation is to
treat the concentric subsurface shells as parallel layers over the coherent
footprint, as discussed in Sec.~\ref{sec:layered} and Appendix~\ref{app:geometry}.

The roughness factor of Eq.~(\ref{eq:rough}) affects only the absolute field
amplitudes and cancels in the coefficient ratios reported below. These ratios
are therefore independent of the surface roughness model. The power ratio
$r/d$ and the reflected pulses in Sec.~\ref{sec:hical} use the validated
roughness parameters of Sec.~\ref{sec:formalism}. Setting the roughness factor
to unity corresponds to a smooth surface.

Reflection coefficients are quoted for $s$ polarization unless stated
otherwise. In the HiCal reflection geometry, the $s$-polarized component
dominates the H-Pol response. The full H-Pol field includes both $s$ and $p$
components, with their relative contributions determined by the projection
weights in Eq.~(\ref{eq:polweight}). These weights arise from projecting the
dipole field onto the local $s$ and $p$ directions at the incident point $Q$
(Fig.~\ref{fig:stack}) and then projecting the reflected field back onto
$\hat{y}$ at the observer, as detailed in Appendix~\ref{app:fields}.

Frequency-dependent layer ratios are computed one Fourier frequency at a time, which is allowed because Maxwell's equations and the boundary conditions are linear, as described in Sec.~\ref{sec:formalism}. Broadband observables, such as band-averaged reflectivities and time-domain
waveforms, are formed only after evaluating the complex field components
across the frequency band.

\subsection{Reduction to the single-boundary limit}
\label{sec:result_reduction}

The first numerical check is the reduction of the layered calculation to the
validated single-boundary result. For the reference two-layer geometry used
below, we set $n_1=n_s=1.35$ and write the buried-layer index as
$n_2=n_s+\Delta n$. As $\Delta n\to0$, the buried layer becomes index-matched
to the surrounding medium and produces no reflection at its interfaces. Its
thicknesses then become irrelevant, and the stack must reduce to a single
boundary between the upper medium and the substrate for any $d_1$ and $d_2$.
This test checks the matrix ordering, phase convention, and normalization of
the stratified calculation.

\begin{figure}[!tb]
\centering
\includegraphics[width=0.95\linewidth]{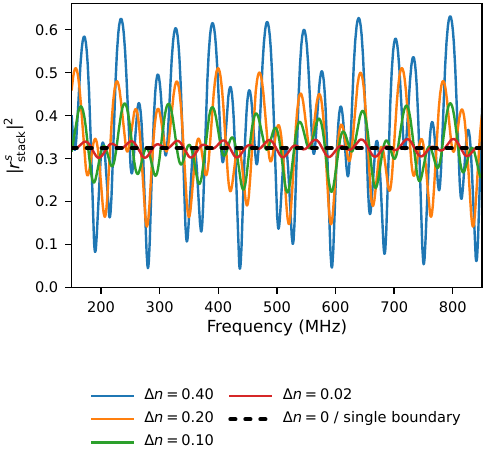}
\caption{Reduction of the layered calculation to the single-boundary limit at
$15^\circ$ elevation. The solid curves show $|r_{\rm stack}^{\,s}|^2$ for the
reference two-layer geometry as the buried-layer contrast is reduced, with
$\Delta n=0.40$ in blue, $0.20$ in orange, $0.10$ in green, and $0.02$ in red.
The layering fringes shrink as the contrast falls, and the result approaches
the single-boundary value of Sec.~\ref{sec:formalism}, drawn as the black
dashed line. At $\Delta n=0$ the two calculations agree to machine precision,
with a largest fractional deviation of $9.99\times10^{-16}$ across the band.
No measured data are used in this calculation.}
\label{fig:reduce}
\end{figure}

Figure~\ref{fig:reduce} shows the result at $15^\circ$ elevation. The same
test was also performed at other elevation angles. As the buried-layer
contrast decreases, the frequency-dependent layering fringes disappear. At
$\Delta n=0$, the layered and single-boundary calculations agree to machine
precision, with a largest fractional deviation of
$9.99\times10^{-16}$ across the band. Repeating the test at the ten
validation elevations in Table~\ref{tab:validation} gives the same agreement,
with a largest fractional deviation of $1.55\times10^{-15}$ over the full
scan.

\subsection{Effect of stratified media on field polarity}
\label{sec:result_q}

We quantify the effect of the subsurface layers on the total reflected field for both
polarizations, using the ratio $\mathcal{Q}^{\,\chi}(f)$ defined in
Eq.~(\ref{eq:Qratio}). The results below are for the $s$-polarized field,
$\mathcal{Q}^{\,s}(f)$. The reference stack has a $2$~m surface layer with
$n_1=1.35$, a $2$~m buried layer with $n_2=1.75$, and a semi-infinite substrate
with $n_s=1.35$. This configuration follows the subsurface-reflector scenario
proposed in Ref.~\cite{Shoemaker2020} and is the two-layer model tested
experimentally in Ref.~\cite{Smith2021}.

\begin{figure*}[!t]\centering
\includegraphics[width=1.0\linewidth]{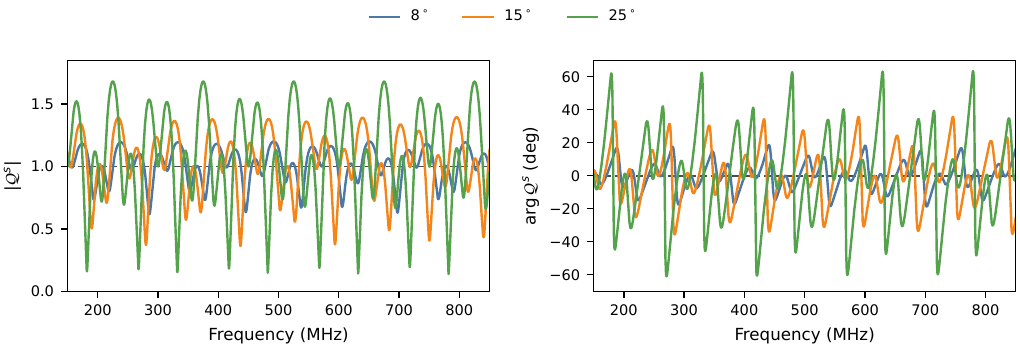}
\caption{Layering signature of the reference firn stack, at $8^\circ$ (blue),
$15^\circ$ (orange), and $25^\circ$ (green) elevation. The curves show the
layering ratio $\mathcal{Q}^{\,s}$ of Eq.~(\ref{eq:Qratio}), evaluated as
$r_{\rm stack}^{\,s}/f_r^{\,s}$ at the specular incidence angle. Within the
factorization of Eq.~(\ref{eq:factor}) this equals the ratio of the reflected
fields to $0.2\%$ at balloon altitude. Left: the amplitude ratio
$|\mathcal{Q}^{\,s}|$, with the black dashed line at unity marking the single
boundary with no subsurface layering. Right: the additional phase
$\arg\mathcal{Q}^{\,s}$ relative to the single-boundary result, with the black
dashed line at zero marking the same reference. Departures from these lines are
the effect of the subsurface layers. The oscillations are Airy
multiple-reflection fringes, produced by interference between waves reflected
from different interfaces in the stack. Across $150$--$850$~MHz,
$|\mathcal{Q}^{\,s}|$ ranges from $0.14$ to $1.68$ and the largest phase
excursion is $63.5^\circ$, so the buried layer changes the reflected amplitude
and phase but does not invert the polarity of the reflected pulse for the layer
indices and elevations considered here. No measured data are used in this
calculation.}
\label{fig:freq}
\end{figure*}

Over $150$--$850$~MHz, $|\mathcal{Q}^{\,s}(f)|$ ranges from $0.62$ to $1.20$ at
$8^\circ$ elevation, from $0.36$ to $1.39$ at $15^\circ$, and from $0.14$ to
$1.68$ at $25^\circ$. The largest phase shift is
$|\arg\mathcal{Q}^{\,s}|=63.5^\circ$ over the frequencies and elevations
considered here (Fig.~\ref{fig:freq}). These values are stable with respect
to the frequency sampling. Increasing the grid from $2500$ to $40\,000$ points
changes the smallest value by only $0.002$ and does not change the quoted
ranges.

Thus, the subsurface layers can substantially change both the reflected
amplitude and phase. Relative to the single-boundary result, the amplitude is
suppressed by up to $38\%$ and enhanced by up to $20\%$ at $8^\circ$, by
$64\%$ and $39\%$ at $15^\circ$, and by $86\%$ and $68\%$ at $25^\circ$.
The layering effect therefore becomes stronger at higher elevation angles,
as the propagation phase accumulated within the layers changes with
incidence angle. The effect also depends on the buried-layer index, which is
scanned in Sec.~\ref{sec:result_threshold}. Neither the amplitude change nor
the phase shift is sufficient to reverse the pulse polarity for the firn
contrasts and elevation angles considered here.

The oscillations in Fig.~\ref{fig:freq} are Airy fringes caused by multiple
reflections within the subsurface layers. Their spacing is set by the
round-trip phase through the layers. Successive returns add in phase whenever
$2\delta_j$ of Eq.~(\ref{eq:Mj}) changes by $2\pi$. For a layer of thickness
$d_j$ and refractive index $n_j$, traversed at the internal angle $\alpha_j$
of Eq.~(\ref{eq:transverse_conservation}), the characteristic frequency
spacing is
\[
\Delta f \sim \frac{c}{2n_jd_j\cos\alpha_j},
\]
where $c$ is the speed of light. For the meter-scale layers considered here,
this is about $50$~MHz, so several fringe periods occur across the
$150$--$850$~MHz band. Averaging over frequency can therefore smooth the
layering structure.

For the $15^\circ$ reference scan, the $|\mathcal{Q}^{\,s}|$ range is
unchanged to the quoted precision for the $200$--$650$~MHz,
$150$--$850$~MHz, $200$--$1200$~MHz~\cite{Gorham2009}, and
$300$--$1200$~MHz~\cite{PUEOtrigger2026} bands. The buried-layer index at
which the reflected field first becomes positive somewhere in the band is
also unchanged, giving $n_2^{\rm th}=2.043$ to three decimal places. This is
the index at which a buried layer can begin to overturn the polarity
inversion produced by the air--firn boundary alone, and it is well above the
firn and ice values considered here. The threshold is defined and scanned in
Sec.~\ref{sec:result_threshold}, where we also show that it is a necessary
but not sufficient condition for a broadband pulse to lose its polarity inversion.

The roughness factor cancels in $\mathcal{Q}$ because, as noted in
Sec.~\ref{sec:layered}, it depends on the surface geometry and not on the
structure below it. At fixed geometry, it is therefore identical in the
layered and single-boundary fields. This applies to the upper air--firn
surface. Roughness of the buried interfaces, described by
Eq.~(\ref{eq:intrough}), is a separate factor that has no counterpart in the
single-boundary calculation and therefore does not cancel. Its effect is
quantified in Sec.~\ref{sec:thinrough}. In the specular factorization of
Eq.~(\ref{eq:factor}), where the same surface roughness factor multiplies the numerator
and denominator, the cancellation is exact.

We verify the cancellation numerically to $2\times10^{-16}$, including for
values of $\sigma_h$ well above any measured Antarctic value, for which the
roughness factor differs strongly from unity across the footprint. In the
full angular integral, the cancellation is not exact because the roughness
factor and the boundary coefficient vary slightly across the
stationary-phase region. At the balloon altitude of $37$~km, this difference
is $0.2\%$ and becomes larger for near-surface sources, as discussed in
Sec.~\ref{sec:validity}. The layering signature and polarity inversion threshold are
therefore insensitive to the surface roughness model, including the
anisotropic and local-slope variants of Ref.~\cite{DasguptaThesis2020}.
Absolute quantities such as the power ratio $r/d$ retain the roughness factor
and are quoted below with it included.

\subsection{A polarity-reversal threshold for buried layers}
\label{sec:result_threshold}

For reflection from a rarer to a denser medium, as for a UHECR-induced radio
signal reflecting from the air--ice boundary, a polarity inversion is
expected. As shown in Eqs.~(\ref{eq:fs})--(\ref{eq:fp}), the $s$-polarized Fresnel coefficient is negative at every incidence angle when $n_1>n_0$. ANITA, however, reported anomalous events dominated by H-Pol that lack this inversion~\cite{ANITA2016,ANITA2018,ANITA2021}. At the near-horizon geometries of these events, the H-Pol response is predominantly $s$ polarized (Appendix~\ref{app:fields}). We therefore determine the refractive index a
buried layer would need for the reflected wave to lose the inversion produced
by the air--firn boundary alone.

Two coherent sums enter the calculation. The first occurs inside the stack.
At the local incidence angle $\alpha_0$ of a given plane wave of
Eq.~(\ref{eq:weyl}), the reflection from the upper surface adds coherently to
the contributions that enter the stack, reflect from the buried interfaces,
and re-emerge into the upper medium, as shown in Fig.~\ref{fig:stack}(b).
Their relative phases are the $\delta_j$ of Eq.~(\ref{eq:Mj}), set by the
layer thicknesses, refractive indices, and $\alpha_0$. The characteristic
matrix sums these contributions exactly into $r_{\rm stack}^{\,s}(f,\alpha_0)$ of Eq.~(\ref{eq:rstack}). The second sum is
the angular integral of Eq.~(\ref{eq:erefstack}), which combines the plane-wave components to give the total reflected field.

\begin{figure}[!tb]
\centering
\includegraphics[width=0.95\linewidth]{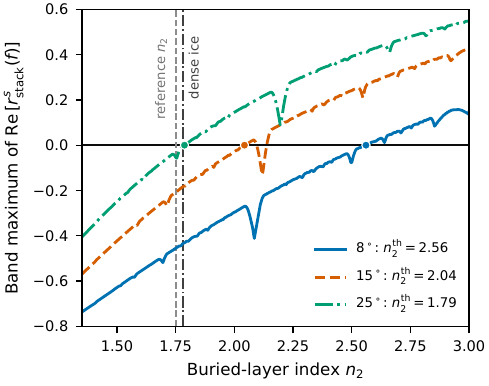}
\caption{Polarity-reversal threshold for the reference two-layer stack, using the air/$n_1$/$n_2$/$n_s$ geometry of Fig.~\ref{fig:freq}. The vertical axis shows the maximum of $\mathrm{Re}\!\left[r_{\rm stack}^{\,s}(f)\right]$ across the $150$--$850$~MHz band. The horizontal line at zero marks the boundary for polarity reversal. A net positive reflection occurs above this line. The threshold index $n_2^{\rm th}$ is defined by each curve's first zero-crossing. A sign change of the coefficient is necessary for a polarity reversal but does not by itself imply one for a broadband pulse. For elevations of $8^\circ$, $15^\circ$, and $25^\circ$, these thresholds are $n_2^{\rm th}\simeq 2.56$, $2.04$, and $1.79$, respectively. Vertical lines denote the reference buried-layer index ($n_2=1.75$) and dense ice ($n_2=1.78$). This calculation uses no measured data.}
\label{fig:threshold}
\end{figure}

The threshold is evaluated from the first sum, using $r_{\rm stack}^{\,s}$ at the specular incidence angle. This is sufficient
because the reflected field is dominated by the neighborhood of the specular direction. The factorization of Eq.~(\ref{eq:factor}) reproduces the full angular integral to $0.2\%$ at balloon altitude (Sec.~\ref{sec:validity}), so the sign of the field follows the sign of the coefficient. We scan the refractive index $n_2$ of the middle layer in the reference stack and find the first value for which $r_{\rm stack}^{\,s}$ becomes positive at any frequency in the $150$--$850$~MHz band
(Fig.~\ref{fig:threshold}). We denote this value by $n_2^{\rm th}$. It is
$\simeq2.56$, $2.04$, and $1.79$ for elevations of $8^\circ$, $15^\circ$, and
$25^\circ$, respectively, showing a strong dependence on geometry.

This criterion is deliberately conservative. A positive reflection
coefficient at one frequency is necessary for the reflected pulse to lose its
polarity inversion, but it is not sufficient because the pulse polarity is
determined by the coherent sum over the full frequency band. Thus,
$n_2^{\rm th}$ is a lower bound on the buried-layer index required for an
actual pulse-polarity reversal. The event-angle waveform test in Sec.~\ref{sec:anita_angles} confirms this.

The threshold decreases with elevation but approaches a finite limit. At
$40^\circ$, $55^\circ$, and $70^\circ$, we find
$n_2^{\rm th}=1.651$, $1.599$, and $1.577$, respectively, approaching
$n_2^{\rm th}=1.569$ at normal incidence. This follows from the decreasing
magnitude of the surface reflection that the buried contribution must
overcome. For $s$ polarization, $|f_r^{\,s}|$ decreases from $0.874$ at
$3.5^\circ$ elevation to $0.737$ at $8^\circ$, $0.406$ at $25^\circ$, and
$0.149$ at normal incidence. Thus, a smaller buried-layer contrast is needed
at higher elevation angles.

The finite limit can be understood from a simple amplitude argument. The
reference stack has two buried interfaces. Since $n_s=n_1$, their reflection
coefficients satisfy $r_{2,s}=-r_{1,2}$ and therefore have the same magnitude.
When the waves reflected from the two buried interfaces add with the most
favorable phase, their combined amplitude is at most approximately
$2|r_{1,2}|$. To cancel the reflection from the upper surface, this requires
\[
|r_{1,2}|\simeq\frac{|f_r^{\,s}|}{2} .
\]
At normal incidence $f_r^{\,s}=(n_0-n_1)/(n_0+n_1)=-0.1489$ for $n_0=1$ and
$n_1=1.35$, so the required buried-interface reflection is
$|r_{1,2}|\simeq0.0745$. Using
\[
|r_{1,2}|=\left|\frac{n_1-n_2}{n_1+n_2}\right| ,
\]
this gives $n_2\simeq1.57$, close to the numerical limit
$n_2^{\rm th}=1.569$. The agreement shows that the finite limit comes from the maximum contribution
that the buried interfaces can provide. This is only an order-of-magnitude estimate because it neglects the transmission
factors on each pass through the upper surface, which shift the estimate by less
than $0.01$ in $n_2$. It also assumes the most favorable phases, so it does not
capture the dependence on elevation, layer thickness, or frequency. These
effects are included in the characteristic-matrix calculation used throughout
this section. The elevation dependence is shown in Fig.~\ref{fig:threshold},
the thickness dependence is quantified below, and the frequency structure is
given by the Airy fringes in Fig.~\ref{fig:freq}. The numbers quoted here are
for $s$ polarization. The $p$-polarized coefficient passes through the Brewster
zero and must be considered separately.

The threshold is therefore not a universal material property but depends on
the reflection geometry. The realistic firn contrast considered here,
$n_2=1.75$, remains below threshold at all three elevations. Dense ice, with
$n\simeq1.78$, is also below threshold at $8^\circ$ and $15^\circ$. At
$25^\circ$, however, the threshold is $n_2^{\rm th}=1.79$, placing a buried
dense-ice layer close to it. The difference is only about $0.01$ in refractive
index, smaller than the spread in reported values for dense ice. A strongly
reflecting buried layer such as water or brine is well above these dielectric
thresholds and could therefore produce a distinct polarity signature,
relevant to subglacial and planetary sounding.

The thresholds quoted above are deterministic results of the forward
calculation and therefore have no statistical uncertainty. They do, however,
depend on the assumed layer structure. We vary the layer thickness and the
upper-layer index at $15^\circ$ elevation, where the reference stack of
Fig.~\ref{fig:threshold} gives $n_2^{\rm th}=2.04$, and then check the
numerical sampling.

The layer thickness matters least. Varying the equal layer thicknesses over
$d_1=d_2=1$, $2$, $3$, $4$, and $8$~m moves the threshold from $2.16$ at
$1$~m to $2.04$ for the thicker layers. Unequal thicknesses between $1$ and
$4$~m change it by less than $0.04$. The threshold is determined by whether
the coherent sum changes sign somewhere in the frequency band. Once the
layers are thick enough to produce several Airy fringes across the
$150$--$850$~MHz band, adding more fringes does not change whether a sign
change occurs. For thinner layers, fewer fringes fall in the band and the
threshold rises.

The upper-layer index matters more. Taking $n_1=n_s$ and varying $n_1$ from
$1.30$ to $1.45$, which brackets reported near-surface firn values, changes
the threshold from $1.87$ to $2.41$. The threshold therefore depends on the
surrounding firn index and is not a property of the buried layer alone. The
value of $n_1$ must therefore be specified when quoting the threshold, as is
done throughout this paper.

The result is converged with respect to frequency sampling. In the $15^\circ$
scan over the $150$--$850$~MHz band, the threshold changes from
$n_2^{\rm th}=2.056$ with $200$ frequency samples to $2.043$ with $2500$,
and is unchanged at $8000$. Taking the thickness dependence together with
this sampling effect, the threshold varies by at most $0.12$ in $n_2$ at
fixed $n_1$, with the largest change occurring for the thinnest layers
tested. The choice of $n_1$ has a substantially larger effect.

\subsection{Thin lenses and rough buried interfaces}
\label{sec:thinrough}

\begin{figure*}[!t]
\centering
\includegraphics[width=0.94\linewidth]{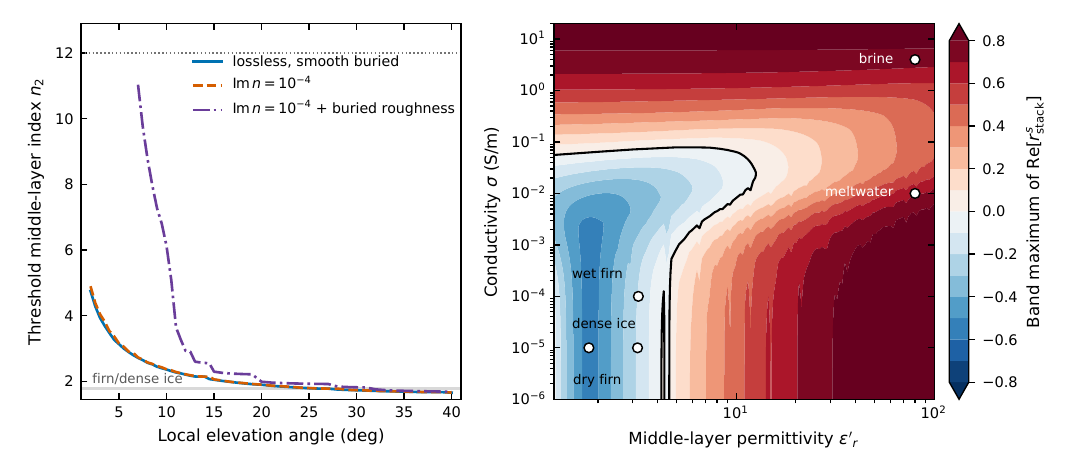}
\caption{Effect of absorption, buried-interface roughness, and buried-layer
material on the polarity threshold, described in Sec.~\ref{sec:robust}.
Left: threshold index $n_2^{\rm th}$ against local elevation for three
cases: real indices with smooth buried interfaces (blue, solid), the same with
absorption added (orange, dashed), and absorption together with rough buried
interfaces (purple, dash-dotted). Markers are the sampled elevations, not
uncertainties. The dotted line is the top of the scanned range, and the gray
band marks realistic firn and dense-ice values. Right: material scan
at $15^\circ$, over the real permittivity $\varepsilon'_r$ and the conductivity
$\sigma$ of the buried layer. The color is the band maximum of
$\mathrm{Re}\!\left[r_{\rm stack}^{\,s}(f)\right]$ over $150$--$850$~MHz. Blue
means the reflected pulse keeps its inversion across the whole band, red means
the sign changes somewhere in it, and the contour marks the boundary. Dry firn,
dense ice, and wet firn lie on the non-inverting side; meltwater and brine lie
on the inverting side.}
\label{fig:robustness}
\end{figure*}

The same matrix calculation applies to layers much thinner than a wavelength.
For a very thin layer, the reflections from its two interfaces have opposite
signs and nearly cancel, leaving a residual set by the round-trip phase through
the layer. The thickness enters through $\delta_j=k_{z,j}d_j$, so no spatial
sampling within the layer is required.

For a firn-like thin lens, a $1~\mathrm{mm}$ dense-ice layer ($n=1.78$)
embedded in $n=1.35$ firn, viewed at the angle refracted from $15^\circ$ local
elevation in air, has a slab reflection of only $0.8\%$ of the air--firn
surface reflection at $300$~MHz and $2.2\%$ at $850$~MHz. This return is too
small to cancel the surface reflection, so an isolated millimeter-scale
density feature cannot remove the polarity inversion. Many such lenses can
behave differently if they are coherently spaced, since their returns then
add in phase at the Bragg frequencies of the stack. We do not consider that
case here.

We also examine whether roughness at the buried interfaces can lower the
polarity threshold. The upper-surface roughness is kept fixed according to
Eq.~(\ref{eq:rough}), while the two buried interfaces in the reference stack
are assigned the same rms height $\sigma_{\rm int}$. Their coherent reflection
amplitudes are damped by Eq.~(\ref{eq:intrough}), with the results shown in
Table~\ref{tab:internalrough}. The range tested is motivated by the measured
surface roughness. The validated surface parameterization of
Eq.~(\ref{eq:rough}) gives an rms height of about $4.6$~cm over the coherent
footprint at $15^\circ$. A buried interface was itself a former surface, so
this is a reasonable scale to assign it, and rms heights of a few centimeters
are therefore physically relevant. At these values, the crossing moves above
the scan limit at $8^\circ$ and shifts to $2.56$ at $15^\circ$.
Buried-interface roughness therefore either raises the required $n_2$ or
removes the crossing within the scanned range. It does not make polarity
reversal easier, so the smooth buried-interface thresholds are lower bounds
within this model.

\begin{table}[!tb]
\caption{Effect of rough buried interfaces on the polarity threshold. The
entries are the first crossing $n_2^{\rm th}$ for the reference stack when
the two buried interfaces are assigned the same rms roughness
$\sigma_{\rm int}$ and the coherent reflected amplitudes are multiplied by
Eq.~(\ref{eq:intrough}). The upper air--firn surface roughness is unchanged.}
\label{tab:internalrough}
\begin{ruledtabular}
\begin{tabular}{cccc}
$\sigma_{\rm int}$ (mm) & $8^\circ$ & $15^\circ$ & $25^\circ$ \\ \hline
0 & 2.56 & 2.04 & 1.79 \\
1 & 2.56 & 2.04 & 1.79 \\
5 & 2.57 & 2.06 & 1.79 \\
10 & 2.58 & 2.16 & 1.80 \\
20 & 2.85 & 2.19 & 1.81 \\
50 & $>3.2$ & 2.56 & 1.94 \\
\end{tabular}
\end{ruledtabular}
\end{table}

\subsection{Lossy and conducting buried layers}
\label{sec:robust}

The reference calculation assumes that the firn and the buried layer are
lossless, and that the buried interfaces are smooth. These are simply input
choices rather than fundamental restrictions of the formalism. Here, we relax
these assumptions within the same characteristic-matrix calculation.

Absorption is included through a complex refractive index. Taking $\mathrm{Im}\,n=10^{-4}$ for the firn layers and substrate, a value appropriate for polar ice at these frequencies, shifts the threshold at $15^\circ$ from $n_2^{\rm th}=2.043$ to $2.054$. A highly conservative value of $\mathrm{Im}\,n=10^{-3}$ shifts it further to approximately $2.17$. This shift
is upward in both cases because the buried contribution traverses the lossy
layers twice, whereas the surface reflection does not. Consequently, absorption
disproportionately attenuates the buried signal, which is the component that
must first cancel and then dominate the surface reflection.

We include buried-interface roughness via Eq.~(\ref{eq:intrough}), setting its scale using the surface roughness model since a buried interface is simply a former surface. Specifically, we compute $\sigma_{\rm int}$ using $\sigma_h$ from Eq.~(\ref{eq:rough}), evaluated at the exact Fresnel footprint for each elevation angle at the $650$~MHz reference frequency. This yields values of $54.6$, $45.9$, and $39.4$~mm at elevations of $8^\circ$, $15^\circ$, and $25^\circ$, respectively. Because the footprint expands at lower frequencies, this parameterization does not represent the most aggressive possible damping from buried roughness. 

The left panel of Fig.~\ref{fig:robustness} shows the elevation dependence of all three cases. When both absorption and buried-interface roughness are included in the computation, the threshold for polarity inversion increases to $8.77$ at $8^\circ$, $2.291$ at $15^\circ$, and $1.922$ at $25^\circ$, as shown by the purple dash-dotted curve in Fig.~\ref{fig:robustness}. This curve diverges at shallow angles, illustrating the rapid increase in the required threshold index. At even shallower local elevations, the combined damping is so strong that no polarity-reversal crossing occurs within the scanned index range. Therefore, incorporating signal loss and rough buried interfaces strictly strengthens the physical constraint rather than weakening it. Each case is evaluated by repeating the same forward calculation at $52$ discrete elevations between $2^\circ$ and $40^\circ$, rather than relying on a continuous fit. Because the threshold is a deterministic output of this calculation, no quantity plotted in Fig.~\ref{fig:robustness} carries an uncertainty.

Instead of assuming a fixed imaginary refractive index, we directly scan the material parameter space, as shown in the right panel of Fig.~\ref{fig:robustness}. By defining the complex permittivity as $n_2^2=\varepsilon'_r+i\sigma/(\omega\varepsilon_0)$ and varying both the real permittivity $\varepsilon'_r$ and the conductivity $\sigma$, we map the boundary where the reflected polarity inverts. For a non-conducting layer ($\sigma=0$), this inversion occurs only above a minimum permittivity of $\varepsilon'_r=6.56$, $4.17$, and $3.19$ at elevations of $8^\circ$, $15^\circ$, and $25^\circ$, respectively. These values exactly recover the real-index thresholds of $n_2=2.561$, $2.043$, and $1.787$ derived in Sec.~\ref{sec:result_threshold}.

The dependence on conductivity is not monotonic at small $\sigma$. At $15^\circ$, the required permittivity threshold rises from $\varepsilon'_r=4.17$ at $\sigma=0$, to $5.12$ at $\sigma=10^{-3}$~S/m, and up to $10.7$ at $\sigma=10^{-2}$~S/m. However, for $\sigma\gtrsim0.1$~S/m, the conducting response becomes dominant. The band maximum of $\mathrm{Re}\!\left[r_{\rm stack}^{\,s}(f)\right]$ turns positive across the entire scanned permittivity range.

Consequently, the inversion threshold effectively categorizes broad classes of materials. We take representative values for dry firn, dense ice, and wet firn from standard radio-frequency parameterizations~\cite{Kovacs1995}. At $15^\circ$, dry firn ($\varepsilon'_r\simeq1.8$), dense ice ($3.17$), and wet firn ($3.2$) all lie inside the non-inverting region, with corresponding band maxima of $\mathrm{Re}\!\left[r_{\rm stack}^{\,s}(f)\right]$ as $-0.563$, $-0.185$, and $-0.191$, respectively. Conversely, meltwater and brine, represented by $\varepsilon'_r=80$ and conductivities ranging from weakly conducting meltwater to saline water~\cite{Stogryn1971}, fall strictly in the polarity inversion region, producing band maxima of $+0.681$ and $+0.756$.

Finally, we note that a sign change in this frequency-domain parameter plane is a necessary, but not sufficient, condition for a broadband pulse to lose its phase inversion. Even if the buried layer is replaced entirely by brine, the fully computed time-domain waveforms remain inverted at all four near-horizon ANITA-IV geometries, as detailed in Sec.~\ref{sec:anita_angles}.

\subsection{Validation with HiCal-1 direct and reflected pulse pairs}
\label{sec:hical}

\begin{figure*}[!t]
\centering
\includegraphics[width=1.0\linewidth]{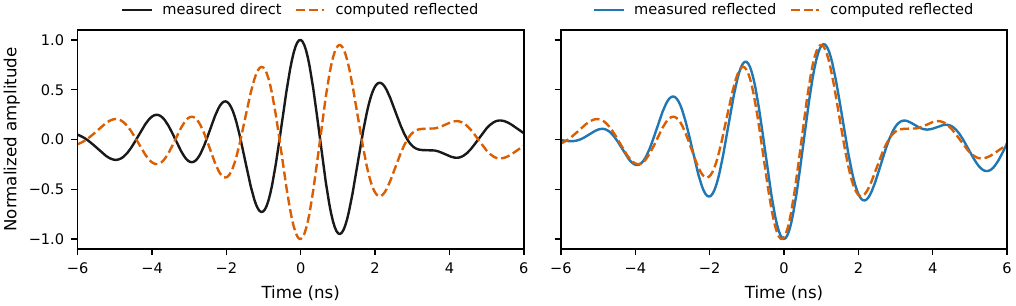}
\caption{A matched HiCal-1 direct and reflected pulse pair at approximately
$5^\circ$ elevation, band-limited to $200$--$650$~MHz. Both panels show
normalized field amplitude against time, with the origin at the alignment lag
defined in the text. Left: the measured direct waveform (black, solid)
and the reflected pulse calculated from it (orange, dashed), using the transfer function developed here. The calculated
pulse is inverted, as the negative reflection coefficient requires.
Right: the calculated pulse (orange, dashed) against the measured
reflected partner of the same pair (blue, solid), with a signed correlation of
$0.83$. The positive value means the calculated and measured pulses share the
same polarity; the opposite polarity would give a negative correlation. This is
the best-correlating example of the $106$ pairs. The full distribution is shown in Fig.~\ref{fig:population}.}
\label{fig:hical}
\end{figure*}

Section~\ref{sec:formalism} validated the single-boundary limit against the
published HiCal-2 reflectivity~\cite{Prohira2018}, a band-averaged power
reflection ratio $r/d$. Here we test the same implementation against measured
HiCal-1 waveforms, testing the phase and shape of the reflected pulse rather
than only its band-averaged power. The layered results of
Secs.~\ref{sec:result_q}--\ref{sec:robust} and the event-angle test of
Sec.~\ref{sec:anita_angles} use the same procedure, so we first test
it in a case where the expected result is known.

We use $106$ HiCal-1 direct and reflected pulse pairs~\cite{Gorham2017}. Each
pair consists of a direct pulse and the corresponding pulse reflected from the
Antarctic surface. The pairs are at approximately $5^\circ$ and $3.5^\circ$
elevation, the two near-grazing geometries sampled during the flight, with an
uncertainty of about $0.5^\circ$ from the elevation-to-incidence
conversion~\cite{Prohira2018}. We compare the H-Pol field component,
corresponding to the HiCal transmitter and the ANITA channels that recorded
these calibration pulses. Its decomposition into $s$ and $p$ components is
given in Appendix~\ref{app:fields}, in particular Eq.~(\ref{eq:apphpol}).
These near-grazing geometries are useful because surface curvature and
roughness have their largest effects there. The HiCal-2 reflectivity~\cite{Prohira2018} test and HiCal-1 waveforms tests
therefore probe different aspects of the framework developed here.

For each pair, we Fourier transform the measured direct waveform, multiply
its spectrum by the reflected-field transfer function
$\mathcal{T}^{H}_{\rm ref}(f)$ of Eq.~(\ref{eq:eref}) over the
$200$--$650$~MHz band, and transform the computed spectrum back to the time domain. This
gives the reflected pulse computed from the measured direct pulse at the
corresponding geometry. We then compare it with the measured reflected pulse
from the same pair.

Before comparison, the two waveforms are aligned in time. We refer to the
required time shift as the lag. Because the pulses are bipolar and band
limited, their cross-correlation oscillates and has positive and negative
peaks of similar size, separated by roughly half a cycle. Choosing the largest
peak by absolute value can therefore shift the waveforms by half a cycle and
reverse the apparent polarity. We instead use the analytic envelope of the
cross-correlation, which tracks the magnitude of the match without these
oscillations. The lag is taken at the maximum of this envelope, and the signed
correlation is evaluated at that lag.

For the best-correlating pair, the calculated and measured reflected waveforms
have a signed correlation of $0.83$ (Fig.~\ref{fig:hical}). Multiplying the
calculated waveform by $-1$ gives $-0.83$ at the same lag, as expected for a
fixed-lag signed correlation. Because the correlation is normalized by the
norms of both waveforms, it tests the sign and shape of the reflected pulse
but not its amplitude. The amplitude is tested separately using the
band-averaged $r/d$ comparison in Table~\ref{tab:validation}.

The elevation does not provide an additional constraint in this waveform
comparison. Over $200$--$650$~MHz, the single-boundary $s$ coefficient is a
real negative constant, $-0.84$ at $5^\circ$ and $-0.88$ at $3.5^\circ$,
while the roughness factor of Eq.~(\ref{eq:rough}) is within one percent of
unity at these grazing angles. The resulting transfer function is therefore
close to a real negative constant, and normalization by the two waveform
norms removes this overall scale. Across all $106$ HiCal-1 pairs, the median
signed correlation is $0.70$.

We repeat the calculation using the reference-stack transfer function of
Sec.~\ref{sec:layered} in place of the single-boundary coefficients of
Sec.~\ref{sec:formalism}. For the reference stack, the median correlation is $0.68$ for the same
HiCal-1 pairs. The small difference is much less than the spread across the sample,
so these HiCal-1 waveforms do not distinguish the bare boundary from the
reference stack at these angles. This follows from the geometry. Near the
horizon, the layered and single-boundary responses are nearly the same. The
HiCal-1 sample therefore provides a waveform-level validation of the
single-boundary calculation, while the stratified-media calculation is tested
through its reduction to the single-boundary limit to machine precision in
Sec.~\ref{sec:result_reduction} and through its application to the ANITA anomalous-polarity event angles in Sec.~\ref{sec:anita_angles}.

\begin{figure*}[!t]
\centering
\includegraphics[width=1.0\linewidth]{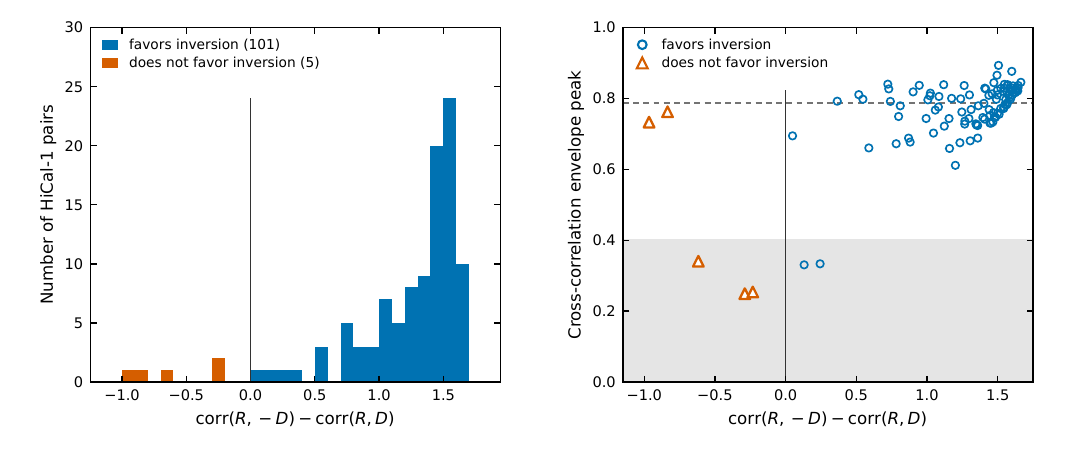}
\caption{Polarity test for the $106$ matched HiCal-1 pairs used in
Refs.~\cite{DasguptaThesis2020,DasguptaJain2021}. For each pair, the measured
direct ($D$) and reflected ($R$) waveforms are band-limited to
$200$--$650$~MHz and aligned using the peak of their cross-correlation
envelope. The horizontal axis shows
$\mathrm{corr}(R,-D)-\mathrm{corr}(R,D)$ at the envelope peak, with positive
values favoring polarity inversion. The solid black line at zero separates
the two classes. Left: $101$ pairs favor inversion and five do not. Right: the
same pairs plotted against the envelope peak, which indicates the quality of
the match. The dashed line marks the median peak of $0.79$, and the shaded
region shows peaks below $0.40$. Three of the five non-inverting pairs lie in
this region, indicating that neither polarity provides a strong match. This
test uses only the measured waveforms and is independent of the reflection
calculation.}
\label{fig:population}
\end{figure*}

Let $D$ and $R$ denote the measured direct and reflected waveforms of a pair.
After aligning them at the envelope lag defined above, we evaluate
$\mathrm{corr}(R,-D)-\mathrm{corr}(R,D)$, shown in
Fig.~\ref{fig:population}. A positive value means that the measured reflected
pulse matches the inverted direct pulse better than the direct pulse itself.
By this criterion, $101$ of the $106$ pairs, or
$(95.3\pm2.1)\%$, favor the expected polarity inversion. The remaining
$5/106=(4.7\pm2.1)\%$ do not. The quoted uncertainty is the binomial counting
uncertainty for a fixed sample of $106$ pairs.

This population test and the waveform correlation above are closely related.
At these angles, the transfer function is nearly a real negative number, so
the calculated reflected pulse is close to an inverted copy of the measured
direct pulse. The two tests therefore quantify essentially the same agreement
in different ways. For example, the best-correlating pair gives $0.83$ in
Fig.~\ref{fig:hical} and about $1.66$ in
$\mathrm{corr}(R,-D)-\mathrm{corr}(R,D)$ in
Fig.~\ref{fig:population}. Taken together, they show that the calculation
predicts a polarity inversion with little change in pulse shape at grazing
incidence, consistent with the measured HiCal-1 sample.

The five pairs that do not favor inversion fall cleanly into two distinct groups, as shown in the right panel of Fig.~\ref{fig:population}. Three pairs exhibit envelope peaks of only $0.25$ to $0.34$ compared with a population median of $0.79$, indicating that neither polarity provides a strong match. Exactly five of the $106$ pairs have envelope peaks below $0.40$, and three of these are among the non-inverting pairs. The remaining two pairs have larger envelope peaks of $0.73$ and $0.76$, which are comparable to the population median, and therefore require inspection on an event-by-event basis. Together these two pairs represent $2$ out of $106$ pairs or $1.9 \pm 1.3\%$ of the sample. We quote this fraction only as a diagnostic of the HiCal-1 data.

The five non-inverting pairs are distributed across four of the six
calibration runs and both near-grazing elevations, with three at $5^\circ$ and
two at $3.5^\circ$. The two pairs with near-median envelope peaks are both at
$5^\circ$. Given the $81$ and $25$ pairs in the two elevation groups, however,
this distribution is consistent with an elevation-independent rate and does
not provide evidence for an angular dependence.

The HiCal-1 sample cannot directly test the effect of shallow layering as a function of elevation. All $106$ pairs originate from two near-grazing elevations of approximately $5^\circ$ and $3.5^\circ$, where the polarity threshold evaluated in Sec.~\ref{sec:result_threshold} is highest with $n_2^{\rm th}\gtrsim3$. The two well-matched non-inverting pairs therefore do not support the shallow-firn layering scenario considered here. Producing a physical polarity reversal at these specific elevations requires a buried-layer index well above any realistic values for firn or dense ice. Surface topography and local slope effects represent distinct possibilities and are not addressed by the present theoretical model. These surface features along with related roughness models and slopes were studied separately in Ref.~\cite{DasguptaThesis2020}.

This section establishes a waveform-level validation of the exact
reflected-field calculation at near-grazing incidence, where the boundary
response and the expected polarity are well constrained.
Section~\ref{sec:anita_angles} applies the same calculation at the reported
ANITA event angles, spanning $1.5^\circ$ to $35.5^\circ$, and determines the
buried-layer contrast a subsurface stack would need to cancel the inversion
produced by the air--firn boundary, as the anomalous-polarity events would require.

\subsection{ANITA event geometries}
\label{sec:anita_angles}

\begin{figure*}[!t]
\centering
\includegraphics[width=1.0\linewidth]{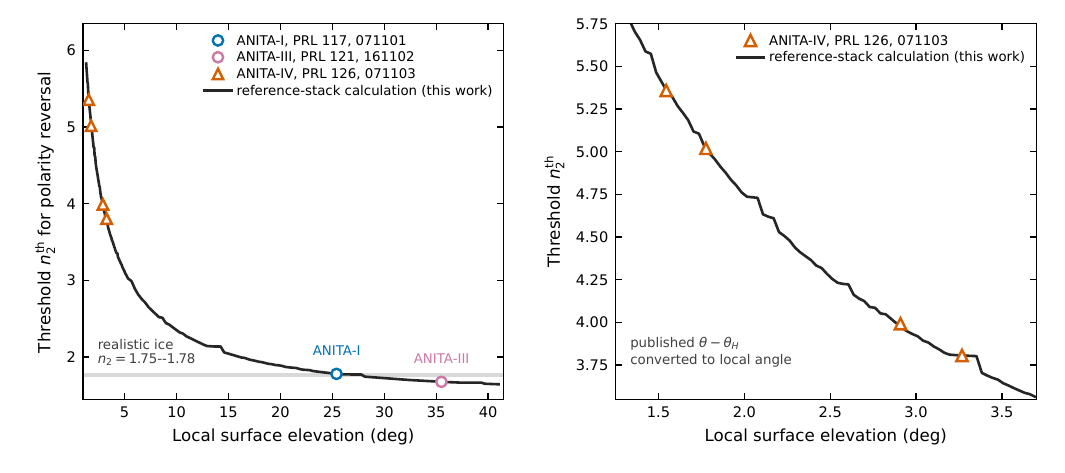}
\caption{Buried-layer index required for a polarity sign change at the angles of
the six ANITA anomalous-polarity events. The solid black curve shows the
calculated threshold $n_2^{\rm th}$, defined as the lowest buried-layer index
for which the band maximum of
$\mathrm{Re}\!\left[r_{\rm stack}^{\,s}(f)\right]$ crosses zero over the
$150$--$850$~MHz band. The curve is obtained by repeating the reference-stack
calculation at $229$ discrete local surface-elevation values between
$1.3^\circ$ and $41^\circ$; it is not a fit. The colored markers show the
published central values for the six ANITA events, evaluated at their
corresponding local surface elevations. Left: the full elevation range,
showing the two steep events (circles) and four near-horizon events (triangles)
reported by ANITA-I~\cite{ANITA2016}, ANITA-III~\cite{ANITA2018}, and
ANITA-IV~\cite{ANITA2021}. The horizontal gray band shows the realistic-ice
reference range $n_2=1.75$--$1.78$. Right: an expanded view of the four
near-horizon ANITA-IV events. Their published angular offsets below the payload
radio horizon~\cite{Prechelt2022} are converted to local surface elevations
using the same spherical geometry as in the rest of this work. A threshold
below the gray band means that the coefficient can change sign for the
reference-layer index, but does not by itself imply a broadband polarity
reversal. The effects of angular uncertainty and the granularity of the
calculated threshold curve are discussed in Sec.~\ref{sec:anita_angles}.}
\label{fig:anita_threshold}
\end{figure*}

The calculation developed here applies directly to the reported ANITA event
geometries. For the two steep ANITA-I and ANITA-III anomalous-polarity events,
the reported payload elevation angles are $-27.4^\circ$ and $-35.0^\circ$,
respectively. The corresponding surface-emergence angles used in tau-lepton
interpretations of the ANITA anomalous-polarity events are $25.4^\circ$ and
$35.5^\circ$, with an uncertainty of about $1^\circ$~\cite{ANITA2016,ANITA2018,RomeroWolf2019}. We adopt these
published values as the local surface elevation angles of the two events. They
are not derived from the reflection geometry. Mapping the reported payload
elevations onto the surface with Eq.~(\ref{eq:alpha0}), at $R=R_\oplus$ and
$h=37$~km, gives $26.8^\circ$ and $34.5^\circ$ instead. The two determinations
differ by $1.4^\circ$ and $1.0^\circ$, comparable to the quoted uncertainty,
and change the thresholds quoted below by less than $0.03$. The published angles
correspond to polarity-reversal thresholds of $n_2^{\rm th}\simeq1.781$ and
$1.677$, respectively, as defined in Sec.~\ref{sec:result_threshold}. For the
$35.5^\circ$ geometry the threshold lies below the reference buried-layer
index, $n_2=1.75$, so the reflection coefficient does change sign somewhere in
the band. As shown below, the fully computed broadband
reflected pulse nevertheless remains inverted at this angle, because a sign
change at a single frequency is necessary but not sufficient to cause
non-inversion of the pulse.

The four ANITA-IV near-horizon events require a different conversion because the reported quantities are offsets below the payload radio horizon, $\theta-\theta_H=-0.19^\circ$ to $-0.81^\circ$~\cite{ANITA2021,Prechelt2022}. Using the same spherical-Earth geometry adopted throughout this work, with $R=R_\oplus$ and $h=37$~km, these offsets correspond to local surface-elevation angles of $1.54^\circ$, $1.78^\circ$, $2.91^\circ$, and $3.27^\circ$. The corresponding polarity-reversal thresholds are $n_2^{\rm th}\simeq5.4$, $5.0$, $4.0$, and $3.8$, respectively (Fig.~\ref{fig:anita_threshold}). We quote these values to two significant figures because the first zero crossing can be located only to a precision of about one Airy fringe, corresponding to an uncertainty of approximately $0.03$--$0.05$ in $n_2$, as discussed below.

The threshold curve in Fig.~\ref{fig:anita_threshold} is not smooth, and the reason is physical rather than numerical. The threshold is
defined as the \emph{first} value of $n_2$ at which the band maximum of
$\mathrm{Re}\!\left[r_{\rm stack}^{\,s}(f)\right]$ crosses zero. This band
maximum inherits the Airy-fringe structure of the stack, so near the threshold
it does not simply pass through zero once. Instead, it oscillates about zero and
changes sign at several closely spaced values of $n_2$. At $2.5^\circ$
elevation, for example, sign changes occur at $n_2=4.272$, $4.351$, and
$4.356$, as resolved with $20\,000$ frequency samples. With a coarser frequency
grid, we find that the first crossing is split into several crossings within a
range of about $0.005$ in $n_2$. Which crossing is identified as the first
depends on the position of the fringe pattern relative to the band edges, and
this position shifts continuously with elevation angle. Consequently, we find
that the threshold advances in steps of roughly one fringe rather than varying
smoothly.

This granularity is therefore intrinsic to the threshold definition rather than
a sampling artifact. We find that it does not diminish when the frequency grid
is refined, with the characteristic step size remaining unchanged between
$2000$ and $20\,000$ frequency samples. It therefore limits the precision with
which individual thresholds can be quoted in this work, which is why the
near-horizon values are given to two significant figures. It does not, however,
affect the underlying elevation dependence, which remains monotonic and is
governed by the optical path length through the layers.

The sensitivity of the threshold to the published event angle differs sharply
between the two event classes, the two steep ANITA-I/III events and the four
near-horizon ANITA-IV events, because their angles are reported in different
ways. Figure~\ref{fig:anita_threshold} plots central values only, with no error
bars. For the steep ANITA-I/III events the published uncertainty is quoted on
the emergence angle itself, so it needs no conversion. Changing the elevation
by $\pm1^\circ$ changes the threshold by only about $0.02$.

For the ANITA-IV events, the published uncertainties are quoted for
$\theta-\theta_H$, namely $-0.19\pm0.10^\circ$, $-0.25\pm0.21^\circ$,
$-0.65\pm0.20^\circ$, and $-0.81\pm0.20^\circ$~\cite{Prechelt2022}, rather than
for the local surface-elevation angle. Two geometric effects then act in
series. First, for a ray leaving the surface close to grazing incidence, a
given interval in $\theta-\theta_H$ maps onto a larger interval in local
surface elevation, with an amplification factor of $2.2$ to $4.3$ over the
angles considered here. Propagating the published uncertainties through this
geometric conversion gives local surface elevations of
$1.54^{+0.37}_{-0.49}$, $1.78^{+0.65}_{-1.07}$,
$2.91^{+0.44}_{-0.51}$, and $3.27^{+0.41}_{-0.45}$ degrees. Second, the
threshold curve itself becomes steeper toward the horizon. Propagating these
elevation intervals through the calculated threshold curve gives threshold
ranges of $4.8$--$6.5$, $4.4$--$8.0$, $3.8$--$4.4$, and $3.6$--$4.1$,
respectively; the upper ends of the first two ranges correspond to elevations
below the range plotted in Fig.~\ref{fig:anita_threshold}. The widths of these ranges therefore arise from the published uncertainties
combined with the near-horizon geometric mapping and the steepness of the
threshold curve, rather than from numerical uncertainty in the forward
calculation, which is deterministic and converged. Two systematics are not
included in this conversion. The reported horizon is the apparent radio horizon
of the payload, which includes atmospheric refraction, whereas the mapping used
here is geometric, and the payload altitude is held fixed at $37$~km rather
than following the flight. Neither changes the conclusion, because the
thresholds stay far above realistic firn and ice indices for any elevation
within a few degrees of those quoted. 

Together with the intrinsic fringe granularity discussed above, this limits the precision with which the
central threshold values are quoted here. They should not be interpreted as
three-significant-figure determinations. However, the conclusion of this study is
unchanged, because even the widest threshold range remains well above realistic
refractive indices for Antarctic firn and ice.

We also apply the frequency-dependent reference-stack transfer function, derived from first principles in Sec.~\ref{sec:layered}, to the $106$ measured HiCal-1 direct H-Pol pulses at each of the six ANITA anomalous-polarity event angles. None of the resulting reflected templates changes to non-inverted polarity. We find $0/106$ non-inverted templates at each angle. The combined $0/636$ count, obtained by computing the reflected pulse for each of the $106$ direct HiCal-1 waveforms at each of the six angles, is quoted only as a correlated diagnostic, since the same $106$ waveforms are reused at all six angles. The steepest ANITA-III angle provides a particularly useful check. At $35.5^\circ$, the reference-stack index $n_2=1.75$ lies just above the coefficient threshold, $n_2^{\rm th}\simeq1.677$, yet the broadband reflected templates remain inverted for all $106$ input waveforms. This result demonstrates that a coefficient sign change at a single frequency, while necessary for a polarity reversal, is not sufficient to produce a non-inverted broadband pulse. The input spectra used in this test are those of the HiCal-1 transmitter rather than of a cosmic-ray air shower. At the four near-horizon geometries the layered coefficient does not change sign anywhere in the analysis band, so the polarity of the reflected pulse there does not depend on the shape of the source spectrum. Only the steepest geometry has an in-band sign change, and the test at that angle therefore uses the $106$ measured spectra as its input ensemble.

A brine layer gives a useful comparison, as shown in Fig.~\ref{fig:robustness}, right panel. It can produce non-inverted templates for the steep ANITA-I/III geometries, but still gives $0/424$ non-inverted templates for the four near-horizon ANITA-IV geometries.

This constraint is one part of a broader picture. IceCube searched for events from the
directions of the ANITA neutrino candidates and found no supporting neutrino
signal~\cite{IceCubeANITA2020}. A dedicated search for upward-going air showers with the
Pierre Auger Observatory also found none. If the anomalous-polarity event rate were
produced by a persistent flux of genuine upward-going showers, the reported rate would
predict several to several dozen Auger events, depending on the assumed spectral
index~\cite{Auger2025}. Other proposed explanations, such as coherent transition radiation
from air-shower currents crossing the surface, address a different physical
channel~\cite{deVriesProhira2019}. The present work constrains one specific
surface-reflection mechanism, shallow and laterally uniform subsurface layering.

It is important to distinguish the coefficient threshold from the waveform test. The threshold we compute tells us the buried-layer index at which the layered reflection coefficient first changes sign somewhere in the frequency band. The waveform test asks whether this frequency-dependent coefficient can reverse the polarity of a broadband pulse. These are not the same condition. The ANITA-III geometry shows this directly. Although the reference stack exceeds the coefficient threshold, all $106$ calculated templates remain inverted. Within the shallow, laterally uniform firn stacks tested here, we therefore find no combination of realistic indices and layer thicknesses that converts a measured direct pulse into a non-inverted reflected pulse at any of the six reported ANITA event geometries. We do not claim that layering explains the anomalous-polarity events. Rather, our results show that this proposed mechanism does not produce the observed polarity over the parameter range tested. The origin of the ANITA anomalous-polarity events therefore remains open.

\section{Applications to other media}
\label{sec:applications}

\begin{figure*}[!t]\centering
\includegraphics[width=1.0\linewidth]{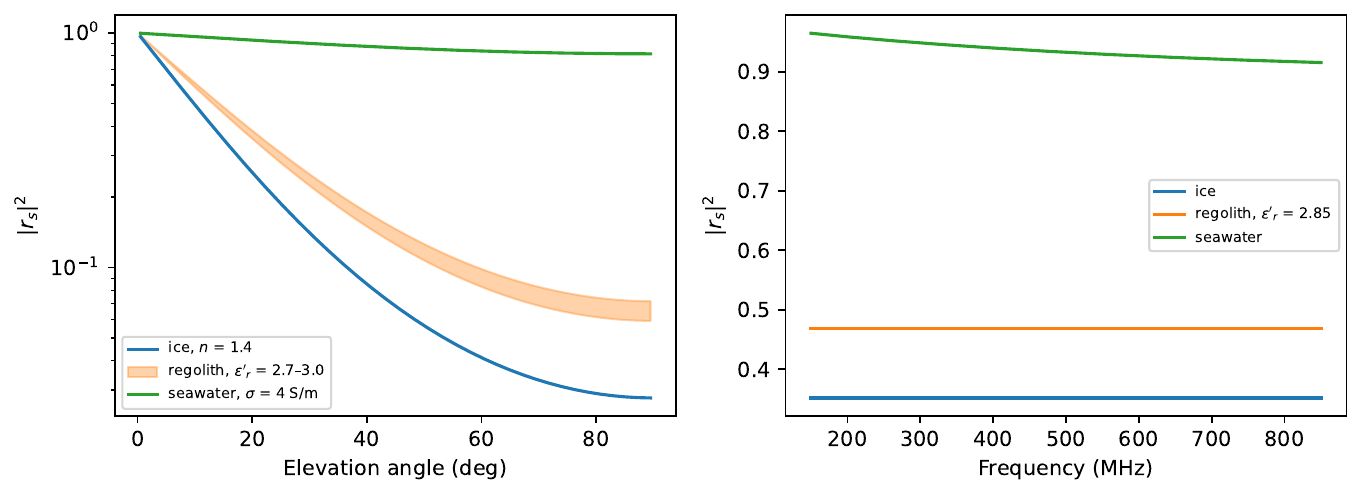}
\caption{Boundary reflection coefficient for three media. Left:
$|r_s|^2$ versus elevation angle at $300$~MHz. Right: $|r_s|^2$ versus
frequency over $150$--$850$~MHz at $15^\circ$ elevation. The three cases
are a lossless ice interface ($n=1.4$, the validated Antarctic
reference), lunar regolith, and conducting seawater, with
$n^2=\varepsilon'_r+i\sigma/(\omega\varepsilon_0)$, using
$\varepsilon'_r=80$ and $\sigma=4$~S/m. The left panel shows the shaded band spanned by
$\varepsilon'_r=2.7$--$3.0$, while the right panel uses the single
midpoint value $\varepsilon'_r=2.85$. At these frequencies, seawater is
conduction dominated, $\sigma/(\omega\varepsilon_0)\gg\varepsilon'_r$,
which produces the strong frequency dependence. The material parameters can
be changed within the formalism to describe different media. The seawater
example is illustrative. No measured data are used.}
\label{fig:media}
\end{figure*}

The formalism also applies to isotropic, non-magnetic media with real or
complex refractive indices. For a conducting medium,
\begin{equation}
n^2=\varepsilon'_r+i\frac{\sigma}{\omega\varepsilon_0},
\end{equation}
where $\varepsilon_0$ is the permittivity of free space and $\sigma$ is the
conductivity. For $\mu_r=1$, the layer admittances in
Eq.~(\ref{eq:qdef}) are
$q_j^{\,s}=n_j\cos\alpha_j$ and
$q_j^{\,p}=\cos\alpha_j/n_j$. For magnetic media, the corresponding
expressions are
$q_j^{\,s}=\sqrt{\varepsilon_j/\mu_j}\cos\alpha_j$ and
$q_j^{\,p}=\sqrt{\mu_j/\varepsilon_j}\cos\alpha_j$~\cite{BornWolf}, which are
not considered here. The superscript labels the polarization and should not
be confused with the substrate admittance $q_s$ in
Eq.~(\ref{eq:rstack}).

The results below use $s$ polarization. The $s$ and $p$
components are defined with respect to the local plane of incidence at $Q$, as shown in Fig.~\ref{fig:stack}, and the same calculation applies to either polarization.

Lunar regolith, with $\varepsilon'_r\simeq2.7$--$3.0$~\cite{OlhoeftStrangway1975},
provides an example beyond ice. Layered regolith is relevant to the Cosmic
Ray Lunar Sounder concept~\cite{CoRaLS,Costello2025} and to reflected
Galactic emission, a target for low-frequency instruments on and around the
Moon~\cite{Burns2017,PRATUSH2023}. The regolith values used here are effective
medium parameters. If rocks, pores, or other structures become comparable to the wavelength, volume scattering must be included in addition to the layered boundary treatment.

Seawater, with $\sigma\simeq4$~S/m~\cite{Stogryn1971}, provides an example of
a conducting medium. Its conductivity makes the refractive index complex and
produces strong frequency-dependent reflection. Figure~\ref{fig:media} shows
the resulting behavior alongside the validated ice calculation using the same
formalism. The seawater result is intended only to illustrate the extension to
conducting media. A detailed treatment of grazing incidence and skin-depth
effects is beyond the scope of this work.

\begin{table*}[!t]
\caption{Geometric quantities used in the exact spherical-wave forward model
developed in Secs.~\ref{sec:formalism}--\ref{sec:layered} for representative
radio-sounding configurations. The last two columns correspond to the criteria
discussed in Sec.~\ref{sec:validity}. The stationary-phase window is
$w\sim\sqrt{2\pi/kL}$, an order-of-magnitude estimate of the angular width of
the coherent region. The last column gives the departure of a concentric shell
from its tangent plane across the first Fresnel zone, expressed in wavelengths
(Appendix~\ref{app:geometry}). The ANITA/HiCal balloon case uses the
$5^\circ$ elevation benchmark. Orbital sounders are evaluated at nadir, and
the near-surface cases at $30^\circ$ elevation. The near-surface rows do not
satisfy the far-zone condition used in Sec.~\ref{sec:formalism}. They are
included to indicate where a near-field extension is required. Dashes indicate
values below $10^{-6}$ at the precision shown. Instrument frequencies and
representative operating altitudes are taken from the MARSIS~\cite{Picardi2009},
SHARAD~\cite{Seu2007}, REASON~\cite{REASON2024}, and RIME
\cite{Bruzzone2013} instrument descriptions.}
\label{tab:bodies}
\begin{ruledtabular}
\begin{tabular}{lccccc}
Configuration & $R$ (km) & $h$ & $f$ (MHz) & $w$ (deg) & sag/$\lambda$ \\ \hline
ANITA/HiCal balloon & 6371 & 37 km & 650 & 0.05 & 0.013 \\
MARSIS (Mars)         & 3390 & 300 km & 4  & 0.64 & 0.022 \\
SHARAD (Mars)         & 3390 & 280 km & 20 & 0.30 & 0.021 \\
REASON (Europa)       & 1561 & 50 km & 60  & 0.41 & 0.008 \\
RIME (Ganymede)       & 2631 & 500 km & 9  & 0.33 & 0.048 \\
Moon lander bistatic  & 1737 & 2 m   & 100 & 35   & --- \\
shallow in-ice pulser & 6371 & 5 m   & 300 & 13   & --- \\
\end{tabular}
\end{ruledtabular}
\end{table*}

Bright reflections from beneath the Martian south polar layered deposits have
been interpreted as possible evidence for basal liquid water, but similar
reflections can also result from constructive interference between thin,
closely spaced layers~\cite{Lalich2022,SmithMars2021}. The frequency dependence
of the reflected power can help distinguish these cases. In the present
formalism, this dependence is described by the layering ratio
$\mathcal{Q}^{\,\chi}(f)$ of Eq.~(\ref{eq:Qratio}), whose frequency scale is set
by the layer thickness. Measurements at only a few frequencies may therefore
miss the interference structure. A similar issue arises for icy moons.
REASON on Europa Clipper operates at $9$ and $60$~MHz, while RIME on JUICE
operates at $9$~MHz. Both instruments can encounter reflections from layered
ice, including possible salt-rich layers, as well as from deeper ice--ocean
interfaces~\cite{REASON2024,Bruzzone2013}.

These applications also probe the limits of the specular approximation, as
summarized in Table~\ref{tab:bodies}. The balloon geometry used here for the
HiCal validation provides one reference case. For orbital sounders, the
region that contributes coherently is generally wider, and the locally planar
approximation becomes less accurate for smaller bodies. The effect is more
pronounced when the source is close to the surface. For a source only a few
meters above the ground, the coherent region can span tens of degrees, so the
stationary-phase approximation used in Eq.~(\ref{eq:factor}) breaks down and
the full angular integral is required. This regime is relevant to future
lander or rover bistatic measurements, where the reflected-to-direct field
ratio derived in Secs.~\ref{sec:formalism}--\ref{sec:layered} could probe
shallow subsurface structure. Plane-wave layered models are already widely
used in radar sounding. In this work, we develop and validate an exact treatment that provides a methodology to assess the accuracy of these approximations for near-surface sources.

The same formalism can be applied to other layered media by changing the
stack and its refractive indices. Examples include layered ice shells of
ocean worlds~\cite{RomeroWolf2015}, embedded dielectric layers in orbital
sounding, and lunar regolith. Salt formations, which have been considered as
radio-transparent media for neutrino detection~\cite{Gorham2002}, and desert
sand can be treated in the same way through their dielectric properties.

The lunar case is particularly relevant to global 21-cm cosmology. The Moon
contributes to the measured sky spectrum through the antenna backlobe, with
both thermal emission from the regolith and reflected Galactic emission.
Proposed analyses treat the lunar reflectivity as a free parameter
~\cite{Burns2017}. A layered-boundary calculation of the type developed here
could instead predict the reflectivity and its frequency dependence from an
assumed regolith structure, providing an input to analyses such as
PRATUSH~\cite{PRATUSH2023}.

\section{Summary and outlook}
\label{sec:summary}

We have developed an exact spherical-wave forward model for broadband
reflection from a spherical boundary with an arbitrary number of laterally
uniform layers beneath it. The method combines the Sommerfeld--Weyl decomposition of the
source field with the characteristic-matrix description of the layered
medium, while retaining the full angular integration over the spherical
surface. It therefore accounts simultaneously for spherical source
geometry, spherical surface geometry, frequency-dependent interference
within the layers, and coherent summation over the angular spectrum.

The calculation reproduces the published single-boundary HiCal-2
reflectivity to within $1.1\%$ at ten elevation angles and approaches the
single-boundary result at machine precision when the layer contrast is
removed. It also reproduces the measured HiCal-1 reflected waveforms, with
a best-pair signed correlation of $0.83$ and the expected polarity inversion
in $(95.3\pm2.1)\%$ of the $106$ matched direct/reflected pulse pairs. These
checks establish the calculation as a validated forward model for the
geometry and frequency range considered here.

Applied to shallow firn layering, the model shows that realistic buried
layers can substantially modify the reflected amplitude and phase without
reversing the polarity of the broadband pulse. For the ANITA event
geometries, the required buried indices rise sharply toward the horizon,
reaching $n_2^{\rm th}\simeq3.8$--$5.4$ for the four ANITA-IV events. The
full waveform calculation gives no non-inverted templates at any of the six
reported event angles for the reference stack. In particular, the steepest
ANITA-III geometry shows that a sign change of the layered reflection
coefficient at one frequency is not sufficient to reverse a broadband
reflected pulse. Within the shallow, laterally uniform parameter space
tested here, subsurface layering therefore does not provide an explanation
for the anomalous-polarity ANITA events.

The framework can be extended to rough buried interfaces, anisotropic
surface structure, conducting media, and other planetary or subglacial
settings. An important next step is a near-field extension for sources close
to a curved boundary, where the stationary-phase and far-zone
approximations used here no longer hold. The resulting forward model could
also be used in inverse problems to infer layer properties from broadband
reflected waveforms. More broadly, the exact treatment provides a way to
quantify when commonly used specular and locally planar layered-media
approximations are adequate and when the full spherical-wave calculation is
required.

\begin{acknowledgments}
The author thanks Pankaj Jain for the ideas, discussions, and guidance that
led to the development of the first-principles spherical-wave treatment of
dipole radiation reflected from a spherical boundary, presented in
Refs.~\cite{Prohira2018,DasguptaJain2021}, which provides the foundation for
the general formalism developed here. The author thanks David Besson and Steven
Prohira for providing the HiCal-1 calibration pulser data used in
Sec.~\ref{sec:hical} and the HiCal-2 reflectivity calculations from
Ref.~\cite{Prohira2018} used in Sec.~\ref{sec:formalism}. The author also
thanks David Besson, John F. Beacom, Amy Connolly, and Austin Cummings for
valuable discussions and input. The author thanks the ANITA Collaboration
for the published measurements and results used to validate the formalism.
This work was supported by the Center for Cosmology and AstroParticle Physics
(CCAPP) at The Ohio State University.
\end{acknowledgments}

\appendix

\section{Geometry of the spherical boundary}
\label{app:geometry}

This appendix gives the geometric relations used in Secs.~\ref{sec:formalism} and~\ref{sec:layered}, the local incidence angle,
the incident point and ray directions, and the bound on treating the concentric shells as parallel layers.

\subsection{Local incidence angle}

Let $C$ be the center of the sphere, the point named but not drawn in
Fig.~\ref{fig:geometry}, $S$ the source at radius $R+h$, and $Q$ the incident
point at radius $R$. A plane wave
leaving $S$ at angle $\alpha$ from the downward vertical $SC$ reaches $Q$ at
the local incidence angle $\alpha_0$, measured from the outward normal. In the
triangle $CSQ$, the interior angle at $S$ is $\alpha$, while that at $Q$ is
$\pi-\alpha_0$. The law of sines gives
\begin{equation}
\frac{R}{\sin\alpha}=\frac{R+h}{\sin(\pi-\alpha_0)}
\;\;\Longrightarrow\;\;
\sin\alpha_0=\frac{R+h}{R}\,\sin\alpha ,
\label{eq:appalpha0}
\end{equation}
which is Eq.~(\ref{eq:alpha0}). In the flat limit $R\to\infty$,
$\alpha_0\to\alpha$. The remaining interior angle at $C$ is the central angle
between $O$ and $Q$,
\begin{equation}
\gamma=\alpha_0-\alpha ,
\label{eq:appgamma}
\end{equation}
which is the tilt of the local normal at $Q$ relative to the vertical
$\hat{z}$ at $O$. For a convex surface, $\gamma>0$ and hence
$\alpha_0>\alpha$.

\subsection{Incident point and ray directions}

The incident point is determined by $\gamma$. Measuring positions from the
origin $O$ on the surface directly below the source, with $\hat{z}$ the
outward vertical there,
\begin{equation}
\vec{Q}(\alpha,\beta)=
\big(R\sin\gamma\cos\beta,\ R\sin\gamma\sin\beta,\ -R(1-\cos\gamma)\big),
\label{eq:appQpos}
\end{equation}
and the outward unit normal at that point is
\begin{equation}
\hat{n}(Q)=(\sin\gamma\cos\beta,\ \sin\gamma\sin\beta,\ \cos\gamma).
\label{eq:appnormal}
\end{equation}
The source-to-surface distance follows from the same triangle and the law of
cosines. The resulting quadratic has discriminant $R^2-(R+h)^2\sin^2\alpha$, giving
\begin{equation}
|SQ|=(R+h)\cos\alpha\mp\sqrt{R^2-(R+h)^2\sin^2\alpha}\, .
\label{eq:appSQ}
\end{equation}
The reflection point is the first intersection of the ray with the spherical surface, so
the minus root is used. The plus root gives the far-side exit point, which is
$3436$~km away at $15^\circ$ elevation.

Equations~(\ref{eq:appQpos})--(\ref{eq:appSQ}) are evaluated for each
$(\alpha,\beta)$ in the numerical calculation and determine the propagation
phase used in Appendix~\ref{app:fields}.

The directions in Eqs.~(\ref{eq:khatr})--(\ref{eq:khatt}) follow from
reflection and refraction at the local tangent plane. Since
$\hat{k}_i\cdot\hat{n}(Q)=-\cos\alpha_0$, the reflected direction is
\begin{equation*}
\hat{k}_r=\hat{k}_i-2(\hat{k}_i\cdot\hat{n})\hat{n}.
\end{equation*}
Using Eq.~(\ref{eq:appgamma}), its horizontal and vertical components reduce to
$\sin(2\alpha_0-\alpha)\cos\beta$ and
$\cos(2\alpha_0-\alpha)$, respectively. Hence
$\theta_r=2\alpha_0-\alpha$. For the transmitted wave, which makes an angle
$\alpha_t$ with the inward normal in the same plane of incidence,
$\hat{k}_t=\sin\alpha_t\,\hat{e}_\parallel-\cos\alpha_t\,\hat{n}(Q)$ with
$\hat{e}_\parallel=(\cos\gamma\cos\beta,\cos\gamma\sin\beta,-\sin\gamma)$,
giving
$\theta_t=\alpha_t-\gamma=\alpha+\alpha_t-\alpha_0$.
Both reduce to the flat-surface results in the limit $\gamma\to0$.

\subsection{Local approximation for the layered boundary}

Curvature in the plane of incidence is treated exactly. Each plane wave
component reflects from the local tangent plane at its corresponding
$\alpha_0(\alpha)$. The approximation enters only in the transverse direction,
where the concentric shells are treated as locally parallel layers over the
surface region that contributes coherently. The rest of the calculation retains
the spherical geometry.

\subsection{Curvature bound for stratified media}
\label{app:cbound}

A sphere of radius $R$ departs from its tangent plane at transverse distance
$\rho$ by the sagitta $\eta$,
\[
\eta=R-\sqrt{R^2-\rho^2}\simeq \frac{\rho^2}{2R}.
\]
The relevant transverse scale is the first Fresnel radius of the reflected
path. For two path lengths $L_1=|SQ|$ and $L_2=|QP|$ meeting at the reflection
point,
\begin{equation}
\rho_F=\sqrt{\frac{\lambda L_1L_2}{L_1+L_2}} .
\label{eq:appfresnel}
\end{equation}
Equation~(\ref{eq:appfresnel}) gives the radius of the coherent region perpendicular to the plane of incidence. In that plane, the coherent region is longer by $1/\cos\alpha_0$, so the footprint is an ellipse with semi-axes $\rho_F$ and $\rho_F/\cos\alpha_0$. The bound below uses the cross-track radius, because curvature in the plane of incidence
is carried exactly by the angular integral.

For the symmetric reflection estimate used here, $L_1=L_2=\ell$ and
$L_{\rm sph}=L_1+L_2=2\ell$, so
\[
\rho_F=\sqrt{\frac{\lambda L_{\rm sph}}{4}}
      =\sqrt{\frac{\lambda \ell}{2}} .
\]
The one-way spherical path length is not the flat-surface value
$h/\cos\alpha_0$. Using Eqs.~(\ref{eq:alpha0}) and~(\ref{eq:appSQ}), it can be
written in terms of the local incidence angle as
\begin{equation}
\ell=|SQ|=
\sqrt{(R+h)^2-R^2\sin^2\alpha_0}-R\cos\alpha_0 .
\label{eq:appsphL}
\end{equation}
The perpendicular distance from the source to the tangent plane at $Q$ is therefore
\begin{equation}
h'=\ell\cos\alpha_0,
\qquad
L_{\rm sph}=2\ell=\frac{2h'}{\cos\alpha_0}.
\label{eq:apphprime}
\end{equation}
Only in the flat limit does $h'\to h$ and $L_{\rm sph}\to2h/\cos\alpha_0$.

For $h=37$~km, $R=R_\oplus=6371$~km, and $f=650$~MHz
($\lambda=0.461$~m), the exact spherical values are
\[
L_{\rm sph}=275~{\rm km},\qquad
\rho_F=178~{\rm m},\qquad
\eta=2.5~{\rm mm}
\]
at $15^\circ$ elevation. This sag is $0.54\%$ of a wavelength. At
$5^\circ$ elevation the corresponding values are
\[
L_{\rm sph}=657~{\rm km},\qquad
\rho_F=275~{\rm m},\qquad
\eta=5.9~{\rm mm},
\]
or $1.3\%$ of a wavelength. These values are smaller than the corresponding
flat-surface estimates because $h'<h$ for a convex surface. The sag is also
much smaller than the rms surface roughness over the same footprint, which
Eq.~(\ref{eq:rough}) gives as $4.6$~cm and $6.1$~cm at the two elevations.

A displacement $\eta$ along the local normal changes the two-way path of a ray
at incidence $\alpha_0$ by $2\eta\cos\alpha_0$, giving
\[
\Delta\phi=2k\eta\cos\alpha_0 .
\]
Substituting $\eta=\rho_F^{\,2}/2R$ and
$\rho_F^{\,2}=\lambda L_{\rm sph}/4=\lambda h'/(2\cos\alpha_0)$ gives
\begin{equation}
\Delta\phi
=2k\cos\alpha_0\,\frac{\rho_F^{\,2}}{2R}
=\frac{\pi h'}{R}.
\label{eq:appsagphase}
\end{equation}
Thus, the wavelength dependence cancels. The remaining elevation dependence
enters only through $h'$. Since $h'<h$ for the convex geometry,
\[
\Delta\phi < \frac{\pi h}{R}=0.018~{\rm rad}
\]
for a $37$~km balloon altitude above Earth. The exact values are
$0.0176$~rad at $15^\circ$ elevation and $0.0141$~rad at $5^\circ$ elevation.
The more conservative normal-incidence estimate, $4\pi\eta/\lambda$, gives
$0.07$~rad and $0.16$~rad at the same two elevations. These phases are small
compared with the layer round-trip phases $2\delta_j$, which can be many
radians.

\section{The reflected and transmitted fields}
\label{app:fields}
Appendix~\ref{app:geometry} sets out the geometry of the spherical boundary, and
Appendix~\ref{app:stack} gives the response of the medium beneath the incident point
where a given plane wave strikes the spherical surface. This appendix connects the two, carrying the Hertzian
dipole potential of Eq.~(\ref{eq:hertz}) through to the double integrals of
Eqs.~(\ref{eq:erefstack}) and~(\ref{eq:etransstack}), in the notation of Refs.~\cite{Prohira2018,DasguptaJain2021}.

Throughout, $r^{\,\chi}$ and $t^{\,\chi}$ denote the boundary coefficients for the medium beneath $Q$, with $\chi=s,p$ labeling the two polarizations. These are the Fresnel coefficients $f_r^{\,\chi}$ of Eqs.~(\ref{eq:fs})--(\ref{eq:fp}) for a single boundary, or the stack coefficients $r_{\rm stack}^{\,\chi}$ and $t_{\rm stack}^{\,\chi}$ of Eqs.~(\ref{eq:rstack})--(\ref{eq:tstack}) for a stratified medium. The construction below is identical in both cases.

\subsection{Plane wave and incident point}
The decomposition of Eq.~(\ref{eq:weyl}) labels each plane wave by a polar angle $\alpha$ and an azimuth $\beta$. A plane wave with direction $\hat{k}_i(\alpha,\beta)$ leaves the source $S$ along $\hat{k}_i$ [Eq.~(\ref{eq:khat})] and meets the surface at $\vec{Q}(\alpha,\beta)$ [Eq.~(\ref{eq:appQpos})]. The mapping
\begin{equation}
(\alpha,\beta)\ \longleftrightarrow\ \vec{Q}(\alpha,\beta)
\label{eq:appmap}
\end{equation}
is therefore one-to-one, and the position on the surface is not an independent variable. Integrating over $(\alpha,\beta)$ in Eq.~(\ref{eq:weyl}) already covers the whole illuminated surface.

\subsection{Hertzian dipole potential and field calculation}

Expressing the Hertzian dipole potential from Eq.~(\ref{eq:hertz}) using Eq.~(\ref{eq:weyl}),
\begin{equation}
\vec{\Pi}_{\rm dir}=\frac{1}{4\pi\epsilon}\,\frac{ik}{2\pi}
\int_{0}^{2\pi}\!\!\!\int_{0}^{\pi/2-i\infty}\!\!\!\!
e^{i\Phi_0}\,\sin\alpha\,d\alpha\,d\beta\;\hat{y},
\label{eq:apppidir}
\end{equation}
each element of the integrand represents a plane wave with direction $\hat{k}_i$ and polarization along $\hat{y}$. Its electric field follows from the relation
$\vec{E}=\vec{\nabla}(\vec{\nabla}\!\cdot\!\vec{\Pi})+k^2\vec{\Pi}$.
Because a plane wave carries a spatial dependence proportional to $e^{i k \hat{k}_i\cdot\vec{r}}$, the differential operator evaluates algebraically as $\vec{\nabla}\to i k \hat{k}_i$, where $k$ is the magnitude of the incident wave vector and $\hat{k}_i(\alpha, \beta)$ is its direction. Substituting this into the field equation gives
\begin{equation}
\vec{E}\propto k^2
\big[\hat{y}-(\hat{y}\cdot\hat{k}_i)\hat{k}_i\big],
\label{eq:apptransverse}
\end{equation}
which is the transverse projection of the dipole orientation $\hat{y}$ onto the plane perpendicular to the propagation direction $\hat{k}_i$.

\subsection{The local polarization basis at $Q$}
\label{app:basis}

Appendix~\ref{app:stack} gives the stack coefficients for a plane wave incident on a flat boundary, separately for $s$ and $p$ polarization. On a spherical surface, each plane wave in Eq.~(\ref{eq:weyl}) intersects the surface at a different point $Q$, and we therefore consider a different tangent plane for each plane-wave component. This tangent plane lies at a perpendicular distance $h'$ from the source [Eq.~(\ref{eq:apphprime})], and the polarization of each plane wave must be resolved in a basis defined locally at $Q$. We construct this basis here and express it in terms of the polar and azimuthal angles $(\alpha,\beta)$ in the global coordinate system $(x,y,z)$, which label the plane wave in Eq.~(\ref{eq:weyl}), so that every factor in the reflected-field integral of Eq.~(\ref{eq:erefstack}) depends only on the integration variables.

The local basis follows directly from the incident propagation direction and the surface normal. The propagation direction is $\hat{k}_i$ of Eq.~(\ref{eq:khat}), fixed by the polar angle $\alpha$ and the azimuth $\beta$. The outward normal is $\hat{n}(Q)$ of Eq.~(\ref{eq:appnormal}), fixed by the angle $\gamma=\alpha_0-\alpha$ of Eq.~(\ref{eq:appgamma}) and by the same azimuth. Here $\gamma$ is the angle subtended at the center $C$ between $O$ and $Q$, so it is also the tilt of the local normal at $Q$ away from the vertical $\hat{z}$ at $O$ (Fig.~\ref{fig:geometry}). It vanishes in the flat limit and is the variable that accounts for the effect of curvature in the calculation without approximation. A plane wave leaving $S$ at a larger $\alpha$ lands further around the surface, picks up a larger $\gamma$, and meets a more strongly tilted tangent plane, as illustrated by the point $Q^\prime$ in Fig.~\ref{fig:geometry}. Since $\hat{k}_i$ and $\hat{n}(Q)$ share the azimuth $\beta$, the source $S$, the center $C$, the origin $O$, and the incident point $Q$ all lie in one plane, the vertical plane at azimuth $\beta$ shown in Fig.~\ref{fig:geometry}.

The local angle of incidence is the angle between the incoming ray and that normal,
\begin{equation}
\cos\alpha_0=-\,\hat{k}_i\cdot\hat{n}(Q),
\label{eq:appalphaloc}
\end{equation}
the minus sign appearing because $\hat{k}_i$ points into the surface while $\hat{n}(Q)$ points out of it. Constructing the dot product component by component, the two horizontal terms share the factor $\sin\alpha\sin\gamma$ and add as
$\cos^2\!\beta+\sin^2\!\beta=1$, while the vertical components contribute
$-\cos\alpha\cos\gamma$. Hence
\begin{equation}
\begin{aligned}
-\,\hat{k}_i\cdot\hat{n}(Q)
&=\cos\alpha\cos\gamma-\sin\alpha\sin\gamma\\[2pt]
&=\cos(\alpha+\gamma)=\cos\alpha_0 ,
\end{aligned}
\label{eq:appalphacheck}
\end{equation}
the last step using $\gamma=\alpha_0-\alpha$. Equation~(\ref{eq:appalphaloc}) therefore returns the closed form $\alpha_0(\alpha)$ of Eq.~(\ref{eq:appalpha0}). The azimuth cancels, so the boundary coefficients depend on $\alpha$ alone.

The plane of incidence is the plane containing $\hat{k}_i$ and $\hat{n}(Q)$. Since $\hat{k}_i$ and $\hat{n}(Q)$ lie in the same vertical plane, this is the vertical plane at azimuth $\beta$. The two transverse directions are
\begin{equation}
\hat{s}_Q=\frac{\hat{n}(Q)\times\hat{k}_i}{|\hat{n}(Q)\times\hat{k}_i|},
\qquad
\hat{p}_Q=\hat{s}_Q\times\hat{k}_i .
\label{eq:appspbasis}
\end{equation}
The cross product is perpendicular to both of its arguments. Thus $\hat{s}_Q$ is perpendicular to both $\hat{k}_i$ and $\hat{n}(Q)$. It is therefore perpendicular to the plane of incidence and, being perpendicular to the normal, lies in the tangent plane at $Q$. The vector $\hat{p}_Q$ is perpendicular to both $\hat{s}_Q$ and $\hat{k}_i$, so it lies in the plane of incidence.

The two vectors can now be obtained directly from the cross products. For the first,
\begin{equation}
\hat{n}(Q)\times\hat{k}_i=
\begin{vmatrix}
\hat{x} & \hat{y} & \hat{z}\\[2pt]
\sin\gamma\cos\beta & \sin\gamma\sin\beta & \cos\gamma\\[2pt]
\sin\alpha\cos\beta & \sin\alpha\sin\beta & -\cos\alpha
\end{vmatrix},
\label{eq:appcrossdet}
\end{equation}
the $\hat{z}$ component vanishes because the horizontal parts of the two vectors are parallel, both pointing along $(\cos\beta,\sin\beta)$. The $\hat{x}$ and $\hat{y}$ components contain the combination $\sin\gamma\cos\alpha+\cos\gamma\sin\alpha=\sin(\alpha+\gamma)$, giving
\begin{equation}
\begin{aligned}
\hat{n}(Q)\times\hat{k}_i
&=\sin(\alpha+\gamma)\,(-\sin\beta,\ \cos\beta,\ 0)\\[2pt]
&=\sin\alpha_0\,(-\sin\beta,\ \cos\beta,\ 0).
\end{aligned}
\label{eq:appcross}
\end{equation}
Since $(-\sin\beta,\cos\beta,0)$ is a unit vector, the norm of $\hat{n}(Q)\times\hat{k}_i$ is $\sin\alpha_0$. Using this in the denominator of Eq.~(\ref{eq:appspbasis}) gives $\hat{s}_Q$, and the second relation in Eq.~(\ref{eq:appspbasis}) then gives $\hat{p}_Q$:
\begin{equation}
\hat{p}_Q=\hat{s}_Q\times\hat{k}_i=
\begin{vmatrix}
\hat{x} & \hat{y} & \hat{z}\\[2pt]
-\sin\beta & \cos\beta & 0\\[2pt]
\sin\alpha\cos\beta & \sin\alpha\sin\beta & -\cos\alpha
\end{vmatrix},
\label{eq:apppdet}
\end{equation}
in which the $\hat{z}$ component is
$-\sin\alpha(\sin^2\!\beta+\cos^2\!\beta)=-\sin\alpha$. The basis at $Q$ is
therefore
\begin{equation}
\begin{aligned}
\hat{s}_Q &= (-\sin\beta,\ \cos\beta,\ 0),\\[2pt]
\hat{p}_Q &= (-\cos\alpha\cos\beta,\ -\cos\alpha\sin\beta,\ -\sin\alpha).
\end{aligned}
\label{eq:appspexplicit}
\end{equation}
Both are unit vectors, and $\hat{s}_Q$, $\hat{p}_Q$ and $\hat{k}_i$ are mutually perpendicular, so any field transverse to $\hat{k}_i$ is fixed by its $\hat{s}_Q$ and $\hat{p}_Q$ components alone. Neither vector in Eq.~(\ref{eq:appspexplicit}) contains $\gamma$. The curvature dependence cancels when $\hat{n}(Q)\times\hat{k}_i$ is normalized in Eq.~(\ref{eq:appcross}). Thus the basis has the same dependence on $(\alpha,\beta)$ as for a flat surface. The curvature enters through $\alpha_0$, which determines the boundary coefficients. Also, $\hat{s}_Q$ is horizontal, whereas $\hat{p}_Q$ is not, since $\hat{p}_Q\cdot\hat{n}(Q)=-\sin\alpha_0$. Thus $\hat{s}_Q$ is the H-Pol direction at $Q$, while the $p$ direction has a component normal to the surface set by the local angle of incidence.

The source Hertz dipole couples to a plane wave only through the component of its dipole axis perpendicular to $\hat{k}_i$. Since $\hat{s}_Q$, $\hat{p}_Q$, and $\hat{k}_i$ are mutually perpendicular, the transverse projection in Eq.~(\ref{eq:apptransverse}) can be written either by removing the component along $\hat{k}_i$ or by resolving it in the local basis,
\begin{equation}
\hat{y}_\perp=\hat{y}-(\hat{y}\cdot\hat{k}_i)\,\hat{k}_i
=(\hat{y}\cdot\hat{s}_Q)\,\hat{s}_Q+(\hat{y}\cdot\hat{p}_Q)\,\hat{p}_Q .
\label{eq:appyperp}
\end{equation}
The two coefficients give the incident $s$ and $p$ amplitudes of the horizontal
dipole in Eq.~(\ref{eq:hertz})~\cite{DasguptaJain2021}:
\begin{equation}
\begin{aligned}
E_s^{(i)} &= \hat{y}\cdot\hat{s}_Q = \cos\beta,\\[2pt]
E_p^{(i)} &= \hat{y}\cdot\hat{p}_Q = -\cos\alpha\,\sin\beta.
\end{aligned}
\label{eq:appproj}
\end{equation}
For $\beta=0$, the dipole axis $\hat{y}$ is perpendicular to the plane of incidence, so the wave is purely $s$ polarized. For $\beta=\pi/2$, the dipole lies in the plane of incidence, so the wave is purely $p$ polarized. For intermediate values of $\beta$, both components are present. Each component enters once in the incident field and again when the reflected field is
projected onto $\hat{y}$, giving the $\cos^2\!\beta$ and $\sin^2\!\beta$ factors in Eq.~(\ref{eq:apphpol}).

\subsection{Reflection and polarization projection}
\label{app:pol}

Each plane wave is reflected by the medium at the local incidence angle $\alpha_0(\alpha)$, and the reflected fields for the two polarizations are
\begin{equation}
E_s^{(r)}=r^{\,s}\big(\alpha_0\big)\,E_s^{(i)},
\qquad
E_p^{(r)}=r^{\,p}\big(\alpha_0\big)\,E_p^{(i)} .
\label{eq:appapply}
\end{equation}
The source, spherical boundary geometry, Sommerfeld--Weyl spherical-wave decomposition, roughness factor, and field reconstruction do not change when the single-boundary coefficient is replaced by the stratified-medium coefficient.

The reflected wave, which is locally a plane wave, leaves $Q$ along $\hat{k}_r$ of Eq.~(\ref{eq:khatr}), as shown in Fig.~\ref{fig:geometry}. The vector $\hat{s}_Q$ is unchanged by reflection, while the reflected $\hat{p}_r=\hat{s}_Q\times\hat{k}_r$ has a horizontal projection $\hat{y}\cdot\hat{p}_r=\cos(2\alpha_0-\alpha)\sin\beta$. The horizontally polarized component therefore carries the weight

\begin{equation}
W_{\rm ref}=r^{\,s}\cos^2\!\beta
-r^{\,p}\cos\alpha\,\cos(2\alpha_0-\alpha)\sin^2\!\beta ,
\label{eq:apphpol}
\end{equation}
which is Eq.~(\ref{eq:polweight}) for a single boundary and Eq.~(\ref{eq:wrefstack}) for a stack. The relative sign follows from the polarization definitions. At normal incidence, $f_r^{\,p}=-f_r^{\,s}$, so Eq.~(\ref{eq:apphpol}) reduces to
$r^{\,s}(\cos^2\beta+\sin^2\beta)=r^{\,s}$, as required by azimuthal symmetry.

\subsection{Computing the total field at the observation point}

The reflected field at $P(x,y,z)$ is the sum of the contributions from all reflected plane waves. Each plane wave accumulates a phase along its path,
\begin{equation}
\Phi_r(\alpha,\beta)=k\big(|SQ|+|QP|\big),
\label{eq:appphaser}
\end{equation}
with $|SQ|$ from Eq.~(\ref{eq:appSQ}) and $|QP|=|\vec{P}-\vec{Q}(\alpha,\beta)|$ from Eq.~(\ref{eq:appQpos}).

In the flat limit, this is equivalent to the image-source construction. The reflected field is that of a source at $-z_0$, so Eq.~(\ref{eq:weylphase}) applies with $(z_0-z)$ replaced by $(z_0+z)$. The latter is the vertical distance traveled by a ray that first descends to the surface and then returns upward. The flat-surface calculation is given in Ref.~\cite{Prohira2018}.

The roughness factor of Eq.~(\ref{eq:rough}) is evaluated at
\[
\rho_\perp=|\vec{Q}(\alpha,\beta)-\vec{Q}_{\rm spec}| ,
\]
the transverse offset of the incident point from the specular point, at which
$|SQ|+|QP|$ is stationary. Since $\rho_\perp$ varies from one plane wave to
another, the roughness factor remains inside the integral.

Restoring the measure $\sin\alpha\,d\alpha\,d\beta$ and the prefactor of Eq.~(\ref{eq:weyl}), and dropping the constants common to Eq.~(\ref{eq:apppidir}), the horizontally polarized reflected field at one frequency is
\begin{widetext}
\begin{equation}
\begin{split}
E^{H}_{\rm ref}(f)=\frac{ik}{2\pi}
\int_{0}^{2\pi}\!\!\!\int_{0}^{\pi/2-i\infty}
&\Big[\,r^{\,s}(f,\alpha_0)\cos^2\!\beta
-r^{\,p}(f,\alpha_0)\cos\alpha\,\cos(2\alpha_0-\alpha)\sin^2\!\beta\,\Big]\\[4pt]
&\times\;e^{-2k^2\sigma_h^2(\rho_\perp)\cos^2\alpha_0^{\rm spec}}\;
e^{\,ik(|SQ|+|QP|)}\;\sin\alpha\,d\alpha\,d\beta ,
\end{split}
\label{eq:apperef}
\end{equation}
\end{widetext}
The boundary coefficients are evaluated at $\alpha_0(\alpha)$ from
Eq.~(\ref{eq:appalpha0}), the azimuthal weights are those of
Eq.~(\ref{eq:appproj}), and the roughness factor is evaluated at $\rho_\perp$
with the fixed specular angle $\alpha_0^{\rm spec}$ of Eq.~(\ref{eq:rough}).
Equation~(\ref{eq:apperef}) is Eq.~(\ref{eq:eref}) when
$r^{\,\chi}=f_r^{\,\chi}$ and Eq.~(\ref{eq:erefstack}) when
$r^{\,\chi}=r_{\rm stack}^{\,\chi}$. This is the quantity evaluated
numerically in Sec.~\ref{sec:results}.

A broadband waveform is obtained by evaluating Eq.~(\ref{eq:apperef}) at each Fourier frequency across the measured band and then taking the inverse Fourier transform. The reflectivity is obtained by normalizing to the field the same source would produce along the same path in the absence of the boundary,
\begin{equation}
E_{\rm direct}(f)=\frac{e^{ikL}}{L},\qquad
L=|SQ_{\rm spec}|+|Q_{\rm spec}P| ,
\label{eq:appdirect}
\end{equation}
written with the same constants dropped as in Eq.~(\ref{eq:apperef}). This removes the geometric spreading of the reflected path, leaving a quantity that depends on the surface rather than on the geometry. The power ratio quoted in Sec.~\ref{sec:formalism} is
\begin{equation}
\frac{r}{d}=\Big\langle\,\big|E^{H}_{\rm ref}(f)\big/E_{\rm direct}(f)\big|^{2}
\Big\rangle_{f} ,
\label{eq:apprd}
\end{equation}
averaged over the measured band. This fixes the absolute scale without a free parameter. In the flat, smooth, single-boundary limit Eq.~(\ref{eq:apprd}) returns the Fresnel power reflection coefficient~\cite{Prohira2018}.

\subsection{The transmitted field}

The same construction gives the field transmitted into the medium below. This is not needed for the balloon-borne reflection geometry considered above, but it is needed for a source inside the ice or regolith. Three things change for the transmitted field. First, the boundary coefficients are $t^{\,\chi}$ instead of $r^{\,\chi}$. Second, the outgoing direction is $\hat{k}_t$ of Eq.~(\ref{eq:khatt}) instead of $\hat{k}_r$, so the $p$ basis vector on the far side is $\hat{p}_t=\hat{s}_Q\times\hat{k}_t$. Its projection on the horizontal is $\hat{y}\cdot\hat{p}_t=-\cos(\alpha+\alpha_t-\alpha_0)\sin\beta$. Third, the phase beyond the boundary accumulates at the wavenumber of the medium below rather than at $k$. Repeating the projection that led to Eq.~(\ref{eq:apphpol}) with these replacements gives
\begin{equation}
\begin{split}
W_{\rm trans}={}&t^{\,s}\cos^2\!\beta\\
&+\frac{n_0}{n_1}t^{\,p}\cos\alpha\,\cos(\alpha+\alpha_t-\alpha_0)\sin^2\!\beta ,
\end{split}
\label{eq:apphpoltrans}
\end{equation}
which is Eq.~(\ref{eq:wtransstack}) with $(n_1,\alpha_t)$ replaced by $(n_s,\alpha_s)$ for a stack, since the wave then emerges into the substrate.

The factor $n_0/n_1$ in the second term on the right-hand side of Eq.~(\ref{eq:apphpoltrans}) comes from the conversion described after Eq.~(\ref{eq:rhoint}). For $p$ polarization, the transmission coefficient is a ratio of tangential magnetic-field amplitudes, while the projection onto $\hat{y}$ uses the ratio of electric-field amplitudes. The two differ by the index ratio. At normal incidence, Eq.~(\ref{eq:qdef}) gives $t^{\,s}=2n_0/(n_0+n_1)$ and $t^{\,p}=2n_1/(n_0+n_1)$. With the factor $n_0/n_1$, Eq.~(\ref{eq:apphpoltrans}) then reduces to $t^{\,s}$ for every $\beta$, as required by the azimuthal symmetry of normal-incidence transmission.

The phase has three parts. From $S$ to $Q$, the wave travels through air over the distance $|SQ|$ and contributes $k|SQ|$. Inside the layers, the phase is already included through the $\delta_j$ of Eq.~(\ref{eq:Mj}) in $t^{\,s,p}_{\rm stack}$, so it is not included again. Below the layers, the wave travels through the substrate with wavenumber $k\,n_s$ over a path $\ell_s$ measured from the base of the stack to the field point. Thus,
\begin{equation}
\Phi_t=k|SQ|+k\,n_s\ell_s .
\label{eq:appphaset}
\end{equation}
For a single boundary there are no layers, $\ell_s$ begins at $Q$ itself, and
the middle piece is absent. The transmitted field is then
\begin{equation}
E^{H}_{\rm trans}(f)=\frac{ik}{2\pi}
\int_{0}^{2\pi}\!\!\!\int_{0}^{\pi/2-i\infty}\!\!\!\!
W_{\rm trans}F^{\,t}_{\rm rough}\,e^{i\Phi_t}\sin\alpha\,d\alpha\,d\beta ,
\label{eq:appetrans}
\end{equation}
with $F^{\,t}_{\rm rough}$ from Eq.~(\ref{eq:roughtrans}).
Equations~(\ref{eq:apphpoltrans})--(\ref{eq:appetrans}) are included for completeness. The numerical results in this paper use Eq.~(\ref{eq:apperef}).

\subsection{Single-boundary limit}
Setting $N=0$ gives $r^{\,s,p}_{\rm stack}=f_r^{\,s,p}$, so Eq.~(\ref{eq:apphpol}) reduces to the H-Pol weight of Refs.~\cite{Prohira2018,DasguptaJain2021}. No other part of the calculation changes, and this reduction is confirmed numerically to machine precision in Sec.~\ref{sec:result_reduction}.

Both polarizations are retained throughout, although the $s$ component dominates the H-Pol response. The reflected field is concentrated near $\beta=0$, where Eq.~(\ref{eq:appproj}) gives pure $s$ polarization. The $p$ contribution enters only at second order in the azimuthal width. This behavior was shown numerically in Refs.~\cite{Prohira2018,DasguptaJain2021} by varying the $\alpha$ and $\beta$ integration ranges around the specular direction and checking the convergence of the integrated field.

\subsection{Approximations in the calculation}
\label{app:approx}
The main parts of the calculation are exact. The Sommerfeld--Weyl decomposition of Eq.~(\ref{eq:weyl}) is an identity, not an expansion. The transverse projection in Eq.~(\ref{eq:apptransverse}) and the reflection of each plane wave from the local tangent plane at the spherical surface follow directly from Maxwell's boundary conditions. The layered reflection coefficient is obtained from the same boundary-value problem as the single-boundary coefficient. None of these steps requires weak layer contrast, thin layers, or a monotonic refractive-index profile. There is also no restriction on layer thickness. It enters analytically through the phase $\delta_j$ of Eq.~(\ref{eq:Mj}), with no spatial mesh required, from the millimeter-scale lens of Sec.~\ref{sec:thinrough} to the meter-scale layers scanned in Sec.~\ref{sec:result_threshold}.

The approximations enter when the curved layers are treated as parallel to the local tangent plane. The resulting phase error is bounded by Eq.~(\ref{eq:appsagphase}). We also neglect the evanescent part of the contour in Eq.~(\ref{eq:weyl}), which is negligible for the far-field geometry considered here. Thus, the stratified calculation uses the same geometric approximation as the single-boundary calculation. Only the boundary coefficient is replaced. The single-boundary validation in Sec.~\ref{sec:formalism} therefore also tests this part of the construction.

The stack considered in this paper need not be the same beneath every incident point. Taking $n_j\to n_j(Q,\omega)$ and $d_j\to d_j(Q)$ evaluates each plane wave using the column of material beneath its own $Q$. Nothing in
Appendix~\ref{app:stack} changes, since the characteristic matrix is built one
column at a time. This requires each column to represent the medium over the
region that contributes coherently. Thus, the lateral scale of the
stratification must exceed the first Fresnel radius, $\rho_F\simeq178$~m at
$15^\circ$ elevation in the balloon geometry [Eq.~(\ref{eq:hierarchy})].

\section{The layered boundary coefficient}
\label{app:stack}

This appendix derives Eqs.~(\ref{eq:Mj})--(\ref{eq:tstack}) from Maxwell's
boundary conditions using the standard characteristic-matrix treatment of
stratified media~\cite{BornWolf}. The same matrix approach is used in
seismology~\cite{Thomson1950,Haskell1953} and X-ray reflectometry
\cite{Parratt1954}. We introduce the notation used in
Secs.~\ref{sec:formalism} and~\ref{sec:layered}, show the reduction to the
single-boundary limit, and specify the sign convention for $p$ polarization.
The construction follows directly from the continuity of the tangential
electric and magnetic fields at each interface.

\subsection{Local geometry and conserved tangential wavenumber}

At the incident point $Q$ of Fig.~\ref{fig:stack}, we describe the layered medium in the local tangent plane. The interfaces are concentric curved shells globally (Fig.~\ref{fig:geometry}), but for each plane wave they are treated locally as parallel layers
in this plane. Surface curvature enters the layer calculation through the plane wave incidence angle $\alpha_0$ of Eq.~(\ref{eq:alpha0}), with the validity of the parallel-layer treatment bounded by Eq.~(\ref{eq:appsagphase}). Let $\zeta$ be the local depth measured from the upper surface into the medium along the inward normal $-\hat{n}(Q)$. For concentric shells, this is the common normal to the layer interfaces at $Q$, and the thickness $d_j$ is measured along it. Beneath $Q$ lie the $N$ layers
of Sec.~\ref{sec:layered}, followed by a semi-infinite substrate of index $n_s$, as in panel (b) of Fig.~\ref{fig:stack}.

Because each layer is uniform parallel to the boundary, the tangential component of the wavevector is conserved across every interface. With $k=2\pi f/c$ as in Eq.~(\ref{eq:hertz}),
\begin{equation}
\kappa=k\,n_0\sin\alpha_0 ,
\label{eq:appkappa}
\end{equation}
marked in panel (b) of Fig.~\ref{fig:stack}, and it has the same value in every layer and in the substrate. This gives Snell's law,
\begin{equation}
n_0\sin\alpha_0=n_j\sin\alpha_j=n_s\sin\alpha_s ,
\label{eq:appsnell}
\end{equation}
so each plane wave in Eq.~(\ref{eq:weyl}) can be treated independently. The wavenumber normal to the boundary in layer $j$ follows from
$k_{z,j}^2+\kappa^2=k^2n_j^2$,
\begin{equation}
k_{z,j}=\sqrt{k^2n_j^2-\kappa^2}=k\,n_j\cos\alpha_j ,
\label{eq:appkz}
\end{equation}
with the branch chosen so that $\mathrm{Im}\,k_{z,j}\ge0$. Thus, a wave entering a lossy medium decays rather than grows. Traversing layer $j$ once gives the phase factor $e^{i\delta_j}$, where
\begin{equation}
\delta_j=k_{z,j}d_j=k\,n_jd_j\cos\alpha_j .
\label{eq:appdelta}
\end{equation}

\subsection{Reflection and transmission at a single interface}

We first consider a single interface, following the treatment of
Refs.~\cite{Prohira2018,DasguptaJain2021}. This shows how the characteristic
variable $q_j$ of Eq.~(\ref{eq:qdef}) follows from the boundary conditions. We
use the local tangent frame at $Q$, with $\hat z'$ along the outward normal
$\hat{n}(Q)$ and the plane of incidence in the $\hat x'$--$\hat z'$ plane.
Thus, $\hat y'$ is perpendicular to the plane of incidence. Let medium $a$ lie
above the interface and medium $b$ below. The three wavevectors are
\begin{align}
\vec k_i &= k n_a\left(\sin\alpha_a,\,0,\,-\cos\alpha_a\right), \nonumber\\
\vec k_r &= k n_a\left(\sin\alpha_a,\,0,\,+\cos\alpha_a\right), \label{eq:appkvecs}\\
\vec k_t &= k n_b\left(\sin\alpha_b,\,0,\,-\cos\alpha_b\right), \nonumber
\end{align}
and their common tangential component gives $n_a\sin\alpha_a=n_b\sin\alpha_b$,
which is Eq.~(\ref{eq:appsnell}). Since the interface is uniform in $\hat x'$
and $\hat y'$, the three waves have the same tangential phase. The boundary
conditions therefore give algebraic relations between their amplitudes.

For $s$ polarization the electric field is along $\hat y'$ and is entirely
tangential. Taking the incident amplitude to be unity,
\begin{equation}
E_{y'} = \left(1+r^{\,s}\right) \;\;\text{above},\qquad
E_{y'} = t^{\,s} \;\;\text{below},
\label{eq:appEs}
\end{equation}
at the interface. The magnetic field follows from
$\vec H=(\hat k\times\vec E)\,n/Z_0$, with $Z_0=\sqrt{\mu_0/\varepsilon_0}$ the
impedance of free space. Its tangential component is
$H_{x'}=\pm(n\cos\alpha)E_{y'}/Z_0$, with opposite signs for the down-going and
up-going waves. Thus,
\begin{equation}
\begin{split}
H_{x'} &\propto n_a\cos\alpha_a\left(1-r^{\,s}\right) \;\;\text{above},\\[2pt]
H_{x'} &\propto n_b\cos\alpha_b\, t^{\,s} \;\;\text{below}.
\end{split}
\label{eq:appHs}
\end{equation}
Continuity of $E_{y'}$ and $H_{x'}$ gives
\begin{equation}
1+r^{\,s}=t^{\,s},\qquad q_a\left(1-r^{\,s}\right)=q_b\,t^{\,s},
\label{eq:appmatchs}
\end{equation}
with $q\equiv n\cos\alpha$. Eliminating $t^{\,s}$ gives
$q_a(1-r^{\,s})=q_b(1+r^{\,s})$, and hence
\begin{equation}
r^{\,s}=\frac{q_a-q_b}{q_a+q_b},\qquad
t^{\,s}=\frac{2q_a}{q_a+q_b}.
\label{eq:appsolves}
\end{equation}

For $p$ polarization the magnetic field is along $\hat y'$ and is entirely
tangential, while the tangential electric field is
$E_{x'}=\mp(\cos\alpha/n)\,Z_0 H_{y'}$. Applying the same boundary conditions to
$H_{y'}$ gives
\begin{equation}
1+r^{\,p}=t^{\,p},\qquad q_a\left(1-r^{\,p}\right)=q_b\,t^{\,p},
\label{eq:appmatchp}
\end{equation}
now with $q\equiv\cos\alpha/n$. The solution is same as Eq.~(\ref{eq:appsolves}). Thus, the two polarizations differ only in the definition of $q$, given by Eq.~(\ref{eq:qdef}). Substituting $q=n\cos\alpha$
into Eq.~(\ref{eq:appsolves}) gives $f^{\,s}_r$ of Eq.~(\ref{eq:fs}).
Similarly, $q=\cos\alpha/n$ gives $f^{\,p}_r$ of Eq.~(\ref{eq:fp}) after
multiplying the numerator and denominator by $n_an_b$.

Because the $p$-polarization matching was written in terms of the magnetic
field, $t^{\,p}$ in Eq.~(\ref{eq:appsolves}) is a ratio of tangential
magnetic-field amplitudes. The corresponding electric-field ratio has the
additional factor $n_a/n_b$, which is the conversion used in
Eq.~(\ref{eq:wtransstack}).

The derivation does not require $n_b>n_a$, nor does it require either index to be real. The same two continuity conditions apply at each of the $N+1$ interfaces in the stack. The following sections combine these interface relations into the characteristic matrix.

\subsection{Interface coefficients of the stack}

Written for a general pair of media, the coefficients of
Eq.~(\ref{eq:appsolves}) are Eqs.~(\ref{eq:fs})--(\ref{eq:fp}) with
$(n_0,n_1,\alpha_0,\alpha_t)$ replaced by
$(n_a,n_b,\alpha_a,\alpha_b)$:
\begin{equation}
\begin{split}
r^{\,s}_{ab}&=\frac{n_a\cos\alpha_a-n_b\cos\alpha_b}
                   {n_a\cos\alpha_a+n_b\cos\alpha_b},\\[4pt]
t^{\,s}_{ab}&=\frac{2n_a\cos\alpha_a}
                   {n_a\cos\alpha_a+n_b\cos\alpha_b},
\end{split}
\label{eq:appfres_s}
\end{equation}
\begin{equation}
\begin{split}
r^{\,p}_{ab}&=\frac{n_b\cos\alpha_a-n_a\cos\alpha_b}
                   {n_b\cos\alpha_a+n_a\cos\alpha_b},\\[4pt]
t^{\,p}_{ab}&=\frac{2n_b\cos\alpha_a}
                   {n_b\cos\alpha_a+n_a\cos\alpha_b}.
\end{split}
\label{eq:appfres_p}
\end{equation}
As in Eq.~(\ref{eq:qdef}), $t^{\,p}_{ab}$ is a ratio of tangential
magnetic-field amplitudes; the corresponding electric-field ratio is
$(n_a/n_b)\,t^{\,p}_{ab}$, the conversion noted after Eq.~(\ref{eq:rhoint}).
Two identities that follow directly from these coefficients and are used below are
\begin{equation}
r_{ba}=-r_{ab},\qquad t_{ab}\,t_{ba}=1-r_{ab}^{\,2},
\label{eq:appstokes}
\end{equation}
for each polarization. Substituting Eqs.~(\ref{eq:appfres_s})--(\ref{eq:appfres_p})
gives $t_{ab}t_{ba}=4q_aq_b/(q_a+q_b)^2$ and
$1-r_{ab}^{\,2}=4q_aq_b/(q_a+q_b)^2$, with $q$ from Eq.~(\ref{eq:qdef}).
The identities therefore also hold for complex $n$. These are the usual Stokes
relations, and their derivation here does not require time reversal. This is
useful for the dissipative media considered in Sec.~\ref{sec:applications}.

\subsection{A single layer}
Consider one layer $(n_1,d_1)$ between the air and the substrate, as drawn in
panel (b) of Fig.~\ref{fig:stack}. The incident wave is partly reflected at the
upper interface, with amplitude $r_{01}$, and partly transmitted into the layer.
The transmitted wave crosses the layer, reflects from the lower interface with
amplitude $r_{1s}$, returns through the layer, and crosses back out with
amplitude $t_{10}$. The two traversals give a phase $e^{2i\delta_1}$. Repeated
internal reflections between the two interfaces add the remaining terms,
\begin{equation}
r_{\rm layer}=r_{01}
+t_{01}t_{10}r_{1s}e^{2i\delta_1}
\sum_{m=0}^{\infty}\left(r_{10}r_{1s}e^{2i\delta_1}\right)^{m}.
\label{eq:appseries}
\end{equation}
The amplitudes add coherently because all paths have the same tangential
wavenumber. For a passive medium the series converges, and summing it gives
\begin{equation}
r_{\rm layer}
=r_{01}+\frac{t_{01}t_{10}\,r_{1s}\,e^{2i\delta_1}}
              {1-r_{10}r_{1s}e^{2i\delta_1}} .
\label{eq:appsum}
\end{equation}
Using Eq.~(\ref{eq:appstokes}), with
$r_{10}=-r_{01}$ and $t_{01}t_{10}=1-r_{01}^{\,2}$, gives
\begin{equation}
r_{\rm layer}
=r_{01}+\frac{\left(1-r_{01}^{\,2}\right)r_{1s}e^{2i\delta_1}}
              {1+r_{01}r_{1s}e^{2i\delta_1}} .
\label{eq:appairymid}
\end{equation}
Combining the two terms over a common denominator gives
\[
r_{\rm layer}
=\frac{r_{01}+r_{01}^{\,2}r_{1s}e^{2i\delta_1}
       +\left(1-r_{01}^{\,2}\right)r_{1s}e^{2i\delta_1}}
      {1+r_{01}r_{1s}e^{2i\delta_1}} ,
\]
in which the two terms in $r_{01}^{\,2}r_{1s}e^{2i\delta_1}$ cancel, so that
\begin{equation}
r_{\rm layer}=\frac{r_{01}+r_{1s}e^{2i\delta_1}}
                   {1+r_{01}r_{1s}e^{2i\delta_1}}
\label{eq:appairy1}
\end{equation}
for $s$ and $p$ polarizations separately. The same sum with a single traversal
of the layer and an exit into the substrate gives the transmitted coefficient,
\begin{equation}
t_{\rm layer}=\frac{t_{01}t_{1s}e^{i\delta_1}}
                   {1+r_{01}r_{1s}e^{2i\delta_1}} .
\label{eq:apptrans1}
\end{equation}

Equation~(\ref{eq:appairy1}) includes all internal reflections. The result
holds whether the layer index is larger or smaller than that of the medium
above. The corresponding sign is set by the Fresnel coefficients in
Eqs.~(\ref{eq:appfres_s})--(\ref{eq:appfres_p}). In the limit $d_1\to0$,
$\delta_1\to0$, and $(r_{01}+r_{1s})/(1+r_{01}r_{1s})=r_{0s}$, as expected.

\subsection{Multiple layers}

For several layers, the same single-layer result can be applied from the
bottom upward. Let $R_j$ denote the reflection coefficient of everything
below the top of layer $j$. At the substrate,
\begin{equation}
R_{N+1}=r_{N,s},
\label{eq:apprecstart}
\end{equation}
and for $j=N,N-1,\dots,1$,
\begin{equation}
R_j=\frac{r_{j-1,j}+R_{j+1}\,e^{2i\delta_j}}
         {1+r_{j-1,j}R_{j+1}\,e^{2i\delta_j}}
\label{eq:apprecursion}
\end{equation}
for $s$ and $p$ separately, with $r_{j-1,j}$ from Eq.~(\ref{eq:rhoint}) and
$\delta_j$ from Eq.~(\ref{eq:appdelta}). The reflection coefficient of the
full stack is
\begin{equation}
r^{\,s,p}_{\rm stack}=R_1^{\,s,p}.
\label{eq:apprstack}
\end{equation}
The recursion includes the internal reflections between all interfaces. For $N=0$, it gives $R_1=r_{0s}$, recovering the single-boundary result of Eqs.~(\ref{eq:fs})--(\ref{eq:fp}).

\subsection{Matrix form}

The same layered solution can be written as a matrix product. Using
$q_j$ from Eq.~(\ref{eq:qdef}), the interface coefficients are
$r_{ab}=(q_a-q_b)/(q_a+q_b)$ and $t_{ab}=2q_a/(q_a+q_b)$. For $s$ polarization
this is the electric-field transmission coefficient. For $p$ polarization it
refers to the tangential magnetic field. The corresponding electric-field
ratio is obtained with the factor discussed above. The reflection coefficient
is unchanged because transmission enters the recursion only through
$t_{ab}t_{ba}=1-r_{ab}^{\,2}$.

Let $A^{+}_j$ and $A^{-}_j$ be the amplitudes of the waves traveling in layer
$j$ toward increasing depth and back toward the surface, and define
\begin{equation}
U_j=A^{+}_j+A^{-}_j ,
\qquad
V_j=q_j\left(A^{+}_j-A^{-}_j\right).
\label{eq:appUV}
\end{equation}
These combinations represent the two tangential field components. For $s$ polarization, the electric field is entirely tangential, so $E_\parallel\propto A^{+}+A^{-}=U$, while the tangential magnetic field is
$H_\parallel\propto q(A^{+}-A^{-})=V$. For $p$ polarization, the magnetic field is entirely tangential, so the roles of $U$ and $V$ are exchanged and $q$ takes the second form of Eq.~(\ref{eq:qdef}). In both cases, $U$ and $V$ are, up to a
common factor, the two tangential fields whose continuity is required by
Maxwell's boundary conditions. Thus, the same pair of variables can be used
across the entire stack.

Within layer $j$ the depth dependence is $e^{\pm ik_{z,j}\zeta}$, with $\zeta$ measured from the top of that layer. Writing the two amplitudes at depth $\zeta$ inside the layer,
\begin{equation}
\begin{split}
U(\zeta)&=A^{+}e^{ik_{z,j}\zeta}+A^{-}e^{-ik_{z,j}\zeta},\\[2pt]
V(\zeta)&=q_j\left(A^{+}e^{ik_{z,j}\zeta}-A^{-}e^{-ik_{z,j}\zeta}\right),
\end{split}
\label{eq:appUVz}
\end{equation}
where the subscript $j$ on $A^{\pm}$ is suppressed. Let $U_{\rm t},V_{\rm t}$ be the values at the top, $\zeta=0$, and $U_{\rm b},V_{\rm b}$ the values at the bottom, $\zeta=d_j$, so that the accumulated phase is $\delta_j=k_{z,j}d_j$ as in Eq.~(\ref{eq:appdelta}). Evaluating Eq.~(\ref{eq:appUVz}) at $\zeta=d_j$ and solving for the amplitudes gives
\begin{equation}
\begin{split}
A^{+}&=\tfrac{1}{2}e^{-i\delta_j}\left(U_{\rm b}+V_{\rm b}/q_j\right),\\[2pt]
A^{-}&=\tfrac{1}{2}e^{ i\delta_j}\left(U_{\rm b}-V_{\rm b}/q_j\right).
\end{split}
\label{eq:appAmp}
\end{equation}
Substituting these into Eq.~(\ref{eq:appUVz}) at $\zeta=0$, where $U_{\rm t}=A^{+}+A^{-}$ and $V_{\rm t}=q_j(A^{+}-A^{-})$, gives
\begin{align}
U_{\rm t}&=U_{\rm b}\cos\delta_j-i\,\frac{V_{\rm b}}{q_j}\sin\delta_j ,
\nonumber\\
V_{\rm t}&=-i\,q_jU_{\rm b}\sin\delta_j+V_{\rm b}\cos\delta_j .
\label{eq:appMjrows}
\end{align}
In matrix form,
\begin{equation}
\begin{split}
\begin{pmatrix}U_{\rm t}\\ V_{\rm t}\end{pmatrix}
&=M_j\begin{pmatrix}U_{\rm b}\\ V_{\rm b}\end{pmatrix},\\[4pt]
M_j&=
\begin{pmatrix}
\cos\delta_j & - i\sin\delta_j/q_j\\[3pt]
- i q_j\sin\delta_j & \cos\delta_j
\end{pmatrix},
\end{split}
\label{eq:appMj}
\end{equation}
which is Eq.~(\ref{eq:Mj}). The matrix therefore transfers the pair
$(U,V)$ from the bottom of a layer to its top using only the phase thickness
$\delta_j$ and the admittance $q_j$. No internal sampling of the layer is
needed, and its thickness enters analytically. Since
$\det M_j=\cos^2\delta_j+\sin^2\delta_j=1$, the product
\begin{equation}
M=\prod_{j=1}^{N}M_j ,
\qquad
\begin{pmatrix}U_0\\ V_0\end{pmatrix}
=M\begin{pmatrix}U_s\\ V_s\end{pmatrix},
\label{eq:appMprod}
\end{equation}
also has unit determinant, providing a useful numerical check.

Writing the four entries of $M$ as $m_{11},m_{12},m_{21},m_{22}$, we impose
the boundary conditions at the two ends of the stack. Above the stack there
is an incident and a reflected wave. Taking unit incident amplitude, so that
$A^{+}_0=1$ and $A^{-}_0=r$, Eq.~(\ref{eq:appUV}) gives
$U_0=1+r$ and $V_0=q_0(1-r)$. Below the stack only the transmitted wave is
present, so $A^{+}_s=t$ and $A^{-}_s=0$, giving $U_s=t$ and $V_s=q_st$.
Defining
\begin{equation}
X=m_{11}+m_{12}q_s ,
\qquad
Y=m_{21}+m_{22}q_s ,
\label{eq:appBC}
\end{equation}
the two rows of Eq.~(\ref{eq:appMprod}) become
\begin{equation}
1+r=Xt,\qquad q_0(1-r)=Yt .
\label{eq:appends}
\end{equation}
These are two linear equations for $r$ and $t$. Dividing the second by the
first eliminates $t$,
\[
\frac{q_0(1-r)}{1+r}=\frac{Y}{X}
\quad\Longrightarrow\quad
q_0X(1-r)=Y(1+r),
\]
Using $t=(1+r)/X$ then gives $t=2q_0/(q_0X+Y)$. In terms of the matrix entries,
\begin{align}
r^{\,s,p}_{\rm stack}
&=\frac{(m_{11}+m_{12}q_s)q_0-(m_{21}+m_{22}q_s)}
       {(m_{11}+m_{12}q_s)q_0+(m_{21}+m_{22}q_s)},
\label{eq:apprmat}\\[4pt]
t^{\,s,p}_{\rm stack}
&=\frac{2q_0}
       {(m_{11}+m_{12}q_s)q_0+(m_{21}+m_{22}q_s)},
\label{eq:apptmat}
\end{align}
where $q_0$ and $q_s$ are evaluated using the angles $\alpha_0$ and $\alpha_s$
from Eq.~(\ref{eq:appsnell}). These are
Eqs.~(\ref{eq:rstack})--(\ref{eq:tstack}) of the main text.

For $N=0$, the matrix product is the identity, so $X=1$ and $Y=q_s$. Equation
(\ref{eq:apprmat}) then gives $(q_0-q_s)/(q_0+q_s)$, which reduces to the
single-boundary Fresnel coefficients in Eqs.~(\ref{eq:fs})--(\ref{eq:fp}).
The recursion in Eq.~(\ref{eq:apprecursion}) and the matrix expression in
Eq.~(\ref{eq:apprmat}) agree to machine precision for the layered media used
in this work. For $p$ polarization we use $q_j=\cos\alpha_j/n_j$, consistent
with Eq.~(\ref{eq:fp}).

\subsection{Lossy and conducting layers}

The refractive indices may be complex. For a medium with relative permittivity $\varepsilon'_j$ and conductivity $\sigma_j$,
\begin{equation}
n_j^2=\varepsilon'_j+\frac{i\sigma_j}{\omega\varepsilon_0},
\label{eq:appcond}
\end{equation}
where $\omega=2\pi f$. Thus $n_j$, $\cos\alpha_j$,
$k_{z,j}$, and $\delta_j$ are generally complex. The factor
$e^{i\delta_j}$ then accounts for both propagation and attenuation,
$|e^{i\delta_j}|=e^{-\mathrm{Im}\,\delta_j}$, with skin depth
$1/\mathrm{Im}\,k_{z,j}$. Equations~(\ref{eq:appfres_s})--(\ref{eq:apptmat})
remain unchanged.

A lossy layer is therefore treated in the same way as a dielectric layer.
Each traversal is attenuated by $e^{i\delta_j}$, and a sufficiently absorbing
layer automatically suppresses contributions from deeper interfaces. No
additional treatment is required. We assume nonmagnetic media, $\mu_r=1$, as
appropriate for ice, firn, regolith, and seawater at radio frequencies.

\subsection{Scope of the approximation}

Within the local tangent plane, the layered calculation is exact. The fields in each layer, the boundary conditions at every interface, and all internal reflections are retained, with no expansion in layer contrast or thickness. Introducing stratification therefore adds no new geometric approximation. The only approximation is the treatment of the concentric shells as locally parallel layers, which is common to the single-boundary calculation of Refs.~\cite{Prohira2018,DasguptaJain2021} and is bounded in Appendix~\ref{app:cbound}.

\section{Stationary-phase window}
\label{app:spa}

At fixed Fourier frequency $f$, the angular integral of
Eq.~(\ref{eq:apperef}) can be written schematically for one polarization as
\begin{equation}
I^\chi=\int g^\chi(\alpha)\,r^\chi(\alpha)\,e^{i\Phi(\alpha)}\,d\alpha ,
\qquad \chi=s,p ,
\label{eq:appspaform}
\end{equation}
where $r^\chi$ is the boundary coefficient, either $f_r^{\,\chi}$ of
Eqs.~(\ref{eq:fs})--(\ref{eq:fp}) or $r_{\rm stack}^{\,\chi}$ of
Eq.~(\ref{eq:rstack}); $\Phi$ is the propagation phase $\Phi_r$ of
Eq.~(\ref{eq:appphaser}); and $g^\chi$ collects everything else in the
integrand, namely the source amplitude, the polarization weight, the measure
$\sin\alpha$, and the roughness factor. The azimuthal integral is suppressed
here so that the argument can be made in $\alpha$ alone.

The specular direction is the one where the phase is stationary. Writing $\Phi'$ and $\Phi''$ for the first and second derivatives of $\Phi$ with respect to $\alpha$, this is
\begin{equation}
\Phi'(\alpha_{\rm spec})=0 ,
\label{eq:appstatpoint}
\end{equation}
which implies that the path length $|SQ|+|QP|$ of Fig.~\ref{fig:geometry}
does not change to first order as the incident polar angle $\alpha$ is varied.
Here $\alpha_{\rm spec}$ is the value of $\alpha$ for the stationary ray,
related to the local incidence angle $\alpha_0^{\rm spec}$ at the specular
point by Eq.~(\ref{eq:alpha0}). Nearby plane waves have nearly the same phase
and therefore add coherently, while contributions farther from the stationary
direction oscillate and tend to cancel. The reflected field is consequently
concentrated around the specular point, defining the coherent footprint shown
in panel (a) of Fig.~\ref{fig:stack}.

We now quantify this range. We define $w$ as its half-width in $\alpha$, the
angular distance from $\alpha_{\rm spec}$ over which the plane waves add
coherently, so that the integral of Eq.~(\ref{eq:appspaform}) is effectively
confined to $|\alpha-\alpha_{\rm spec}|\lesssim w$. The width follows from the
next-order term in the phase. Expanding about the stationary point and using
Eq.~(\ref{eq:appstatpoint}) to remove the linear term,

\begin{equation}
\Phi(\alpha)\simeq\Phi_{\rm spec}
+\tfrac{1}{2}\Phi''(\alpha_{\rm spec})\,(\alpha-\alpha_{\rm spec})^2 .
\label{eq:appphaseexp}
\end{equation}
The contributions remain approximately coherent while the phase differs from
$\Phi_{\rm spec}$ by less than about $\pi$. Setting
$\tfrac12|\Phi''|\,w^2=\pi$ gives
\begin{equation}
w=\sqrt{\frac{2\pi}{|\Phi''(\alpha_{\rm spec})|}} .
\label{eq:appwindow}
\end{equation}
The scale of $\Phi''$ follows from the phase itself. Since
$\Phi=k(|SQ|+|QP|)$ and the path length changes over an angular scale of order
unity, $|\Phi''|$ is of order $kL$, where $L$ is the total path length. Thus,
\begin{equation}
w\sim\sqrt{\frac{2\pi}{kL}} ,
\label{eq:appwindowscale}
\end{equation}
as quoted in Sec.~\ref{sec:validity}. This width sets the angular range over
which the boundary coefficient needs to be evaluated. If the coefficient
varies little across this range, it can be evaluated at the specular angle,
leading to the factorization tested below.

The azimuthal integral has the same form, with its stationary point at
$\beta=0$, the azimuth of the observer. The concentration of the integrand
around the stationary direction in both angles is demonstrated numerically in
Refs.~\cite{Prohira2018,DasguptaJain2021}, by varying the ranges of the
$\alpha$ and $\beta$ integrations and following the convergence of the
integrated field.

The single-boundary and layered calculations have the same $g^\chi$ and $\Phi$.
If the ratio $r_{\rm stack}^{\,\chi}(f,\alpha)/f_r^{\,\chi}(f,\alpha)$ varies
little over this window, it can be taken outside the angular integral, giving
Eq.~(\ref{eq:factor}). This approximation is made separately at each frequency,
so the frequency dependence remains inside the Fourier sum for a broadband
waveform.

The error in this factorization is set by the variation of
$r_{\rm stack}^{\,\chi}/f_r^{\,\chi}$ across the stationary-phase window. For
the balloon geometry considered as an example, the full angular calculation
gives an accuracy of $0.2\%$, as shown in Sec.~\ref{sec:validity}. For source
heights close to the surface, the stationary-phase window becomes broader, so
the angular factorization must be checked directly. True near-field sources
also require the near-field source terms omitted here.

\FloatBarrier
\bibliography{paperA}

\end{document}